

\catcode`\@=11
\def\slash{\mathpalette\make@slash}
\def\make@slash#1#2{\setbox\z@\hbox{$#1#2$}%
  \hbox to 0pt{\hss$#1/$\hss\kern-\wd0}\box0}
\catcode`\@=12 


\magnification=\magstep1
\vsize=24truecm
\interlinepenalty=1000
\hsize=16.5truecm
\baselineskip=18pt
\hoffset=1truecm
\voffset=.5cm

 at 14.4 truept

\def\ov{\over}

\def\ha{{1\over 2}}
\def\tr{\,{\rm tr}\,}
\def\pr{\prime}
\def\Box{\Delta}

\def\pa{\partial}
\def\al{\alpha}

\def\lam{\lambda}
\def \ref #1{\lbrack#1\rbrack}
\def\be{\bar\eta}

\def\gam{\gamma}
\def\om{\omega}

\def\cd{{\cal D}}

\def\cs{{\cal S}}

\def\R{{\cal R}}

\def\>{\rangle}
\def\<{\langle}
\def\pb{\bar{\psi}}

\def\pd{\psi^{\dagger}}
\def\gf{\gamma_5}
\def\gamfive {\gamma _{\ss 5}}
\def\pan{\par\noindent}
\def\qw{\qquad\hbox{where}\quad}

\def\de{\delta}
\def\gd{g_{\mu\nu}}
\def\gu{g^{\mu\nu}}
\def\dl{\overrightarrow{\delta}}
\def\dr{\overleftarrow{\delta}}
\def\pl{\pi_\lam}

\def\pp{\pi_\psi}
\def\si{\sigma}\def\var{\varphi}
\def\Rind{\hat{\cal R}}
\def\gind{\hat g_{\mu\nu}}
\def\g{{g_{\mu\nu}}}
\def\lap{\triangle}
\def\ilap{{{1\over{\lap}}}}
\def\ibox{{{1\over{\nabla^2}}}}
\def\lapind{{\hat \lap}}
\def\parn{\hfill\break}

\def\dmu {\partial _{\mu}}
\def\dnu {\partial _{\nu}}
\def\spinmu {\omega _{\mu}}

\def\gammu {\gamma ^{\mu}}
\def\gamnu {\gamma ^{\nu}}
\def\beinmui {e^{\mu} _a }

\def\gaa { g_{{\scriptscriptstyle 1}}}
\def\gb {g_{{\scriptscriptstyle 2}}}
\def\gc {g_{{\scriptscriptstyle 3}}}

\def\ss {\scriptscriptstyle }
\def\gamtia { \tilde \gamma ^a }
\def\gamtib { \tilde \gamma ^b }

\def\fiveti{\tilde \gamma _{\ss 5}}

\def\mtx#1{\quad\hbox{{#1}}\quad}
\def\es{\!=\!}

\input lecproc.cmm

\contribution{Generalized Gauged Thirring Model on Curved Space-Times}
\author{A. Dettki@1, I. Sachs@2 and A. Wipf@2}
\address{@1 Max-Planck Institut f{\"u}r Physik, Werner-Heisenberg
Institut f{\"u}r Physik, P.O. Box 40 12 12, Munich, Germany
@2Institute for Theoretical Physics, Eidgen\"ossische Technische
Hochschule, H\"onggerberg, CH-8093 Z\"urich, Switzerland}
\topinsert\vbox{\hfill\hbox{ETH-TH/93-14}\smallskip\hrule}
          \vbox{\hfill\hbox{MPI-PH/93-18}\smallskip\hrule}\endinsert
\abstract{We analyse the interacting theory of charged fermions, scalars,
pseuso-scalars and photons propagating in 2-dimensional curved spacetime
in detail. For certain values of the coupling constants the theory
reduces to the gauged Thirring model and for others
the Schwinger model incurved spacetime. It is shown that the
interaction of the fermions with the pseudo-scalars shields the
electromagnetic interaction, and that the non-minimal coupling of the
scalars to the gravitational field amplifies the Hawking radiation. We
solve the finite temperature and density model by using functional
techniques and in particular derive the exact equation of state. The
explicit temperature and curvature dependence of the chiral condensate
is found. When the electromagnetic field is switched off the model reduces
to a conformal field theory. We determine the physically relevant
expectation values and conformal weights of the fundamental fields in
the theory.}

\titleb{1.}{Introduction}

The response of physical systems to a change of external conditions
is of emminent importance in physics. In particular the dependence
of expectation values on temperature, the particle density,
the space region, the imposed boundary conditions or external fields
has been widely studied [1]. Despite all these efforts we are still
unable to understand, for example, the mechanism leading to the spontaneous
symmetry breaking of the $SU_A(N)$ in low temperature $QCD$ [2].
Clearly such subtle effects require a better understanding of the
nonperturbative effects and in particular nonperturbative vacuum sector
of gauge theories. From our experience with $2$-dimensional gauge
theories [3] which we suppose to mimic one-flavour $QCD$ [4],
we are lead to believe that gauge fields with windings are
responsible for the non-vanishing chiral condensate and in particular
its temperature dependence. A related problem is how quantum systems
behave in a hot and dense environment as it exists or existed in
heavy ion collision, neutron stars or the early epochs of the universe
[2].\par
On another front there has been much effort to quantize selfinteracting
field theories in a background gravitational field [5].
For example, one is interested whether a black hole still emits thermal
radiation when self-interaction is included. Due to general arguments by
Gibbons and Perry [6] this question is intimately connected with universality
of the second law of thermodynamics.\par
Rather than seeking new partial results for more general and realististic
$4$-dimensional systems we have chosen an idealized
$2$-dimensional model with self-interaction to investigate
the questions mentioned and others. It is a
theory containing photons\fonote{Althouth photons in $1\!+\!1$ dimensions
possess no transversal degrees of freedom we still call them photons.
However, thru their interaction with charged fermions
they may become dynamical fields as exemplified by the Schwinger
mechanism.}, charged massless fermions, scalars and pseudoscalars in
interaction with themselves and a gravitational background field.
The model has the action
$$
\eqalign{S=\int\sqrt{-g}\Big[&-{1\over4}
F_{\mu\nu}F^{\mu\nu}+i\pb \gamma^\mu (\nabla_\mu- ig_1\pa_\mu\lam+ig_2
\eta_\mu^{\;\;\nu}\pa_\nu\phi )\psi\cr
&+g^{\mu\nu}(\pa_\mu\phi\pa_\nu\phi+\pa_\mu\lam\pa_\nu\lam )-g_3
{\cal R}\lam\Big], \cr}\eqno(1.1)
$$
where $F_{\mu\nu}$ is the electromagnetic field strength, the
gamma-matrices in curved space are related to the flat ones as
$\gamma^\mu\es e^\mu_{\,a}\hat\gamma^a$, $\nabla_\mu
\es\pa_\mu\!+\!i\omega_\mu\!-\!ieA_\mu$ is the generally and gauge
covariant derivative containing the $U(1)$ gauge
potential and spin connection,
$\eta_{\mu\nu}\es \sqrt{-g}\,\epsilon_{\mu\nu}$ denotes the totally
antisymmetric tensor and ${\cal R}$ the Ricci scalar. The gravitational
field $g_{\mu\nu}$ (or rather the $2$-bein $e^a_{\,\mu}$, since the
theory contains fermions) is treated as classical background field,
whereas the 'photons' $A_\mu$, 'electrons' $\psi$, scalars $\lambda$
and pseudoscalars $\phi$ are fully quantized. In two dimensions
the electric charge $e$ has the dimension of a mass. The other $3$
couplings are dimensionless. The classical theory is invariant
under $U(1)$ gauge- and axial transformations and correspondingly
possesses conserved vector and axial-vector currrents.
\par
We have chosen this
model since it allows to address the above raised questions and
since it relates to known soluble models for certain values
of the coupling constants. For example, it contains the
{\it gauged Thirring model}, the {\it Schwinger model in curved
space time} and the {\it minimal models} in conformal field theory
as particular limits. Its vacuum
structure is non-trivial, i.e. it has $\theta$-vacua like
more realistic $4$-dimensional gauge theories [7].
The 'photon' aquires a mass $m_\gam^2=e^2/(\pi+\ha g_2^2)$ via
the Schinger mechanism. For finite volumes the theory possesses
instantons which mimimize the euclidean action. These instantons
lead to chirality violating vacuum expectation values. For example,
a non-zero chiral condensate develops which only for high temperature
and large curvature vanish exponentially.\par
For $e\es 0$ all coupling
constants are dimensionless and the theory becomes conformally invariant.
In this limit the vacuum structure becomes trivial.
Despite its complexity the general model (1.1) is solvable for
arbitrary classical backgrounds $g_{\mu\nu}$ and allows for an
analytical treatment. This in turn enables the entire stress tensor
in {\it any} curved space, the induced currents, their correlators
and the equation of state to be constructed.\par
The physical role of the coupling constants is the following:
The coupling of $\phi$ to the transversal current decreases the effective
electromagnetic interaction between fermions. For example, the electric
charge becomes renormalized to $e_{\ss R}=e/\sqrt{1+g_2^2/2\pi}$, the
chiral condensate decreases as $\sim (2\pi\!+\!g_2^2)^{-\ha}$.
The mass in the bosonised theory depends on $g_2$.
In the ungauged sector the Kac-Moody central extension, conformal
weights and $U(1)$ charges dependend on $g_2$.
The coupling constant $g_3$ amplifies the Hawking radiation which
remains thermal for the interacting model. It is $(3+24\pi g_3^2)$ times
as strong as that of a free massless scalar field. The central
charge and conformal weights depend on $g_3$. Actually, the
weights of the fermionic fields become complex for $g_3\!\neq\! 0$.
However, $g_3$ does not enter in the finite size effects.
The coupling constant $g_1$ to the longitudinal current weakens the
long range gauge invariant electron-electron correlators
in the one-instanton sector (see 6.27).
In the ungauged sector it enters in expectation values of local
operators and in
particular in the short distance expansions of the fermionic
fields and energy momentum tensor. It does not influence the
thermodynamics of the model.
\par
Since for particular choices of the coupling constants the
model reduces to wellknown and wellstudied exactly soluble
models there are many earlier works which are related
to ours. Some of them concentrated more on the gauge sector
and investigated the renormalization of the electric charge
in the gauged Thirring model by the four-fermi interaction
[8] or the non-trivial vacuum structure in the
Schwinger model [3,9]. Others concentrated
on the ungauged conformal sector. Freedman and Pilch calculated the
partition function of the ungauged Thirring model on arbitrary
Riemann surfaces [10]. We do not agree with their
result and in particular show that there is no
holomorphic factorization for general
fermionic boundary conditions. Also we deviate from Destri and deVega [11]
which investigated the ungauged model on the cylinder
with twisted boundary conditions.
We shall comment on the discrepancies in sections $3$ and $7$.
Other papers which are relevant and are dealing with
different aspects of certain limiting cases of (1.1) are
[12], where the thermodynamics of the Thirring model
has been studied or [5] in which the Hawking radiation
has been derived.\par
The paper is organized as follows: In section $2$ we analyse the
classical model to prepare the ground for the quantization. In particular
we derive the general solution of the field equations, discuss the
conservation laws and investigate the limiting theories. By employing the
graded structure we derive the classical Poissson (anti) commutators
of the fundamental fields with the energy momentum tensor.
In the following section we quantize
the finite temperature model. To avoid infrared problems
we assume space to be finite. Together with the finite temperature
boundary conditions we are lead to considering the theory on
the $2$-dimensional {\it euclidean torus}. Due to the twists in the
fermionic boundary conditions, the non-trivial vacuum structure
and the associated instantons and fermionic zero-modes the quantization
is rather subtle. Actually we show that some of the results
in the literature are incorrect. In section $4$ the general
results are applied to derive the partition function of the
gauged model. Its dependence on the spatial size, temperature
and gravitational field is explicitly found.
In section $5$ we show that for equal couplings the gauged model on curved
spacetime can be bosonized. It turns out that only the non-constant parts
of the currents can be bosonized and that for this part
the wellknown flat spacetime rules need just be covariatized. In the
following section the chiral symmetry breaking is studied. The exact form
of the chiral condensate is found. On the flat torus the
formula simplifies to (6.13). Various limits, e.g. $L\to\infty,$
$T\to 0$, $T\to \infty$ or $g_2\to \infty$ are investigated.
By comparing the temperature and curvature dependence of the
condensate we derive an effective curvature induced temperature.
In section $7$ the thermodynamics of the ungauged model is
studied. We derive the ground state energy and its dependence
on the coupling constants, size of the system and boundary
conditions. We compute the equation of state and our result
does not agree with [12].
In the last section we investigate the conformal sector of
(1.1), that is the ungauged model in flat spacetime.
Besides the Virasoro algebra the model contains an $U(1)$ Kac-Moody
algebra. We calculate the important commutators and in particular
determine the conformal weights and $U(1)$-charges
of the fundamental fields from first principles. Also we show
that the finite size effects are in general not proportional
to the central charge as has been conjectured by Cardy [13].
The appendix $A$ contains our conventions and scaling formulae for the
various geometrical objects. In appendix $B$ we collected some useful
variational formulae which we have used in this work.
In appendix $C$ we derive the partition function within
the canonical approach.

\titlea{2.}{Classical theory}
\titlec{2.1}{Equations of Motion}
The field equations of the model (1.1) are
$$
\eqalign{
&i\gam^\mu (\nabla_\mu-i\gaa\pa_\mu\lam+i\gb\eta_{\mu\nu}
\pa^\nu\phi )\; \psi \equiv i\gam^\mu D_\mu= \; 0\cr
&2\;\nabla^2 \lam =-\gc\R- \gaa\nabla_\mu j^\mu\cr
& 2\;\nabla^2\phi=-\gb \nabla_\mu j^{{\ss 5}\mu} \cr
& \nabla_\nu F^{\mu\nu}=e\,j^{\mu}
 \; , \cr}
\eqno (2.1)
$$
which are the Dirac equation for massless charged fermions
propagating in a curved space-time and interacting with
the scalar and pseudoscalar-fields, Klein Gordon type of equations
and Maxwell equation. Here $j^{{\ss 5}\mu}$ is the axial vector current
which is defined by
$$
j^{{\ss 5}\mu}\es
\pb \gam^\mu \gamfive \psi \es\eta^\mu_{\;\;\nu}j^\nu.\eqno(2.2)
$$
When one decomposes the gauge field as
$$
A_\mu=\dmu\al-\eta_{\mu \rho} \partial^{\rho} \var
\quad\quad\hbox{so that}\quad F_{01}=\sqrt{-g}\nabla^2 \var,\eqno(2.3)
$$
and choses isothermal coordinates for which
$g_{\mu\nu}\es e^{2\si}\eta_{\mu\nu}$, then the
generalized Dirac operator reads
$$\eqalign{
\slash D =&\;e^{iF-i\gamfive G-{3\over 2}\si}\;\slash\partial\;
e^{-iF-i\gamfive G+\ha\si},\mtx{where}\cr
&F=\gaa\lam+e\;\al\mtx ,G=\gb\phi+e\;\var\;.\cr} \eqno(2.4)
$$
Hence, if $\psi_{\ss 0} (x)$ solves the free Dirac equation
in flat Minkowski space time, then
$$
\psi (x) \equiv e^{ iF + i \gamfive G - \ha  \si}
\psi_0 \eqno (2.5)
$$
solves the Dirac equation of the ineracting theory on curved
spacetime. The vector currents are related as
$$
j^\mu=\pb \gam^\mu\psi  = \pb_{\ss 0} \hat \gam^\mu\psi_{\ss 0}
e^{-2\si}\equiv  {1 \over \sqrt{- g}}\;j^\mu_{\ss 0}     \; .
\eqno (2.6)
$$
The same relation holds for the axial vector current. From
$\sqrt{-g}\nabla_\mu j^\mu \es \dmu \sqrt{-g}j^\mu $
the conservation of the vector and axial currents follow at once,
$$
\nabla_\mu j^\mu = \nabla_\mu j^{{\ss 5}\mu} = 0 \; ,  \eqno (2.7)
$$
expressing the classical $U(1)\times U_A(1)$ invariance of the model.
Using (2.7) in (2.1) we conclude that
$$
2\nabla^2\lam=-\gc\R\mtx{and}\nabla^2\phi=0\eqno (2.8)
$$
or that there is no backreaction from fermions onto scalars.
Finally the conservation laws imply that the currents are free
fields
$$
\nabla^2 j^\mu = \;\nabla^2 j^{{\ss 5}\mu} = \; 0 \; ,\eqno(2.9)
$$
which is the reason which accounts for the solubility
of the model [14], even in the presence of photons and an
external gravitational field.
As is wellknown, for any gauge invariant regularization the
axial current possesses an anomalous divergence in the quantized
model and (2.9) is modified. Thus the normal $U_A(1)$ Ward identities
in the ungauged Thirring model [8] become anomalous when the fermions
couple to a gauge field.

In isothermal coordinates the {\it general solution of the field equations}
can be expressed in terms of $6$ chiral functions as follows:
Introducing lightcone coordinates $x^\pm\es x^0\!\pm\! x^1$ so that
$ds^2\es e^{2\sigma}dx^+dx^-$, the solutions of (2.8) read
$$
\lam=g_3\sigma+\lam_+(x^+)+\lam_-(x^-),\mtx{and}
\phi=\phi_+(x^+)+\phi_-(x^-)\eqno(2.10)
$$
and depend on $4$ chiral functions which are fixed by
the initial data on some spacelike hypersurface.
The solutions of the free Dirac equations depend on $2$ chiral
functions as
$$
\psi_0=\pmatrix{\psi_-(x^-)\cr\psi_+(x^+)\cr}.\eqno(2.11)
$$
In these coordinate system the Maxwell equations (2.1) can
easily be integrated and one finds
$$
\pa_+\pa_-\phi=F_{01}=2e\,^{2\sigma}\Big[\int\limits^{x^-}
\psi^\dagger_-(\xi)\psi_-(\xi)d\xi
-\int\limits^{x^+}\psi^\dagger_+(\xi)\psi_+(\xi)d\xi\Big]
.\eqno(2.12)
$$
To go further we must fix the gauge. Conveniently one chooses the
Lorentz gauge such that $\al\es 0$ in (2.3) and thus $\phi$ in (2.12)
determines $A_\mu$. We see that in isothermal coordinates
and this gauge the general solution of (2.1)
is given by (2.10), (2.12) and (2.5), that is in terms
of $6$ chiral functions.\par

Besides the currents the symmetric energy-momentum tensor
of the matter fields
$$
T^{\mu\nu}\equiv -  {2\ov\sqrt{g}}{\de\cs\ov\de g_{\mu\nu}}
 \eqno (2.13)
$$
plays an important role in any theory in curved space time.
Applying the variational identities in Appendix B one
obtains after a lengthy but straightforward computation
$$\eqalign{
T^{\mu\nu}
=\;&{1\over4} g^{\mu \nu} F^{\si \rho}F_{\si \rho}
-F^{\si \nu}F_{\si}^{\;\;\mu}+
{i\over 2}  \bigl [ \pb\gam^{(\mu}D^{\nu)}\psi -
(D^{(\mu}\pb)\gam^{\nu)}\psi\bigr ]\cr &
+2 \nabla^\mu\phi\nabla^\nu\phi -\gu \nabla^\al\phi\nabla_\al\phi
\quad + \quad (\phi \leftrightarrow \lam ) \cr &
-2 \gc ( g^{\mu\nu}\nabla^2 - \nabla^{\mu}\nabla^{\nu})\lam \cr&
+\ha j^{\mu}\;(g_1\nabla^{\nu}\lam-g_2\eta^{\nu \al}\nabla_\al\phi
 ) \quad + \quad (\mu \leftrightarrow\nu)\cr
&+\gb \gu j^\al\eta_{\al\beta}\nabla^\beta\phi-2g_2 j^\al\eta_\al\,^{(\mu}
\nabla^{\nu)}\phi \; ,
\cr}\eqno (2.14)
$$
where we have introduced the symmetrization $A^{(\mu }B^{\nu) }\es
\ha (A^\mu B^\nu + A^\nu B^\mu)$.
The first two lines are just the energy momentum
of the electromagnetic field, charged
fermions and free neutral (pseudo-) scalars.
The terms containing second derivatives of $\lam$ are
the improvement terms [15] which
are always present when one couples scalars non-minimally to a background
curvature. The remaining terms reflect the
interaction between the fermionic and auxiliary fields.\par
On shell $T^{\mu\nu}$ is conserved as required by general covariance.
Using the field equations for $\psi$ and $\lam$ its trace
reads
$$
T^{\mu}_{\;\mu}=
\; \gc ^2 \R - \ha F^{\si \rho}F_{\si \rho}\; .\eqno (2.15)
$$
In particular for $\gc \es 0 $ and $A_\mu \es 0$
it vanishes, and the theory becomes Weyl-invariant.
As a consequence it reduces to a conformal field theory in the
flat spacetime limit [16]. It is remarkable that
it can be made Weyl invariant even when
$\gc\!\neq\!0$. Indeed, without changing the
flat spacetime limit we may add a nonlocal Wess-Zumino-type
term to the action, namely
$$
S^\pr= S-{\gc^2\ov 4}S_p\mtx{where} S_p=\int \sqrt{-g}\R\ibox\R
 \eqno (2.16)
$$
the variation of which is
$$
\eqalign{
\delta S_p= \int\Big\{& 4\lbrack g^{\mu\nu}\R
-\nabla^{\mu}\nabla^{\nu}\ibox\R\rbrack +2\nabla^{\mu}\left(
\ibox \R\right)\nabla^{\nu }\left(\ibox\R\right)\cr
&-g^{\mu\nu}\nabla_{\al}\left(\ibox \R\right)\nabla^{\al}
\left(\ibox\R\right)\Big\}\;\sqrt{-g}\delta g_{\mu\nu}\cr}\eqno(2.17)
$$
The trace of the modified energy momentum tensor is now zero,
and for $g_{\mu \nu}\! \to \! \eta _{\mu \nu}$ the Lagrangian
corresponds to a conformal field theory in Minkowski spacetime
. \par \noindent
Choosing the coupling constants appropriately, the model
reduces to various well known exactly solvable models:
\item -  For the special choice
$\gaa \es \gb \es e \es 0 $ and for vanishing gauge field
the $\lam $- dependent
part of  (1.1) is just the Lagrangian of scalar fields coupled
to a background charge and and for imaginary $g_3$
describes the {\it minimal models} of conformal field theory [17].
\medskip
\item - For $\gc \es 0 $ and $\gaa^2\es -\gb^2 \es g^2 $
the fermionic sector reduces to the  gauged version of the {\it Thirring
model} [18] in curved space time.
To see that we solve the Klein Gordon equations in
(2.1) for the $U(1)$ current which yields
$$
j_{\mu}=-{2\over g_1}\dmu\lam-{2\over g_2}\eta_\mu^{\;\;\nu}\dnu\phi.
\eqno(2.18)
$$
Inserting this into the Dirac equation we find
$$
i\gam^\mu \nabla_\mu
\psi - {g^2\ov 2} j^{\mu}\gam_{\mu}\psi =0 \; ,\eqno (2.19)
$$
which is the field equation of the gauged Thirring model
in curved spacetime with Lagrangean
$$
{\cal L}[\pb,\psi ] = \pb i\gam^\mu \nabla_\mu  \psi -
{g^2\ov 4} j^{\mu}j_{\mu}- {1\over 4} F_{\mu \nu} F^{\mu \nu}\; .
\eqno (2.20)
$$
\item - If we further specialize to $g\es 0$ we recover
the {\it Schwinger model in curved spacetime} [9].
\par
In the following sections we are lead to consider the {\it euclidean
version} of the model. Then one must replace the lorentzian
$\gam^\mu,g_{\mu\nu}$ and $\om _\mu$ by there euclidean
counterparts. For example, with our conventions (see appendix A)
the relation (2.2) becomes
$$
j^{{\ss 5}\mu}=-i\eta^\mu_{\;\;\nu}\,j^\nu\eqno(2.21)
$$
and as a consequence the generalized Dirac operator in euclidean
spacetime becomes
$$
\slash D = \; e^{ iF +\gamfive G - {3\over 2} \si}\;
\slash \partial \;e^{-iF+\gamfive G +\ha \si}
\eqno(2.22)
$$
instead of (2.4). Also, to recover the euclidean Thirring model
as particular limit of (1.1) we must
set $g_3=0$ and $\gaa^2\es\gb^2\es g^2$.

\titlec{2.2}{Hamiltonian formalism and classical conformal structure}
In this subsection we investigate the Hamiltonian structure of
the model (1.1) in the conformal limit, i.e. in flat Minkowski space
and for vanishing gauge field. In the presence of both
fermions and bosons it is convenient to exploit the graded
Poisson structure [19]. We recall, that the equal time
Poisson brackets are
$$
\{A(x),B(y)\}\equiv \sum_O\int dz^1
\Bigl({A(x)\dr\ov\de \pi_O(z)}{\dl B(y)\ov\de O(z)}
\mp{A(x)\dr\ov\de O(z)}{\dl B(y)\ov\de\pi_O(z)}\Bigr)
\Big|_{x^0=y^0}\;.\eqno(2.23a)
$$
The sum is over all fundamental fields $O(x)$ in the theory .
The sign is minus if one or both of the fields $A$ and $B$ are
bosonic (even) and it is plus if both are fermionic (odd).
The momentum densities $\pi_O(x)$ conjugate to the $O$-fields are
given by functional left-derivatives
$$
\pi_O(x)={\dl S\ov \delta\pa_0 O(x)}.\eqno(2.23b)
$$
A simple calculation yields the following momenta
$$
\pi_\psi=-i \psi^\dagger,\quad
\pi_\phi=\gb j^{\ss 5}_0+2 \pa_0\phi\mtx{and}
\pi_\lam=\gaa j_0+2\pa_0\lam\eqno(2.24)
$$
which form the fundamental Poisson brackets with the fields
$$\eqalign{
\{\psi^{\ss \dagger}_\al(x),\psi_\beta(y)\}&=i\delta_{\al\beta}
\delta(x^1-y^1),\cr\{\pi_\lam(x),\lam(y)\}=\delta(x^1-y^1),\quad&\quad
\{\pi_\phi(x),\phi(y)\}=\delta(x^1-y^1).\cr}\eqno(2.25)
$$
For the Hamiltonian we obtain
$$\eqalign{
H&=\int dx^1\Big[\pa_0\psi\pp+\pa_0\lam\pl+\pa_0\phi\pi_\phi-{\cal L}\Big]\cr
&=\int dx^1 \Big[\pp\gam_5\pa_1\psi-ig_1\pa_1\lam\pp\gam_5\psi
-ig_2\pa_1\phi \pp\psi+(\pa_1\lam)^2\cr &
\qquad\qquad+(\pa_1\phi)^2+{1\ov 4}(\pl-ig_1\pp\psi)^2 +{1\ov 4}
(\pi_\phi-ig_2\pp\gam_5\psi)^2\Big].\cr}\eqno(2.26)
$$
It can be checked that the corrresponding Hamiltonian equations
are just the field equations (2.1) with flat metric and vanishing
gauge potential, as required. Since $T^\mu_{\;\;\mu}\es 0$ (see
2.15) the only non-zero components of $T^{\mu\nu}$ are the
lightcone components $T_{++}$ and $T_{--}$. To continue it is convenient
to introduce adapted light cone coordinates
$$
x^\pm=x^0\pm x^1\mtx{so that} \pa_{\pm}=\ha(\pa_0\pm \pa_1),\eqno(2.27)
$$
and the chiral components of the Dirac spinor $\psi_\pm=\ha(1\pm\gamfive)
\psi$. Then $T_{--}$ in (2.14) simplifies to
$$\eqalign{
T_{--}=& -{1\ov 2}(\pi_{\psi_+}\pa_-\psi_+-\pa_-\pi_{\psi_+}\psi_+)
+2(\pa_-\lam)^2+2(\pa_-\phi)^2\cr &\qquad\quad
+2g_3\pa_-^2\lam+i\pa_-(g_1\lam+g_2\phi)\pi_{\psi_+}\psi_+.\cr}
\eqno(2.28)
$$
Using the equations of motion one shows explicitly that it is a chiral
field, i.e. depends only on $x^-$. With (2.28) we can
now find the conformal weights of the fundamental fields which
determine their transformations under infinitesimal
conformal symmetry transformations.
For that we must calculate the commutator of the symmetry generators
$T_f=\int dx^-f(x^-)T_{--}$ with the fields. The result
is
$$\eqalign{
\{T_f,\phi\}&=f\pa_-\phi\cr
\{T_f,\lam\}&=f\pa_-\lam-{g_3\ov 2}\pa_-f\cr
\{T_f,\psi_+\}&=f\pa_-\psi_++\ha(1-ig_1g_3)\psi_+\pa_-f\cr
\{T_f,\psi_+^{\ss\dagger}\}&=f\pa_-\psi_+^{\ss\dagger}+\ha(1+ig_1g_3)
\psi_+^{\ss\dagger}\pa_-f.\cr}
\eqno(2.29)
$$
Whereas $\phi$ and $\psi_+$ are primary fields, $\lam$
is not. Actually, the non-primary character of $\lam$ is very much
linked with the $g_3$-dependent term in the transformation of the Dirac
field. To see that more clearly we note that under an infinitesimal
left conformal transformation generated by $\bar T_f=\int
dx^+f(x^+)T_{++}$ the scalar and fermi field transform as
$$
\{\bar T_f,\lam\}=f\pa_-\lam-{g_3\ov 2}\pa_-f\mtx{and}
\{\bar T_f,\psi_+\}=f\pa_-\psi_+-ig_1g_3\psi_+\pa_-f.\eqno(2.30)
$$
Since $\psi_+$ is not any longer a scalar under left transformation
the term
$$
\int dx^+dx^-\Big(2i\psi_+^{\ss \dagger}(\pa_+-ig_1\pa_+\lam)\psi_+
\Big)
$$
appearing in the action is only conformally invariant because
$\lam$ transforms inhomogenously like a spin connection.
It may be surprising that the symmetry transformations depend on the
coupling constant $g_3$ which is not present in the flat space time
Lagrangean. However, the same happens for example in $4$ dimensions
if one couples a scalar field conformally, that is non-minimally, to
gravity. Although the Lagrangeans for the minimally and conformally
coupled particles are the same on Minkowski spacetime, their
energy momentum tensors are not. The same happens for the conformally
invariant nonabelian Toda theories wich admit several energy
momentum tensors and hence several conformal structures [20].
\pan
The current transforms as
$$
\{T_f,j_-\}=f\pa_-j_-+j_-\pa_-f\eqno(2.31)
$$
and the energy momentum tensor as
$$
\{T_f,T_{--}\}=f\pa_-T_{--}+2T_{--}\pa_-f-g_3^2\pa_-^3f.\eqno(2.32)
$$
Recalling that a primary field $O$ with weight $h$ transforms as
$$
\{T_f,O\}=f\pa_-O+hO\pa_-f
$$
and comparing with the above results we have found the following
structure:\parn
\item{-} The pseudoscalar field $\phi$ is primary with $h_\phi\es 0$.
The scalar field $\lam$ is only primary for $g_3\es 0$ in which case
$h_\lam\es 0$.
\item{-} The Dirac field $\psi_+$ is primary with $h_{\psi_+}
\es \ha(1-ig_1g_3)$. The conformal weight is real for imaginary $g_3$.
\item{-} The current is primary with weight $1$.
\item{-} Already on the classical level the energy momentum tensor
is only quasi-primary. The corresponding Virasoro algebra (2.32)
has central charge $c\es 24\pi g_3^2$.



\def\s3{\sigma^3}
\def\bt{\bar\tau}
\def\mtx#1{\quad\hbox{{#1}}\quad}
\titlea{3.}{Quantization of the generalized gauged Thirring model}
In this section we quantize the general model
(1.1) in curved space-times. The results are then applied
in the following sections, where we calculate
the {\it partition function}, {\it ground state energy}, {\it equation
of state} and certain {\it correlators} of interest
and their dependence on the chemical potential,
volume of space, temperature and background metric.
To do that we couple the conserved U(1)-charge to a chemical
potential $\mu$. We enclose the system in a
box with length $L$ to avoid infrared divergences. To investigate the
temperature dependence the time is taken to be purely imaginary
in the functional approach [21]. The imaginary time $x^0$ varies
then from zero to the inverse temperature $\beta$ and
we must impose periodic- and antiperiodic boundary conditions
for the bosonic- and fermionic fields, respectively.
Thus to study the finite temperature model we must assume
that space-time is an euclidean torus $[0,\beta]\times[0,L]$.\par
To see how the partition function and correlators depend
on the gravitational field we assume that the torus is equipped
with an arbitrary metric with euclidean signature
or equivalently with a $2$-bein $e_{\mu a}$. The curved gamma
matrices are $\gam_\mu\es e_{\mu a}\hat\gam^a$ and in particular
$\gamfive\es-{i\over 2}\eta_{\mu\nu}\gam^\mu\gam^\nu\es
\sigma_3$ is constant (see appendix A for our conventions). We can always
choose (quasi) isothermal coordinates and a Lorentz frame such that
$$
e_{\mu a}= e^\sigma\,\hat e_{\mu a}\equiv
e^\sigma\pmatrix{\tau_0&\tau_1\cr 0  &1\cr}
\mtx{or}
g_{\mu\nu}=e^{2\sigma}\,\hat g_{\mu\nu}\equiv
e^{2\sigma}\pmatrix{\vert\tau\vert^2 & \tau_1\cr \tau_1&1\cr},
\eqno(3.1)$$
where $\tau\es\tau_1+i\tau_0$ is the Teichmueller
parameter and $\sigma$ the gravitational Liouville field. Space-time
is then a square of length $L$ and has volume
$V\es\int_0^L d^2x \sqrt{g}$. We allow for
the general twisted boundary conditions for the fermions
$$\eqalign{
\psi(x^0+L,x^1)&= -e^{2\pi i(\al_0+
\beta_0\gf)}\psi(x^0,x^1)\cr
\psi(x^0,x^1+L)&=- e^{2\pi
i(\al_1+\beta_1\gf)}\psi(x^0,x^1).\cr} \eqno(3.2)
$$
The parameters $\al_i$ and $\beta_i$ represent
vectorial and chiral twists, respectively. We could allow
for twisted boundary conditions for the (pseudo) scalars as well, e.g.
$\phi(x^0\!+\!nL,x^1\!+\!mL)=\phi(x^1,x^0)+2\pi(m\!+\!n)$. However,
to recover the Thirring model for certain values of the
couplings we assume that these fields are periodic. For $\sigma\es 0$,
$\tau\es i\beta/L$ and $\al_0\es \beta_0\es0$ the partition
function has then the usual thermodynamical interpretation.
Its logarithm is proportional to the free energy at temperature
$T\es 1/\beta$.\par

\titleb{3.1}{Fermionic path integral}
Twisted boundary conditions as in (3.2) require some care in the
fermionic path integral. Indeed the fermionic determinant
is not uniquely defined when one allows for such twists.
The ambiguities are not related to the unavoidable ultra-violett
divergences but to the transition from Minkowski- to Euclidean
space-time. To see that more clearly let ${\cal{S^\pm}}$
denote the set of fermionic fields
in {\it Minkowski space-time} with chirality $\pm1$.
Since both the commutation
relations and the action
do not connect ${\cal{S^+}}$ and ${\cal{S^-}}$ we
can consistently impose different boundary conditions
on ${\cal{S^+}}$ and ${\cal{S^-}}$.
On the other hand, in the {\it euclidean pathintegral}
for the generating functional
$$
Z_F[\eta,\bar\eta]=\int \cd \pd\cd \psi \;
e^{\int\sqrt{g}\,\pd i\slash D \psi+\int \sqrt{g} \,\big(\bar\eta
\psi+\pd\eta\big)},\eqno(3.3)
$$
the Dirac operator
$$
\slash D =\pmatrix{0&D_-\cr D_+&0\cr}
$$
exchanges the two chiral components of $\psi$, i.e.
$\slash D :{\cal{S^\pm}}\rightarrow {\cal{S^\mp}}$. Thus, in contrast
to the situation in Minkowski space the two chiral
sectors are related in the action. Of course, the eigenvalue
problem for $i\slash D$ is then not well defined. This
is the origin of the ambiguity in the definition of
the determinant. It is related to the ambiguities one
encounters when one quantizes chiral fermions [22]. To solve
this problem we shall analytically continue the
well-defined determinants in the untwisted sector $\beta\es 0$
to $\beta\!\neq\! 0$. The resulting determinants do not
factorize into (anti-) holomorphic pieces and differ
from previous ones in the literature [10].
In appendix C we give further arguments in favour of our result
by calculating the determinants in a different way.
\par
Let us now study the generating functional for fermions in an external
gravitational and gauge field and coupled to the auxiliary fields.
For that we observe that on the torus the decompositon (2.2)
of the gauge potential generalises to
$$
A_\mu=A^I_\mu+{2\pi\over L}t_\mu+\pa_\mu\al-\eta_{\mu\nu}\pa^\nu\var,
\eqno(3.4a)
$$
where the last $3$ terms are recognized as Hodge decomposition
of the single valued part of $A$ in a given topological
sector, that is the harmonic-, exact- and coexact pieces.
In arbitrary coordinates the toron field $t_\mu$ obeys the
harmonic conditions $\nabla_\mu t^\mu\es t_{[\mu,\nu]}\es 0$.
It follows then that in isothermal coordinates $t_\mu$ must
be constant. The role of the toron fields has recently
been emphasized within the canonical approach [23].
In the Hamiltonian formulation they are quantum mechanical
degrees of freedom wich are needed for an understanding
of the infrared sector in gauge theories. Also, in
[24] it has been demonstrated that the $Z_N$-phases of hot
pure Yang-Mills theories [25] should correspond to the
same physical state if one takes care of the toron fields.\par
The first term in (3.4a) is an {\it instanton potential}
which gives rise to a nonvanishing quantized flux $\Phi$ or integer-valued
instanton number $k$:
$$
\Phi=e\int F_{01}\equiv e\int E =e\int E^I= 2\pi\,k.\eqno(3.5)
$$
As representative in the $k$-instanton sector we choose the,
up to gauge transformations, {\it unique absolute minimum}
of the Maxwell action in (1.1).
It has field strength $e\,E^I=\sqrt{g}\,\Phi/ V$.
As instanton potential we choose
$$
e A_\mu^I=e\hat A_\mu^I-\Phi\,\eta_\mu^{\;\;\nu}\pa_\nu\chi,\mtx{where}
e\hat A^I=-{\sqrt{\hat g}\ov \hat V}\big(x^1,0\big)\eqno(3.4b)
$$
is the instanton potential on the flat torus with the same
flux but field strength $\sqrt{\hat g}\,\Phi/ \hat V$. The function
$\chi$ is then determined (up to a constant) by
$$
\sqrt{g}{\Phi\ov V}-\sqrt{\hat g}{\Phi\ov\hat V}=\sqrt{g}\lap\chi.
\eqno(3.4c)
$$
The solution of this equation is given by
$$
\chi(x)=-{1\ov \hat V}\big({1\over\lap}e^{-2\sigma}\big)(x)
={1\ov \hat V}\int d^2y \sqrt{g(y)}G_0(x,y)\,e^{-2\sigma (y)},
\eqno(3.4d)
$$
where
$$
G_0(x,y)=\langle x\vert {1\ov -\lap}\vert y\rangle=\sum_{\lam_n>0}
{\phi_n(x)\phi^\dagger_n (y)\ov \lam_n}\eqno(3.6a)
$$
is the Greenfunction belonging to $-\lap$. In deriving (3.4d) we
have used that ${1\ov\lap}(\Phi/V)\es 0$ which follows from the
spectral resolution (3.6a) for the Green function in which the
constant zero mode $\phi_0\es 1/\sqrt{V}$ of $\lap$ is missing.\par\noindent
Note that $2$-dimensional gauge theories are not scale or Weyl
invariant as $4$-dimensional ones are. For that reason the instantons
on conformally flat spacetimes are not just the 'flat' instantons.\par
To be more explicit
we relate $G_0$ to the Greenfunction $\hat G_0$ on the flat
torus with the hatted metric [26]
$$
\hat G_0(x,y)=-{1\over 4\pi}\log\vert {1\ov \eta(\tau)}
\Big[{\ha+{\xi^0\ov L}\atop
\ha+{\xi^1 \ov L}}\Big](0,\tau)\vert^2,\mtx{where}
\xi=x-y.\eqno(3.6b)
$$
For that  we note that due to the missing zero-mode in (3.6a)
the usual flat spacetime
equations for the Greenfunctions are modified to
$$
-\lap_x G_0(x,y)={\delta(x\!-\!y)\ov \sqrt{g}}-{1\ov V}\quad,\quad
-\hat\lap_x \hat G_0(x,y)={\delta(x\!-\!y)\ov \sqrt{\hat g}}-{1\ov \hat V}.
\eqno(3.7a)
$$
Furthermore one sees at once
that both Green functions annihilate the corresponding constant zeromodes
$$
\int d^2y \sqrt{g(y)}G_0(x,y)=\int d^2y \sqrt{\hat g}\hat G_0(x,y)=0.
\eqno(3.7b)
$$
{}From these two equations one concludes that Greenfunction
on the curved torus is related to the flat one (3.6b) as
$$\eqalign{
G_0(x,y)&=\hat G_0(x,y)
+{1\ov V^2}\int d^2ud^2v \sqrt{g(u)g(v)}\hat G_0(u,v)\cr
&-{1\ov V}\int \hat G_0(x,u)\sqrt{g(u)}d^2u
-{1\ov V}\int d^2u \sqrt{g(u)} \hat G_0(u,y)\cr}\eqno(3.8)
$$
and this replaces the infinite space relations
$G_0=\hat G_0$ [27].\par
Our choice for the instanton potential (3.4)
corresponds to a particular trivialization of the $U(1)$-bundle
over the torus [3]. In other words, the gauge potentials
and fermion fields at
$(x^0,x^1)$ and $(x^0,x^1\!+\!L)$ are necessarily related
by a {\it nontrivial gauge transformation} with windings
$$\eqalign{
&A_{\mu}(x^0,x^1+L)-A_\mu(x^0,x^1)=\pa_\mu\al (x)\cr
&\psi(x^0,x^1+L)=-e^{ie\al (x)}\,e^{2\pi i(\al_1+\beta_1\gamfive)}
\,\psi(x^0,x^1).\cr}\eqno(3.9a)
$$
For the choice (3.4b) we find
$$
e\al(x)=-{\Phi\over L}\,x^0.\eqno(3.9b)
$$
Note that $A$ is still
periodic in $x^0$ with period $L$ and $\psi$ still obeys the
first boundary condition in (3.2). Our trivialization differs
from the one chosen in [28] and so do our instantons and
fermionic zero modes.
\par
Similarly as for the gauge potential we must add a harmonic piece to the
auxiliary vector field $B_\mu$ to which the fermions couple in (1.1),
so that
$$
B_\mu={2\pi\over L}g_0 h_\mu+g_1\pa_\mu\lam-g_2\eta_{\mu\nu}\pa^\nu\phi
\eqno(3.10)
$$
appears in the Dirac operator in (1.1) on the torus. $\lam$ and $\phi$
couple to the divergence of the vector and axial vector currents. The
harmonic fields $h_\mu$ couple to the harmonic part of the current and are
needed to recover the Thirring model in the limit
$g_0^2\es g_1^2\es g_2^2$. Also,
we shall see that $t_\mu$ and $h_\mu$ are essential to obtain the correct
answer for the thermodynamic potential.
Note that $B_\mu$ contains no instanton part since it
couples to the gauge invariant fermionic current.\par
Finally we introduce a {\it chemical potential} for the conserved
$U(1)$ charge. In the euclidean functional approach
this is equivalent to coupling the fermions to a constant
imaginary gauge potential $A_0$ [29].
\par
Inserting the above decompositions and the chemical potential
into the Dirac operator finally yields in isothermal coordinates
$$\eqalign{
&\slash D =\gamma^\nu D_\nu=e^{iF+\gamfive (G+\Phi\chi)
-{3\over 2}\sigma}\,\slash {\hat D} \,
e^{-iF+\gamfive (G+\Phi\chi)+\ha\sigma},\mtx{where}\cr
&\qquad\slash {\hat D} =\gam^\mu\big(\pa_\mu+i\hat\omega_\mu
-ie \hat A^I_\mu-{2\pi i\over L}[H_\mu+\mu_\mu]\big),\cr
&\qquad H_\mu=e\, t_\mu+g_0 h_\mu\mtx{and} \mu_\mu=
-i{\tau_0L\over2\pi}\mu\;\delta_{\mu 0}.\cr}\eqno(3.11)
$$
Here $\hat\omega$ is the spin connection belonging to
$\hat e_{\mu a}$. It vanishes for our choice of the
reference zweibein. $\hat A^I$ is the instanton potential
(3.4b) on the flat torus.
The scalar and pseudoscalar functions
$F$, $G$ and $\chi$ have been introduced in (2.4) and (3.4d).
In the chosen coordinates $t$ and $h$ and hence $H$ are all
constant. In [3] it has been shown that $\slash D$ possesses
$\vert k\vert$ zero-modes of definite chirality and
their chirality is given by the sign of $k$.
They are crucial in any correct quantization. For example,
if one would leave out instanton sectors in which $i\slash D$
has zero-modes then the cluster property would be violated.\par
In a first step we quantize the fermions in the flat instanton
and harmonic background and reference metric $\hat g_{\mu\nu}$,
that is we assume $\slash D\to\slash {\hat D}$ in (3.3).
The dependence on the remaining fields $F,G,\chi$ and $\sigma$,
that is the relation between $Z_F$ and $\hat Z_F$,
is then found by integrating the chiral and trace
anomalies [30] and exploiting
the relation (3.11) between $\slash D$ and $\slash {\hat D}$.\par\noindent
We expand the fermionic
field in a orthonormal basis of the Hilbert space
$$
\eqalign{
\psi(x) &= \sum_n a_n\psi_{n+}(x)+ \sum_n b_n\psi_{n-}(x)\cr
\psi^\dagger(x) &= \sum_n \bar a_n\chi_{n+}^\dagger (x)+\sum_n
\bar b_n\chi_{n-}^\dagger(x),\cr}\eqno(3.12)
$$
where $a_n,b_n,\bar a_n,\bar b_n$ are independent Grassmann variables.
\titlec{}{Topologically trivial sector}
For $k=0$ or vanishing instanton potential
we can immediately write down a basis
$$
\psi_{n\pm}(x)= {1\ov\sqrt{V}}\;e^{i(p^\pm_{n},x)}\,e_\pm
,\mtx{where} (p^\pm_n)_i={2\pi\ov L}
(\ha+\al_i\pm\beta_i+n_i),\eqno(3.13)
$$
and $e_\pm$ are the eigenvectors of $\gamfive$. The $\psi_{n+}$
and $\psi_{n-}$ must obey the ${\cal S}^+$ and ${\cal S}^-$
boundary conditions, respectively. These boundary conditions
fix the admissable momenta $p_n^\pm$ in (3.13).
Since the Dirac operator maps ${\cal S}^\pm$ into ${\cal S}^\mp$ the
$\chi_{n\pm}$ must then obey the same boundary conditions as the
$\psi_{n\mp}$. Thus $\chi_{n\pm}(x)$ is obtained from
$\psi_{n\pm}(x)$ by exchanging $p_n^+$ and $p_n^-$. It follows
then that
$$
i\slash {\hat D}\psi_{n\pm}=\lam_n^\pm\chi_{n\mp}\eqno(3.14a)
$$
with
$$\eqalign{
\lam_n^+=& {2\pi\ov
\tau_0L}[\bt(\ha+a_1+\beta_1+n_1)-(\ha+a_0+\beta_0+n_0)]\cr
\lam_n^-=& {2\pi\ov
\tau_0L}[\tau(\ha+a_1-\beta_1+n_1)-(\ha+a_0-\beta_0+n_0)]
.\cr}\eqno(3.14b)
$$
Here we have introduced $a_\mu \equiv \al_\mu \!-\!H_\mu\!-\!\mu_\mu$.
Substituting (3.12,14) into the generating functional
(3.3) and applying the standart Grassmann integration
rules we arrive at
$$\eqalign{
&\hat Z_F[\eta,\be]= \det\,i\slash {\hat D}
\ e^{-\int\be(x) \hat S(x,y)\eta(y)},\quad
\det i\slash {\hat D}=\prod_n\lam_n^+\lam_n^-,\cr
&\qquad\hat S(x,y)=\sum\limits_n\big({\psi_{n+}(x)\chi_{n-}^\dagger(y)\ov
\lam_n^+}+{\psi_{n-}(x)\chi_{n+}^\dagger (y)\ov \lam_n^-}\big).\cr}
\eqno(3.15)
$$
$\hat S$ is the fermionic Green function in the $0$-instanton sector.
Note that both the 'eigenvalues' and the Green function
depend on the Teichmueller parameter, harmonic potentials,
twists and chemical potential.\par
We proceed to calculate the
infinte product or generalized determinant in (3.15).
This is one of the central points of our article and
for non-zero chiral twists and chemical potential our
result deviates from previous ones [10].
Actually the twists and chemical potential
are related as one can see from (3.14).
One may be tempted so identify
$$
\det (D_+D_-)\sim\prod\lam_n^+\lam_n^-\mtx{and}
\det D_+\det D_-\sim\prod\lam_n^+\prod\lam_m^-\eqno(3.16)
$$
and thus conclude that the determinant is a product,
$f(\tau)\bar f(\tau)$, that is factorizes into holomorphic
and anti-holomorphic pieces (the overall factor $\sim 1/\tau_0 L$
in the eigenvalues (3.14b) drops in the infinite product, since the torus
has vanishing Euler number).
However, the infinite product in (3.15) must be
regularized and the two expressions in (3.16) may differ.
In conformal field theory [26] one is naturally lead to
consider the individual chiral sectors
and thus finds holomorphic factorization. For Dirac fermions
one uses $\slash D^2$ to regularize the product and this
leads to the determinant of the product $D_+D_-$.\par
To continue we recast the infinite product in the form
$$
\prod^\infty\lam_n^+\lam_n^-=\prod_{{\vec{n}}\in Z^2}
\Big({2\pi\ov L}\Big)^2\hat g^{\mu\nu}(\ha+c_\mu+n_\mu)(\ha+c_\nu+n_\nu)
\eqno(3.17a)$$
where $\hat g^{\mu\nu}$ is the inverse of the reference metric
(3.1) and
$$
c_\mu=a_\mu+i\hat\eta_\mu\,^\nu\beta_\nu,\mtx{where}
(\hat\eta_\mu\,^\nu)=-{1\ov\tau_0}\pmatrix{\tau_1&-|\tau|^2\cr 1&-\tau_1\cr}.
\eqno(3.17b)
$$
The point is that for real $c_\mu$, that is for vanishing chiral
twists $\beta_\mu$ and chemical potential (see the definitions
of $a_\mu$ below (3.14b) and $\mu_\mu$ in (3.11)) the
zeta function defined by
$$
\zeta(s)=\sum_n\big(\lam_n^+\lam_n^-\big)^{-s}\eqno(3.17c)
$$
has a well defined analytic continuation to $s\!<\!1$ via
a Poisson resummation. An explicit calculation yields [3,31,38]
$$\eqalign{
&\det(i\slash {\hat D})\equiv \big(\prod_n\lam_n^+
\lam_n^-\big)_{reg}=e^{-\zeta^\pr(s)|_{s=0}},\qw\cr
&\zeta^\pr(s)\big|_{s=0}= -\log\Big[ {1\over |\eta(\tau)|^2}
\Theta\Big[{-c_1\atop
c_0}\Big](0,\tau)\ \bar\Theta\Big[{-c_1\atop
c_0}\Big](0,\tau)
\Big].\cr} \eqno(3.18)
$$
However, for complex $c_\mu$ the Poisson resummation
is not applicable and $\zeta^\pr (0)$ cannot be calculated
by direct means. To circumvent these difficulties
we note that the infinite product (3.17c) defining the
$\zeta$-function for $s\!>\!1$ is a meromorphic function in $c$.
Thus we may first continue to $s\!<\!1$ for real $c_\mu$
and then continue the result to complex values.
Using the transformation properties of theta functions
the resulting determinant can be written as
$$\eqalign{
\det(i\slash {\hat D})  =& e^{2\pi(\sqrt{\hat
g}\hat\gu\beta_\mu\beta_\nu-2i\beta_1a_0)}\cr
&\cdot {1\over |\eta(\tau)|^2}\Theta\Big[{-a_1+\beta_1\atop
a_0-\beta_0}\Big](0,\tau)\bar\Theta\Big[{-\bar a_1-\beta_1\atop
\bar a_0+\beta_0}\Big](0,\tau).\cr}\eqno(3.19)
$$
It can be shown that this
determinant is {\it gauge invariant}, i.e. invariant under
$\al_\mu\rightarrow \al_\mu\!+\!1$, but not invariant under chiral
transformatins, $\beta_\mu\rightarrow \beta_\mu\!+\!1$, as expected.
Furthermore it transforms covariantly under modular
transformations $\tau\to\tau+1$ and $\tau\to -1/\tau$.
In other words, $\det i\slash { \hat D}$ is invariant under modular
transformations if at the same time
the boundary conditions are transformed accordingly.
The exponential prefactor is needed for modular covariance and
is not present in the literature [10].
It correlates the two chiral sectors and will have important
consequences. In the appendix C we confirm
(3.19) with operator methods.

\titlec{}{Topologically nontrivial sectors}
For definiteness we assume $k>0$. Then
$i\slash { \hat D}$ possesses $k$ zero-modes $\hat\psi^p_{0+}$, $p\es
1,\dots,k$ with positive chirality and ${\cal S}^+$ boundary
conditions. Together with the excited modes
$\psi_{n+}$ they form a basis of ${\cal S}^+$.
Thus we must add the zero-mode contribution $\sum c_n\hat\psi^p_{0+}$
to $\psi$ in (3.12).
Similarly we must add $\bar c_n\hat\chi^{p\dagger}_{0+}$ to
$\psi^\dagger$ in (3.12). The zero modes $\hat\chi^p_{0+}$ in $\psi^\dagger$
must obey
${\cal S}_-$ boundary conditions (see the discussion below (3.13)).
Thus the zero-modes in the expansions of $\psi$ and $\psi^\dagger$ have
the same chirality but obey different boundary conditions.
This is required for the zero- and excited modes to form a complete
basis and is consistent
since $i\slash D$ does not relate the zero-mode sector of ${\cal S}^+$
with ${\cal S}^-$. The Grassmann integral over the variables
belonging to the excited modes is performed as in the trivial
sector. Also, the integration over the $c_n$ and $\bar c_n$ can
easily be done and one obtains
$$\eqalign{
&\quad\hat Z_F[\eta,\be]=\prod_{p=1}^{\vert k|}(\bar\eta,\hat\psi^p_{0+})
(\hat\chi^p_{0+},\eta)\;{\det}^\pr i\slash { \hat D} \
e^{-\int\be(x) \hat S_e(x,y)\eta(y)},\cr
&{\det}^\pr i\slash { \hat D}=\prod_{\lam^\pm_n\neq 0}\lam^+_n\lam^-_n,\quad
\hat S_e(x,y)= \sum_{\lam_n\neq 0}\Big({\psi_{n+}(x)
\chi_{n-}^\dagger (y)\ov \lam_n^+}+(+\leftrightarrow -)\Big).\cr}
\eqno(3.20)
$$
Note that the excited Green function $S_e$ anticommutes with $\gamfive$.
\par
To calculate the determinant we observe that
$$
\slash D^2=\pmatrix{D_-D_+&0\cr 0&D_+D_-\cr}=
{1\over \sqrt{g}}D_\mu\sqrt{g}g^{\mu\nu}D_\nu-{1\over 4}\R
+{e\ov 2}\eta^{\mu\nu}F_{\mu\nu}\gamfive\eqno(3.21)
$$
simplifies in the instanton background $\hat A^I$ and on the flat torus
to
$$
-\slash { \hat D}^2=-\hat g^{\mu\nu}\hat D_\mu\hat D_\nu -{\Phi\over \hat V}
\gamfive.\eqno(3.22)
$$
In other words, it is the same in
the two chiral sectors, up to the constant $2\Phi/\hat V$.
This observation allows one to reconstruct the spectrum
of $-\slash { \hat D}^2$ as follows:\par\noindent
First note that we can define two sets of
normalizable zero-modes in the positive chirality
$(\gamfive\es 1)$ sector. One containing $k$ modes
obeying the ${\cal S}^+$ boundary conditions
and the other consisting of $k$ modes with the ${\cal S}^-$ boundary
conditions. The first set is admissable and are just the $k$ zero
modes $\hat\psi^p_{0+}$ appearing in (3.20). The other
$k$ zeromodes $\hat\chi^p_{0+}$ appear also in (3.20) but
the Dirac operator does not act on them.
But because of (3.22) they are at the
same time eigenmodes of $-\slash { \hat D}^2$ in the negative chirality
sector with the correct ${\cal S}^-$ boundary conditions and eigenvalues
$2\Phi/\hat V$. Of course, the $\hat\psi^p_{0+}$ are also
eigenmodes in the $\gamfive\es -1$ sector but with the
wrong boundary conditions. However, applying
$\hat D_-$ to them produces $k$ eigenmodes in the positive chirality
sector with the correct boundary conditions and eigenvalues $2\Phi/\hat V$.
This procedure may now be iterated and one ends up with the
following spectrum of $-\slash { \hat D}^2$:
$$
\lam_n^2=\cases{0 & deceneracy $=k$ \cr 2n\Phi/\hat V&
deceneracy $=2k.$\cr}\eqno(3.23)
$$
With the explicit spectrum at hand we can
compute the zero-mode truncated determinant
with zeta-function methods and find [3]
$$
{\det}^\pr (i\slash { \hat D})=\Big({\pi\hat V\over\Phi}\Big)^{\Phi/4\pi}.
\eqno(3.24)
$$
We proceed with computing the {\it zero modes} of $\slash { \hat D}^2$. For
that
we note that the operator commutes with the time translations
which leads to the ansatz
$$
\tilde\chi_p=e^{2\pi ic_px^0/L}\,e^{2\pi iH_1x^1/L}\;\xi_p(x^1)\,e_+,
\quad c_p=p+(\ha+\al_0+\beta_0),\eqno(3.25a)
$$
where we have assumed $k> 0$. The choice of $c_p$ is
dictated by the time-like boundary conditions in (3.2).
Inserting this ansatz into the zero mode equation
$\slash { \tilde D}^2\tilde\chi_p=0$ yields
$$\eqalign{
&\big(|\tau|^2{d^2\over dy^2}-{\Phi^2\over L^4}y^2-
2i\tau_1{\Phi\ov L^2}y{d\over dy}-i\tau{\Phi\over L^2}\big)
\xi_p =0,\cr
&\mtx{where} y=x^1+{L\over k}(c_p-H_0-\mu_0).\cr}
$$
This is just the differential equation
for the ground state of a generalized harmonic
oscillator to which it reduces for
$\tau=i\tau_0$. The solution is given by
$$
\xi_p=\exp\Big[-{\Phi \over {2i\bar\tau L^2}}
\big\{x^1+{L\over k}(c_p-H_0-\mu_0)\big\}^2
\Big].
$$
These functions do not obey the boundary condition (3.9),
but the correct
eigenmodes can be constructed as superpositions of them.
For that we observe that
$$
\tilde\chi_p(x^0,x^1\!+\!L)=e^{-i\Phi x^0/\beta}\,e^{2i\pi H_1}\;
\tilde\chi_{p+k}(x^0,x^1)
$$
so that the sums
$$\eqalign{
\hat\psi^p_{0+}&={(2k\tau_0)^{1\over 4}\over \sqrt{|\tau|\hat V}}
e^{\pi\mu_0^2\ov k\tau_0}\,e^{2\pi i(H_0-\al_0-\ha)\beta_1}\,
\cr&\cdot
\sum_{n\in Z}\,e^{-2i\pi (n+p/k) (\ha+\al_1+\beta_1-H_1)}\tilde\chi_{p+ n k}
\,e_+,\cr}\eqno(3.25b)
$$
where $p\es 1,\dots,k$, obey the boundary conditions and thus are the
$k$ required zero-modes. We have chosen the phase
such that the accompanying zero-modes in (3.20) are just
$$
\hat\chi^p_{0+}(x,\beta,\dots)=\hat\psi^p_{0+}(x,-\beta,\dots).\eqno(3.25c)
$$
Actually the product $\hat\chi^p_{0+}\hat\psi^p_{0+}$ is only
determined up to a (possibly $\al$ and $\beta$-dependent) phase.
But gauge invariance requires that
$$
\chi^{p\dagger}_{0+}(x)\,\psi^p_{0+}(x)\mtx{and}
\chi^{p\dagger}_{0+}(x)\,\exp\big[ie\int\limits_y^x A\big]\,\psi^p_{0+}(y)
$$
are both invariant under $\al_\mu\to\al_\mu\!+\!n_\mu$ and
$et_\mu\to et_\mu\!+\!n_\mu$, where the $n_\mu$ are integers. This almost
fixes the relative phases of the zero-modes in (3.25b) and (3.25c).
Also, the overall factor normalizes these functions to one. Modes
with different $p$ are orthogonal to
each other, so that the system (3.25b) forms an
orthonormal basis of the zero-mode subspace.
For $k\!<\!0$ the zero-modes are the same if one replaces
$\beta_\mu$ by $-\beta_\mu$ and $e_+$ by $e_-$.
\titlec{}{Integrating the chiral and trace anomalies}
To relate the determinants of $i\slash { \hat D}$ and $i\slash D$ we
introduce the one-parameter family of Diracoperators
$$
\slash D_\tau=e^{\tau[iF+\gamfive (G+\Phi\chi)
-{3\over 2}\sigma]}\,\slash { \hat D}\,
e^{\tau[-iF+\gamfive (G+\Phi\chi)+\ha\sigma]}\eqno(3.26)
$$
which intepolates between $\slash { \hat D}$ and $\slash D$ [30]. The
$\tau$-derivative of the corresponding determiants
is determined by the chiral and trace anomaly. An
explicit calculation yields
$$
\log{{\det}^\pr i\slash D\over {\det}^\pr i\slash { \hat D}}=
{1\over 4\pi}\int\limits_0^1d\tau\int \sqrt{g^\tau}
\tr a_1^\tau \Big(2\gamfive[G+\Phi\chi]-\sigma\Big)+
\log\det{{\cal N}_\psi\over \hat{\cal N}_\psi}.\eqno(3.27)
$$
Here $g^\tau$ is the determinant of the deformed
metric $g^\tau_{\mu\nu}=e^{2\tau\sigma}\hat g_{\mu\nu}$,
and
$$
a_1^\tau=-{1\over 12}R^\tau+\gamfive \tau\triangle^\tau G
+{1\over \sqrt{g^\tau}}\,\Big[(1-\tau)\sqrt{\hat g}\,{\Phi\over \hat V}
+\tau\sqrt{g}\,{\Phi\over V}\Big]\gamfive\eqno(3.28)
$$
is the relevant Seeley-deWitt coefficient of $\slash D^2_\tau$.
Furthermore, $\hat{\cal N}_\psi$ is the norm-matrix of the
zero-modes $\hat\psi^p_{0+}$ in (3.25b). Since those
are orhonormal it is just the $k$-dimensional
identity matrix. ${\cal N}_\psi$ is the norm-matrix of the
zero-modes of $i\slash D$ which are related to the
$\hat\psi^p_{0+}$ as
$$
\psi^p_{0+}=
e^{iF-\gamfive (G+\Phi\chi)-\ha\sigma}\hat\psi^p_{0+}\eqno(3.29)
$$
as follows from (3.11). Inserting (3.28) into (3.27)
one finds the following formula for the determinant
in arbitrary background gravitational and gauge fields:
$$\eqalign{
&{\det}^\pr i\slash D = \det{{\cal N}_\psi
\ov \hat{\cal N}_\psi}\;{\det}^\pr (i\slash { \hat D})
\exp\Big({S_L\over 24\pi}\Big)\cr
&\quad\cdot\exp\Big({1\over 2\pi}\int\sqrt{\hat g}
G\hat\lap G+{2k\ov V}\int\sqrt{g}G+{\Phi^2\ov 2\pi\hat V}\int \sqrt{\hat g}
\chi\Big),\cr}\eqno(3.30a)
$$
where
$$
S_L=\int\sqrt{\hat g}\big[\hat\R\sigma-\sigma\hat\lap\sigma\big]
\eqno(3.30b)
$$
is the {\it Liouville action.} In deriving this result we used
that $\int \sqrt{g}\chi\es 0$.
Actually, for our reference metric the Ricci scalar $\hat\R$
vanishes and the Liouville action simplifies to
$-\int\sqrt{\hat g}\sigma\hat\lap\sigma$.
However, as it stands the formula (3.30) holds for arbitrary
reference metrics and arbitrary Riemannian surfaces.
\par
As expected for a gauge-invariant regularisation,
the function $F$ and thus the pure gauge part
of the vector potential does not appear in the
determinant. \par
For later use we also give the analogous
formula for the zero-mode truncated scalar determinant [32]
$$
{\det}^{\pr\ha}(-\lap)
={\det}^{\pr\ha} (-\hat\lap)\big({V\ov \hat V}\big)^\ha
\exp\Big(- {1\over 24 \pi} S_L\Big).\eqno(3.31)
$$
This completes the computations of the determinants.\par
The {\it generating functional for the full theory} is then obtained
as follows:\par\noindent
First one notes that the formulae (3.15) and
(3.20) for the fermionic functionals still
hold without hats. Thus to calculate the functionals in
arbitrary gauge-, auxiliary- and gauge fields we need
to know the Greenfunctions, determinants and zero-modes
in these backgrounds.\par\noindent
To relate the fermionic Greenfunctions $S$ in the different
topological sectors to the hatted ones we define
$$
S_1(x,y)=e^{-g(x)}\;\hat S(x,y)\,e^{-\bar g(y)},\quad
g=-iF+\gamfive (G+\Phi\chi)+\ha\sigma.\eqno(3.32a)
$$
On the infinite space we would have $S=S_1$ [27]. However,
if the Diracoperator possesses zero modes this simple
relation is modified to
to
$$\eqalign{
&S(x,y)=S_1(x,y)+\int P_0(x,u)S_1(u,v)P_0(v,y)\sqrt{g(u)g(v)}d^2ud^2v\cr
&\;\;-\int S_1(x,u)P_0(u,y)\sqrt{g(u)}d^2u
-\int P_0(x,u)S_1(u,y)\sqrt{g(u)}d^2u,\cr}\eqno(3.32b)
$$
and this formula should be compared with the analogues one
for scalars (3.8). Here
$P_0$ is the orthonormal projector onto the zero modes.
For gauge fields with vanishing flux $S\es S_1$.
Together with the relation (3.30) between the full and hatted
determinant and the explicit form (3.18,19) for $\det i\slash { \hat D}$
this yields the fermionic generating functional
in the various toplogical sectors.\par\noindent
In the {\it trivial sector} one finds explicitly
$$ \eqalign{
Z_F[&\eta,\be]={1\over |\eta(\tau)|^2}\Theta\Big[{-c_1\atop
c_0}\Big](0,\tau)\ \bar\Theta\Big[{-\bar c_1\atop
\bar c_0}\Big](0,\tau)\cr
&e^{-\int\be(x) S(x,y)\eta(y)}\cdot\exp \Big ( {1\over 24 \pi} S_L
+{1\over 2\pi}\int \sqrt{g}G\lap G\big]\Big).\cr}
\eqno(3.33)
$$
By using the scaling properties of the Ricci-scalar
and Laplacian (see appendix B) the exponent can be rewritten
as
$$
-{1\over 96\pi}\int\sqrt{g}\R{1\over \lap}\R+{1\over 2\pi}
\int \sqrt{g} G\lap G,
$$
which makes clear that the resulting functional is
diffeomorphism invariant. Here we used that
$\R$ integrates to zero or that the Euler number of the torus vanishes.
On the sphere or higher genus surfaces the last formula is
modified.\par\noindent
To relate the hatted and full functionals in the
{\it non-trivial sectors} one recalls that in the formula
(3.20) for the full partition function (without
hats) one must use
orthonormal zero-modes. These can be expanded in terms of the
un-normalized modes $\psi^p_{0+}$ and $\chi^p_{0+}$ defined in
(3.29). Inserting these expansions
into (3.20) yields the inverse square roots of the determinants
of the corresponding normmatrices ${\cal N}_\psi$
and ${\cal N}_\chi$ which partly chancel $\det{\cal N}_\psi$
in (3.20). Thus one ends up with
$$\eqalign{
&Z_F[\eta,\be]={{\det}^\ha {\cal N}_\psi\ov {\det}^\ha {\cal N}_\chi}\,
\Big({\pi\hat V\ov \Phi}\Big)^{\Phi\ov 4\pi}\,
e^{\Phi^2/2\pi\hat V\cdot\int\sqrt{\hat g}\chi}\,
\prod_{p=1}^{\vert k|}(\bar\eta,\psi^p_{0+})(\chi^p_{0+},\eta)\cr
&\cdot e^{-\int\be(x) S_e(x,y)\eta(y)}
\exp \Big({S_L\over 24\pi}+{1\over 2 \pi} \int \sqrt{\hat g}
G\hat\lap G+{2k\ov V}\int\sqrt{g}G\Big),\cr}\eqno(3.34)
$$
where the $\psi^p_{0+}$ and $\chi^p_{0+}$ are the un-normalized
zero-modes (3.25).\par
\titleb{3.2}{Bosonic path integral}
To arrive at the generating functional for the complete theory
we must finally quantize the photon and auxiliary fields $A_\mu$
and $B_\mu$ (see (3.10)). For that we insert the
decomposition (3.4a) into the bosonic
part of the (euclidean) action (1.1). This results in
$$\eqalign{
S_B&={\Phi^2\over 2e^2 V}+(2\pi)^2\sqrt{\hat g}\hat g^{\mu\nu}
h_\mu h_\nu\cr
&+\int\sqrt{ g}\Big(\ha\varphi\lap^2\varphi-\lam\lap\lam-\phi\lap\phi-
g_3R\lam\Big).\cr}\eqno(3.35)
$$
The term quadratic in the $h$ field is not present in the
action (1.1) on Minkowski space-time. But on the torus $h$
is part of the Hodge decomposition of $B_\mu$ and thus
on the same footing as $\pa\lam$ and $\eta\pa\phi$.
Since $S_B$ and the fermionic determinants are both gauge
invariant and thus independent of the pure gauge mode
$\al$ in (3.4a), it is natural to change variables from
$A_\mu$ to $(\varphi,\al,t_\mu,\Phi)$ in each topological
sector. One can show [3] that this transformation is
one to one, provided
$$
\int\sqrt{g}\varphi=\int\sqrt{g}\al =0\mtx{and}
et_\mu\in [0,1].\eqno(3.36)
$$
The measures are related as
$$
\cd A_\mu=J\sum_k\;dt_0dt_1\cd\varphi\cd\al,\mtx{where}
J=(2\pi)^2{\det}^\pr (-\lap).\eqno(3.37)
$$
The Jacobian $J$ is independent of the dynamical fields.
In expectation values of gauge invariant and thus $\sl$-independent
operators the $\al$-integration cancels
against the normalization. This is of course related to the fact
that in $QED$ the ghosts decouple in the Lorentz gauge.\par
Finally observe that via the derivative couplings to the
fermionic current [33] we introduced artificial degrees of
freedom. The relation between $B_\mu$ in (3.10) and the fields
$(\phi,\lam,h_\mu)$ is only one to one if we impose the
conditions similar to (3.36), namely
$$
\bar\phi\equiv {1\ov V}\int\sqrt{g}\phi=0,\qquad
\bar\lam=0\mtx{and}h_\mu\in [-\infty,\infty].\eqno(3.38)
$$
There is no restriction on the harmonic part of the
auxiliary field, since $B_\mu$ is not a gauge field.
The constraints are imposed in the functional integral as
$$
\int dh_0dh_1\cd\phi\cd\lam\delta(\bar\phi)\delta(\bar\lam)\cdots .
\eqno(3.39)
$$
The normalization by the volume in (3.38) is needed such that the
constraints and hence the partition function are both dimensionless.
For example, expanding $\phi$ in eigenmodes of the Laplacian
as
$$
\phi=a_0\phi_0+\sum_{n>0}a_n\phi_n\;,\mtx{where} \phi_0={1\ov \sqrt{V}}
$$
is the zeromode, one finds the dimensionless partition function
$$
\int \cd\phi\;\delta(\bar\phi)\,e^{\phi\lap\phi}=\sqrt{V}{1\ov {\det}^{\pr\ha}
(-\lap)}\eqno(3.40)
$$
for free bosons.\par\noindent
Constraining the mean field to zero as in (3.40) is equivalent
to fixing the field at an arbitrary point $\xi$ on the torus to zero [34]
$$
\int \cd\phi\;\delta(\bar\phi)\cdots
=\int \cd\phi\;\delta\big(\phi(\xi)\big)\cdots.\eqno(3.41)
$$
This can be seen as follows:
$$
\int{\cd}\phi\;\delta\big(\phi(\xi)\big)\cdots=
\int du\delta\big(\bar\phi-u\big)\cd\phi\;\delta(\phi(\xi))\cdots.\eqno(3.42a)
$$
Now one shifts the field as $\phi\to\phi+u$. Using that the
action is left invariant by this shift, the measure becomes
$$
\int du\cd\phi\;\delta(\bar\phi)\delta(\phi(\xi)+u\big)\cdots
=\int\cd\phi\;\delta(\bar\phi)\cdots\eqno(3.42b)
$$
which shows that the two constraints are the same.
When integrating over the auxiliary fields it is always
understood that the divergent zeromodes are suppressed as
in (3.39).

\titlea{4.}{Partition function of the generalized Thirring
model}
As a first application of our general results we
calculate the partition function of the theory (1.1).
To compute it we must put the sources
$\eta$ and $\bar\eta$ in (3.3) to zero. Then it is evident
from (3.34) that the non-trivial sectors do not contribute
and hence we may assume $\Phi\es 0$.
Thus the partition function is given by
$$
Z_0 = J\int d^2td^2h\cd \varphi\cd\phi\cd\lam
\;Z_F[0,0]\; e^{-S_B[\Phi=0]},\eqno(4.1)
$$
where $J$ is the Jacobian of the transformation (3.37).
$Z_F$ is the fermionic partition function (3.33) in the trivial
sector and the integration is over fields obeying the
conditions (3.36,38). Now we perform the various integrals in turn.
\par\noindent
{\it Integration over the harmonics:}\par\noindent
By using the series representation of the theta functions
one computes
$$\int\limits_0^1 d^2(et)\Theta\Big[{-c_1\atop
c_0}\Big](0,\tau)\ \bar\Theta\Big[{-c_1\atop
c_0}\Big](0,\tau)={1\over \sqrt{2\tau_0}}.\eqno(4.2)
$$
Since the result appears always together with
the $\eta$-function factor in (3.33) it is convenient to introduce
$$
\kappa :={1\over \sqrt{2\tau_0}}{1\over
\vert\eta (\tau)\vert^2}
$$
in the following expressions.
The result (4.2) does not depend on the $h$-field and hence the
$h$-integration in (4.1) becomes Gaussian. It yields
a factor $1/4\pi$ so that
$$
Z_0=\pi\kappa\,{\det}^\pr(-\lap)\;e^{S_L/24\pi}\int
\cd_\delta (\varphi\phi\lam)\;
e^{{1\over 2\pi}\int \sqrt{g}G\lap G-S_B[h=0]},\eqno(4.3)
$$
where $G$ has been defined in (2.4).
We inserted the explicit expression (3.37) for the Jacobian.
It is interesting to note that already
the toron-integration in (4.2) washes out the dependence on the
boundary conditions and chemical potential. We shall
comment on this point later on.
\par\noindent
{\it Integration over $\lam$ and $\phi$:}\par\noindent
The integral over
$\lam$, subject to the condition (3.38), modifies the
Liouville factor and yields one inverse square-root
of the determinant of $-2\lap$ in (4.3).
To continue we recall the scaling formula for
the determinant of $\lap$ [35]:
$$
\log{{\det}^\pr(-a\lap)\over {\det}^\pr(-\lap)}=
\log a\cdot\zeta(0)=\log a\cdot\big[{1\over 4\pi}\int a_1-p\big],
\eqno(4.4a)
$$
where $p$ is the number of zero modes of the
operator. On the torus $\int a_1\es 0$ and we
find
$$
{\det}^\pr \big(-a\lap\big)=
{1\over a}\;{\det}^\pr (-\lap).\eqno(4.4b)
$$
Using this scaling property the $\lam${\it -integral}
together with (3.40) we obtain
$$
Z_0=\sqrt{2V}\kappa\pi{\det}^{\pr\ha}(-\lap)\;e^{(g_3^2+1/24\pi)S_L}\int
\cd_\delta (\varphi\phi)\;e^{{1\over 2\pi}\int \sqrt{g}G\lap G
-S_B[h=\lam=0]},\eqno(4.5)
$$
To quantize the $\phi$ field we need to recall that
$G\es e\varphi\!+\!g_2\phi$. Since $\varphi\lap\varphi
\sim (A^T,A^T)$, the anomalous term $\sim \int G\lap G$ in the
exponent contains an explicit photon mass term with
bare-mass $e/\sqrt{\pi}$. However, when quantizing the $\phi$
field this mass is renormalized. This can be seen explicitly
in the resulting expression for the partition function
after the $\phi${\it -integration} has been performed
$$
Z_0={2\sqrt{\pi}\kappa e V\ov m_\gam}\;
e^{(g_3^2+1/24\pi)S_L}
\int\cd\varphi e^{-\ha\int\sqrt{g}\varphi\big(
\lap^2-m_\gam^2\lap\big)\varphi},\eqno(4.6)
$$
where the renormalized photon mass is
$$
m_\gam^2={e^2\over \pi}{2\pi\over 2\pi+g_2^2}.\eqno(4.7)
$$
{\it Integration over $\varphi$:}\par\noindent
The zeta-function regulated determinant which one obtains when
performing the integral (4.6) factorizes
$$
{\det}^\pr (\lap^2-m_\gam^2\lap)={\det}^\pr(-\lap)\cdot
{\det}^\pr (-\lap+m_\gam^2).
$$
This factorization property is not obvious since all
determinants must be regulated. But it holds for commuting operators
and in the zeta-function scheme.
Then the partition function simplifies to
$$
Z_0={2\sqrt{\pi}\kappa e V\ov m_\gam}
\big({\det}^\pr(-\lap){\det}^\pr (-\lap+m_\gam^2)\big)
^{-\ha}\,\exp\Big((g_3^2+{1\over 24\pi})S_L\Big).\eqno(4.8)
$$
We can go further by using (3.31) and the known result for the
determinant of $\hat\lap$ [26] which together yield
$$
{\det}^{\pr\ha}(-\lap)=\tau_0L\vert\eta(\tau)\vert^2
\sqrt{{V\over \hat V}}\exp\Big(-{1\over 24\pi}S_L\Big)\eqno(4.9)
$$
which finally leads to
$$
Z_0=\sqrt{2\pi V}\;{e\ov m_\gam}{1\ov \tau_0\vert\eta(\tau)
\vert^4}
\;{1\over {\det}^{\pr\ha} (-\lap+m_\gam^2)}
\,\exp\Big(({1\ov 12\pi}+g_3^2)\,S_L\Big)
\eqno(4.10)
$$
for the partition function of the general
model (1.1) on curved spaces. It shows explicitly
that in the topologically trivial sector the theory
should be equivalent to a theory of free massless
and massive bosons with mass $m_\gam$.\par\noindent
It is interesting to follow the various contributions
to the explicit dependence on the gravitational field
since they contribute to the {\it Hawking radiation}.
For that we recall that when one quantizes a
conformal field theory with central charge $c$
in an external gravitational fields one ends up
with the Liouville term, $Z\sim \exp[c\,S_L/24\pi]$ [32].
Thus the fermions contribute with $c\es 1$, as
expected. The $\phi$ and $\lam$ field contribute with $1$ and
$1+24\pi g_3^2$, respectively. However, the Jacobian
combined with the conformal part of the gauge sector
contribute with $c\es -1$ and we are left with
a total central charge $c\es 2+24\pi g_3^2$.
Of course, the gauged model is not conformally invariant
and the breaking is manifest in the massive determinant in (4.10). The
partition function of the {\it ungauged theory} is (4.6) multiplied by
an inverse determinant (the missing Jacobian) and without
$\varphi$-integration. In this limit one obtains
a conformal theory with central charge $c\es 3+24\pi g_3^2$.\par
By using an elegant result of Christensen and Fulling
[36], that relates the conformal anomaly to the asymptotic
Hawking flux, one concludes that the Hawking radiation
of the ungauged model is $3\!+\!24\pi g_3^2$ times
that of free massless scalars. For the gauged model
the Hawking radiation is still thermal and consists of massless
and massive particles.
\par\noindent
The appearance of $m_\gam$ in (4.6) should be interpreted as
{\it renormalization of the electric charge} induced by the interaction
of the auxiliary fields with the fermions. After summing
over all fermion-loops this leads to an effective coupling
between the photons and the $\phi$-field and in turn to
a modified effective mass for the photons in (4.6). In
the limit $g_2\to 0$ this mass tends to the well-known
Schwinger model result, $m_\gam\to e/\sqrt{\pi}$ [37].\par\noindent
We have already mentioned that the
chemical potential coupled to the electric
charge has completely disappeard
from the partition function. This does not come as a
surprise since the only particle in the gauged Thirring model is a
massive neutral boson. This has no charge
which may couple to the chemical potential.
If the partition function would depend on $\mu$ then the
expectation value of the charge would not vanish, in
contradiction to the integrated Gauss law.\par\noindent
Also note that the final result is independent
of the chiral and non-chiral twists. The normal twists
have been wiped out by the toron integration.
Since the chiral twists are equivalent to chemical potentials
the partition function should not depend on them in
order not to violate gauge invariance.
If one would assume holomorphic factorization for the
fermionic determinant [11] then the partition function
would depend on the chiral twists.
\par\noindent
We conclude this subsection with deriving an explicit formula for
the partition function on the flat torus.
Applying the results in [38] one obtains for the
massive determinant
$$
{\det}^\pr(-\hat\lap+m_\gam^2)^\ha ={1\ov m_\gam}
e^{-\ha\zeta^\pr(0)},\eqno(4.11a)
$$
with
$$
\zeta^\pr(0)= \sum\limits_{n\neq 0}{1\over \pi L}{
\hat V m_\gam\ov\sqrt{(n,n)}}K_1\big(m_\gam L\sqrt{(n,n)}\big)
-{\hat V m^2_\gam\ov 4\pi},\eqno(4.11b)
$$
where $(n,n)\es \hat g_{ij}n^in^j$ is the inner product taken with
the reference metric, and the sum is over all $(n^i)\in Z_2$
with the origin excluded.
For $\gd=\delta_{\mu\nu}$, in which case the partition function has the
usual thermodynamical interpretation,
the result reduces to one derived previously by Ambjorn
[39]. In addition, if $L$ approaches infinity we recover a
result in [29]. The free energy for $\tau_1=0$ and on flat space
simplifies then to
$$
F=-{1\ov\beta}\log Z = {1\ov 2\beta}\;\zeta^\pr(0).\eqno(4.12)
$$
with $\zeta^\pr (0)$ from (4.11b) and the particular choice
for the parameters.
\titlea{5.}{Bosonization of the gauged Thirring model}
In the classical analysis we have already seen that in the limiting
case $g_3\es 0$ and $g_1\es g_2\es g$ the general model reduces
to the gauged Thirring model. In this section we show
that the same is true for the quantized theory on the torus if in
addition we set $g_0\es g$. More precisely, the Hubbard-Stratonovich
transform [40] of the Thirring model is just the derivative coupling
model (1.1) with identical couplings.
In the process of showing that we shall arrive at the bosonization
formulae for the gauged Thirring model on the curved torus.
We shall see that only the non-harmonic part of
the fermion current can naively be bosonized and that for
this part the rules of the ungauged model on flat space time [41]
need just be covariantized.
\par
For that we calculate the partition function (4.1a)
in a different order. First we integrate out the auxiliary
fields. In order to understand the role of $\lam$ and $\phi$ we
introduce sources for them. Thus we study the generating
functional for the correlators of the auxiliary fields
$$
Z[\xi,\zeta]=\int \cd (\lam\phi h\psi A_\mu)
e^{\,-S +\int\sqrt{g}\big[\xi\lam+\zeta \phi\big]}.\eqno(5.1a)
$$
Here
$$
S=-i\int\sqrt{g}\psi^\dagger \slash D\psi+S_B[g_3\es 0]\eqno(5.1b)
$$
is the action of the full theory. $\slash D$ is the
Dirac operator in (3.11) with all couplings set equal
and $S_B$ the bosonic action (3.35). Since $\lam$ and
$\phi$ integrate to zero (see (3.38)) we may assume the same property
to hold for the sources. Also, since there are no fermionic sources
only configurations in the trivial sector contribute, so that
there is not instanton potential in (3.11) and hence $\Phi\es 0$
in (3.35). The integration over the auxiliary fields is gaussian and
yields
$$
Z={\cal N}_0\int\cd (\psi A_\mu)\,e^{-S_T}\,
\exp\int\sqrt{g} \Big[-{1\over 4}(\xi{1\over\lap}\xi+
\zeta{1\over\lap}\zeta)
+{g\over 2}(\xi{1\over \lap}j^\mu_{;\mu}+\zeta
{1\over\lap}j^\mu_{5;\mu})\Big]\eqno(5.2)
$$
where
$$
S_T=-{1\over 4}\int\sqrt{g}\Big(F_{\mu\nu}F^{\mu\nu}-i\psi^\dagger
\slash D\psi -{g^2\over 4}j^\mu j_\mu\Big)\eqno(5.3a)
$$
is the action of the {\it gauged Thirring model} on curved space-time
and
$$
{\cal N}_0={V\ov 2\pi {\det}^\pr (-\lap)}\eqno(5.3b)
$$
comes from the integration over the auxiliary fields.
\par\noindent
Let us first consider the partition function, that is set the
sources to zero. Comparing (5.2) with (4.6) and using
(4.9) we easily find
$$
\int \cd (\psi t)e^{-S_T}=\sqrt{\ha+{g^2\ov 4\pi}}
\;e^{-{1\ov 4}\int
F_{\mu\nu}F^{\mu\nu}}\int\cd \gam\,\delta(\bar\gam)\;e^{-S_\gam},
\eqno(5.4a)
$$
where $\bar\gam$ is the mean field (see 3.38) and
we used (3.37) and (4.4b). The action for the neutral scalar field
$\gam$ is found to be
$$
S_\gam =\ha\int\sqrt{g} \pa_\mu\gam\pa^\mu\gam - {ie\over \sqrt{\pi}}
{1\ov\sqrt{1+g^2/2\pi}}\int\sqrt{g}\gam\lap\varphi.\eqno(5.4b)
$$
Since (5.4) holds for any $\varphi$ (and thus for the
non-harmonic part of any
$A_\mu$, because of gauge-invariance) we read off the following
{\it bosonization rules}:
$$
j^{\pr\mu}\longrightarrow {i\over\sqrt{\pi}}{1\ov\sqrt{1+g^2/2\pi}}
\eta^{\mu\nu}\pa_\nu\gam
\mtx{and} j^{\pr\mu}_5\longrightarrow -{i\over\sqrt{\pi}}
{1\ov\sqrt{1+g^2/\pi}}\pa^\mu\gam,\eqno(5.5)
$$
where prime denotes the non-harmonic part of the currents.
Thus, only the non-harmonic parts of the currents can be
bosonized in terms of a single valued scalar field.
To bosonize their harmonic parts one would have to allow for
a scalar field $\gamma$ with windings as $\phi$ below (3.2).
On the infinite plane the harmonic part is not present
and we may leave out the primes in (5.5). If we further assume
space time to be flat we recover the wellknown bosonization
rules in [41]. What we have shown then, is that for
the gauged model on curved space time the bosonization
rules are just the flat ones properly covariantized and
with the omission of the zero-modes.
\par
Since (5.4a) holds for any gauge field the current correlators
in the Thirring model are correctly reproduced by the
bosonization rules (5.4,5).
To see that more clearly we calculate the two-point functions of the
auxiliary fields in the Thirring model (5.2,3). For that we
differentiate (5.2) ($\varphi$ is treated as external field)
with respect to the sources and find
$$
\eqalign{
&\langle \lam(x)\lam(y)\rangle =\ha G_0(x,y)+
{g^2\over 4}\int\langle G_0(x,u)j^\mu_{;\mu} (u)G_0(y,v)
j^\nu_{;\nu} (v)\rangle_{\ss T}\cr
&\langle \phi(x)\phi(y)\rangle =\ha G_0(x,y)+
{g^2\over 4}\int \langle G_0(x,u)j^\mu_{5;\mu} (u)G_0(y,v)
j^\nu_{5;\nu}(v)\rangle_{\ss T},\cr}\eqno(5.6)
$$
where $G_0$ is the free massless Greenfunction (3.6a,3.8)
in curved space-time and the integrations are over the variables
$u$ and $v$ with the invariant measure on the curved torus.
Here $\langle\dots\rangle_{\ss T}$ are vacuum expectation values
of the Thirring model (5.3a).
Alternatively we can calculate these expectation values
from (4.3) and (4.5), where the fermionic integration has
been performed and find
$$
\eqalign{
&\langle \lam(x)\lam(y)\rangle =\ha G_0(x,y)\cr
&\langle \phi(x)\phi(y)\rangle ={\pi m_\gam^2\over 2 e^2}\,
G_0(x,y)+{m_\gam^2\over 2}\big(1-{\pi m_\gam^2\over e^2}\big)\varphi (x)
\varphi (y).\cr}\eqno(5.7)
$$
Comparing the results (5.6) and (5.7) we see at once that
$$\eqalign{
&\int\langle G_0(x,u) j^\mu_{;\mu} (u)G_0(y,v)
j^\nu_{;\nu} (v)\rangle_{\ss T} =0\cr
&\int\langle G_0(x,u)j^\mu_{5;\mu} (u)G_0(y,v)
j^\nu_{5;\nu} (v)\rangle_{\ss T} ={m_\gam^2\over e^2}\big(
m_\gam^2\varphi(x)\varphi(y)-G_0(x,y)\big).\cr}
\eqno(5.8)
$$
The correlators (5.8) express the gauge invariance and the
axial anomaly $\langle j^\mu_{5;\mu}\rangle\es -m_\gam \lap\varphi$
in the gauged Thirring model. They can be correctly reproduced with the
bosonisation rules (5.5).
\titlea{6.}{Chiral condensate}
We have seen that the partition function of the gauged model
does not depend on the chiral twists $\beta$. As we pointed
out this property is very much linked to gauge invariance.
The same should then hold in the topological non-trivial
sectors. For that reason we shall set $\beta\es 0$ in this
section. This assumption will simplify the manipulations
below considerably. For example, ${\cal N}_\psi\es{\cal N}_\chi$
in (3.34).
\par
Recalling that $S_e$ in (3.34) anti-commutes with
$\gamfive$ one sees at once that only configuration supporting
one fermionic zero-mode with positive chirality contribute to the chiral
condensate
$$
\langle \psi^\dagger P_+\psi\rangle=-{J\ov Z_0}
{\delta^2\over \delta \eta_+(x)\delta\be_+(x)}\int \cd (\dots)
Z_F[\eta,\be]\vert_{\eta=\be=0}\,e^{-S_B},
\eqno(6.1)
$$
where $\eta_+\es P_+\eta$. Earlier we have
seen that these are the gauge fields with flux $\Phi\es 2\pi$
or instanton number $k\es 1$. Thus (6.1) reads
$$
\langle \psi^\dagger P_+\psi\rangle=-{J\ov Z_0}
\sqrt{\hat V\over 2}\int\cd (\dots)\, \psi^\dagger_{0+}(x)\psi_{0+}(x)
\exp(\dots)\,e^{-S_B[k=1]},
\eqno(6.2)
$$
where $\exp(\dots)$ is the last exponential factor in (3.34).
Here we have used that the chiral twists vanish, such that
$\chi_{0+}\es \psi_{0+}$ (see (3.25)).
First we integrate over the toron field $t$.
The $t$ dependence enters only through the zero mode
and more specifically $\hat\psi_{0+}$ in (3.29) and (3.25) with
$p\es 1$. Using the series representation for the theta functions
one finds
$$
\int d^2 t\,\hat\psi_{0+}^\dagger (x)\hat\psi_{0+}(x)={1\over \hat V}.
\eqno(6.3)
$$
Note that the result does not depend on the
chemical potential similarly to our
calculation of the partition function. \par
To continue we observe that the term $\int \sqrt{g}G$ in $\exp(\dots)$
vanishes because of our conditions (3.36) and (3.38)
on the fields $\varphi$ and $\phi$. Also note, that
the fermionic Green function does not enter in the
expression for the chiral condensate.
It follows that the fermionic functional (3.33) in the
trivial sector and (3.34) in the one-instanton sector
are the same, up to the factors in the fist lines.
{}From (6.3) and (4.3) we see that the
toron integral of the first line in (3.34) is
$\vert \eta\vert^2\sqrt{\tau_0/\hat V}\exp(2\pi\int\sqrt{g}\chi/\hat V)$
times the toron integral over the factor in (3.33). Also, since
$$
S_B[k=1]=S_B[k=0]+{2\pi^2\over e^2V}
$$
the functional integral and normalizing
partition function in (6.2) are the same, up to these factor
and the field-dependent factors which relate the
hatted and unhatted zero-modes in (3.29). Finally
note that the $\lam$ integrals in (6.2) and in the
normalizing partition function cancel so that
we end up with the following formula for the condensate
$$
\langle \psi^\dagger P_+\psi\rangle=
\sqrt{\tau_0\ov \hat V}\,\vert \eta (\tau)\vert^2
e^{-2\pi^2/e^2V+2\pi/\hat V\int\sqrt{\hat g}\chi}
\Big\langle e^{-2(g\phi+e\varphi)(x)-\sigma(x)}
\Big\rangle_{ \phi\varphi}.\eqno(6.4a)
$$
The expectation value is evaluated with
$$
S_{eff}=\int\sqrt{g}\Big[\ha\varphi (\lap^2-{e^2\over\pi}\lap)\varphi
-{e^2\over\pi m_\gam^2}\phi\lap\phi-{eg_2\over\pi}\phi\lap\varphi
\Big].\eqno(6.4b)
$$
A formal calculation of the resulting gaussian integrals yield
$$\eqalign{
\langle \psi^\dagger P_+\psi\rangle&=
\sqrt{\tau_0\ov \hat V}\vert \eta (\tau)\vert^2 e^{-2\pi^2/e^2V
+2\pi/\hat V\int\sqrt{\hat g}\chi}\,
e^{-\sigma(x)-2\Phi\chi(x)}\cr
&\cdot\exp\big[{2\pi^2m_\gam^4\ov e^2}\,K(x,x)\big]
\exp\big[{2\pi g_2^2\over 2\pi+g_2^2}\,G_0(x,x)\big],\cr}\eqno(6.5)
$$
where
$$
K(x,y)=\langle x\vert{1\over \lap^2-m_\gam^2\lap}\vert y\rangle
={1\ov m_\gam^2}\big(G_0(x,y)-G_{m_\gam}(x,y)\big) \eqno(6.6)
$$
and $G_{m},G_0$ are the massive and massless Greenfunctions.\par
Here we encounter ultraviolett divergences since
$G_0(x,y)$ is logarithmically divergent when $x$ tends to $y$.
To extract a finite answer we need to renormalize the operator
$\exp(\al \phi)$. This wave function renormalization is equivalent
to the renormalization of the fermion field in the Thirring model
and thus is very much expected [33,41].
In order to do that we first determine the short distance
behaviour of the massless Greenfunction (3.6b).
Using the idendity
$$
\big|\Theta\Big[{\ha+{\xi^0\ov L}\atop
\ha+{\xi^1\ov L}}\Big](0,\tau)\big|^2=
\big|e^{i\pi\tau (\xi^0/L)^2}\Theta_1
\big({\tau \xi^0+\xi^1\ov L},\tau\big)\big|^2\eqno
$$
and the small $z$ expansion
$$
\Theta_1(z,\tau)=2\pi\eta(\tau)^3z +O(z^2),
$$
we see that $\hat G_0$ possesses the expected logarithmic short distance
singularity
$$
\hat G_0(x,y)=-{1\over 4\pi}\log {\hat g_{\mu\nu}\xi^\mu\xi^\nu
\ov \hat V}-{1\over 4\pi}\log \big(4\pi^2\tau_0\vert\eta(\tau)\vert^4\big)
+O(\xi).\eqno(6.7)
$$
{}From the relation (3.8) between the full and hatted Greenfunction
and the definition of $\chi$ in (3.4c) it follows
that $G_0$ has the short distance expansion
$$
G_0(x,y)\sim \hat G_0(x,y)+2\chi (x)-{1\ov\hat V}
\int\sqrt{\hat g}\chi+O(\xi)\eqno(6.8)
$$
\par
To continue we need to regularize the composite operator
$\exp(\al\phi)$ appearing in (6.4a). The normal ordering prescription
$$
:e^{\al\phi(x)}:= {e^{\al\phi(x)}\ov\<e^{\al\phi(x)}\>}.\eqno(6.9)
$$
works well on the whole plane [33,41]. On the curved torus
we must be more careful when renormalizing this operator.
The required wave function renormalization is not unique
but it is very much restricted by the following
requirements: First we take as reference system
(the denominator in 6.9) one with
a minimal number of dynamical degrees of freedom since we do not want to
loose information by our regularization. Second, the renormalized
operator should have a well-defined infinite volume limit
and its expectation values should cluster. Finally, the regularization
should respect general covariance.
These requirements
then force us to take as reference system the infinite plane
with metric $g_{\mu\nu}$. The flat metric $\delta_{\mu\nu}$
is not permitted since it leads to a ill-defined expression
for $\langle\exp(\al\phi)\rangle$. With these choice
the normal ordering in (6.9) is equivalent to replacing
the massless Greenfunction in (6.5) by
$$
G_0^{reg}(x,y):= G_0(x,y)+{1\over 4\pi}\log\big[\mu^2 s^2(x,y)\big].
\eqno(6.10)
$$
Here $s(x,y)$ denotes the geodesic distance between $x$ and $y$.
The occurence of the arbitrary mass scale $\mu$ comes from the
ambiguities in the required ultraviolett regularization. On the
flat torus $\hat G_0^{reg}$ has now the finite coincidence limit
$$
\hat G_0^{reg}(x,x)=
-{1\over 4\pi}\log \Big({4\pi^2\tau_0\vert\eta(\tau)\vert^4\ov
\mu^2\hat V}\Big).\eqno(6.11)
$$
To determine the chiral condensate we also need to determine
$K(x,y)$ on the diagonal. In a first step we shall obtain
it for the flat torus. Its $\sigma$-dependence is then determined
in a second step. For $\sigma\es 0$ and $\tau\es i\tau_0$
the Greenfunction $\hat K$ has been computed in [3].
The generalization to arbitrary $\tau$ is found to be
$$\eqalign{
m_\gam^2 \hat K(x,x)&=-{1\over 2m_\gam L
\tau_0}\coth\big({\pi\tau_0 a\ov \vert\tau
\vert^2}\big)+{1\over m_\gam^2\hat V}\cr &+{1\over 2\pi}
\Big(-\log\vert\eta\big({-1\over\tau}\big)\vert^2+F(L,\tau)-H(L,\tau)\Big)
,\cr}\eqno(6.12a)
$$
where we introduced the dimensionless constant
$a=Lm_\gam\vert\tau\vert/2\pi$ and the functions
$$\eqalign{
F(L,\tau)&=\sum_{n>0}\Big[{1\over n}-{1\ov \sqrt{n^2+a^2}}\Big]\cr
H(L,\tau)&=\sum_{n>0}{1\ov\sqrt{n^2+a^2}}\Big[{1\ov e^{-2\pi iz_+(n)}-1}
+{1\ov e^{2\pi i z_-(n)}-1}\Big].\cr}\eqno(6.12b)
$$
We used the abbreviations
$$
z_\pm={1\ov \vert\tau\vert^2}\big(n\tau_1\pm i\tau_0\sqrt{n^2+a^2}\big).
\eqno(6.12c)
$$
Substituting (6.12) and (6.11) into (6.5) with $\sigma\es 0$
we obtain the following {\it exact formula for the chiral condensate}
on the torus with flat metric $\hat g_{\mu\nu}$:
$$\eqalign{
\langle\psi^\dagger P_+\psi\rangle_{\hat g}&={1\ov L\vert\tau\vert}\,
\Big({m_\gam L\vert\tau\vert\ov 2\pi}\Big)^{g_2^2\ov 2\pi+g_2^2}
\exp\Big({\pi^2 m_\gam\ov e^2 L\tau_0}\coth{Lm_\gam\tau_0\ov 2
\vert\tau\vert}\Big)\cr
&\cdot\exp\Big[{\pi m_\gam^2\ov e^2}\Big(
F(L,\tau)-H(L,\tau)\Big)\Big],\cr}
\eqno(6.13)
$$
where we used that on the flat torus $\chi\es 0$ and $V\es\hat V$.
Futhermore, we identified $\mu$ with the natural mass scale
$m_\gam$ of the theory.\par
To study the finite temperature behaviour of the chiral condensate
we must assume that $\tau\es i\beta/L$ and then $\beta\es 1/T$
is just the inverse temperature. Furthermore we perform the
thermodynamic limit $L\to \infty$. Then
$\coth(\dots)\to 1$, $\;H\to 0$ and the expression for the chiral
condensate simplifies to
$$
\langle\psi^\dagger P_+\psi\rangle_{\beta}=-
T\Big({m_\gam\ov 2\pi T}\Big)^{g_2^2\ov 2\pi+g_2^2}
\exp\Big[-{\pi^2m_\gam\ov e^2}T+{2\pi \ov 2\pi+g_2^2}F\Big].
\eqno(6.14)
$$
Let us now investigate the low and high temperature limits in turn.
To study the {\it low temperature limit} we use that
$$
F(\beta)\to \gam+\log{a\ov 2}+{1\ov 2 a}\quad\mtx{for} a\to\infty,
$$
where $\gam\es 0.57721\dots$ is the Euler number. Inserting
this expansion into (6.14) yields
$$
\langle\psi^\dagger P_+\psi\rangle=
-{m_\gam\ov 4\pi}\,2^{g_2^2/(2\pi+g_2^2)}\,
\exp\Big({2\pi\ov 2\pi+g_2^2}\gamma\Big)\quad\mtx{for}
T\to 0.\eqno(6.15)
$$
For temperatures large compared to the induced photon mass $F$
vanishes. Thus we obtain the {\it high temperature behaviour}
$$
\langle\psi^\dagger P_+\psi\rangle_{T}=
-T\Big({m_\gam\ov 2\pi T}\Big)^{g_2^2\ov 2\pi+g_2^2}\,
\exp\Big(-{\pi^2 m_\gam\ov e^2}T\Big)\mtx{for} T\to\infty \eqno(6.16)
$$
It is instructive to discuss the various limiting cases.
For all $g_i\es 0$, i.e. the Schwinger model limit, the
exact result (6.14) simplifies to
$$
\langle\psi^\dagger P_+\psi\rangle_T=
-T\,e^{-{\pi\ov m_\gam}T+F(\beta)}\longrightarrow
\cases{-{m_\gam\ov 4\pi}e^\gam &for $\quad T\to 0$\cr
-T\,e^{-\pi T/m_\gam}&for $\quad T\to\infty$,\cr}
\eqno(6.17)
$$
where now $m^2_\gam=e^2/\pi$ is the induced photon mass in the
Schwinger model. This formula for the temperature
dependence of the chiral condensate in $QED_2$ agrees with the earlier
results in [3].\par
Next we wish to investigate how the selfinteraction of the
fermions affect the breaking. For large coupling $g_2$
and fixed termperature the exponent in (6.14) vanishes
so that
$$
\langle\psi^\dagger P_+\psi\rangle_T\sim {1\ov \sqrt{2\pi+g_2^2}}\quad
\mtx{for}T\quad\hbox{fixed},\quad g_2\to\infty.\eqno(6.18)
$$
Hence, for very large current-current coupling the chiral condensate
vanishes. Or in other words, the electromagnetic
interaction which is responsible for the chiral condensate, is
shielded by the pseudoscalar-fermion interaction.\par\noindent
For intermediate temperature and coupling $g_2$ we must retreat
to numerical evaluations of the sums defining the chiral condensate
in (6.14). The results of the numerical calculations are depicted in
Figure 1.\par
The study of the influence of the gravitational field is
complicated by the presence of the massive Green function
$G_{m_\gam}$ in (6.5,6). This Green function is known only
for very particular curved spaces. Fortunately we only need
the coincidence limit for which we can use its short distance
expansion [42]. For simplicity we assume infinite volume and
zero temperature. Then [43]
$$
G_m(x,y)\sim {1\ov 4i}\sum_{j=0}^\infty a_j(x,y)
\big(-{\pa\ov \pa m^2}\big)^j\,H_0^{(2)}(ms),\eqno(6.19)
$$
for small geodesic distances $s\es s(x,y)$. Here $H_0^{(2)}$ denotes
the Hankel function of the second kind and order zero. In particular
$$
H_0^{(2)}(z)\to {2\ov i\pi}\big[\log{z\ov 2}+\gam\big]
\mtx{for} z\to 0.\eqno(6.20)
$$
Inserting that into (6.19) we find with $G_0\es\hat G_0$ from (6.7)
the following short distance expansion
$$\eqalign{
G_0(x,y)-G_m(x,y)\sim &-{1\ov 2\pi}\Big[\log\big({2\pi\vert\eta(\tau)\vert^2
\ov m_\gam L e^{\sigma(x)}}\big)-\gam\Big]\cr
&+{1\ov 4\pi}\sum_{j=1}^\infty a_j(x)\big(-{\pa\ov \pa m^2}\big)^j\log(m^2).
\cr}\eqno(6.21)
$$
We have used that $a_0(x)\es 1$ and $s\sim e^{\sigma(x)}\hat s$,
where $\hat s$ is the geodesic distance on the flat spacetime with
hatted metric, $\hat s^2=\hat g_{\mu\nu}(x\!-\!y)^\mu(x\!-\!y)^\nu$.
Finally, substituting (6.21) into (6.5) we end up with
$$
\langle\pd P_+\psi\rangle_\sigma=\langle\pd P_+\psi\rangle_{\sigma=0}\cdot
\exp\Big[-\ha \big({\pi m_\gam\ov e}\big)^2 \sum_1^\infty
a_j(x){(j-1)!\ov m^{2j}}\Big].\eqno(6.22)
$$
The Seeley-deWitt coefficients $a_j$ have been computed up to $j\es 5$
[44]. They are of order $j$ in the curvature and its derivatives. The first
two are
$$
a_0(x)=1\quad\mtx{and}\quad a_1(x)={1\over 6}{\cal R}.
$$
For ${\cal R}<<m^2$ and slowly varying ${\cal R}$ we conclude that the
chiral condensate decreases with increasing curvature as
$$
\langle \psi^{\ss \dagger}P_+\psi\rangle\sim \exp\big[-{\pi^2\ov 12 e^2}
{\cal R}\big].\eqno(6.23)
$$
If we compare this with the temperature dependence (6.16) we are lead
to define a curvature induced effective temperature
$$
T_{eff}={{\cal R}\ov 12 m_\gam}.\eqno(6.24)
$$
For this formal identification of curvature with temperature no horizon
is needed as in black hole physics where the temperature
is related to the surface gravity at the horizon. Note that
contrary to the temperature the curvature may become
negative. Then the condensate is amplified.\par
Finally we consider the {\it chiral two point function} for non-coinciding
points. The gauge invariant form reads
$$
S_+(x,y)\equiv\langle \pd (x)\;e^{ie\int_y^xA_\mu dx^\mu}
P_+\psi(y)\rangle.
$$
It is related to a bound state between a static external charge
and a dynamical fermion [45].\parn
The integration over the various
fields is similar to (6.3-12). The result takes a simple
form in the infinite volume and zero temperature limit:
$$\eqalign{
S_+(x,& y)=S_+(x)^{1\ov 4}S_+(y)^{1\ov 4}\;
\exp\Big[\ha ({\pi m_\gam\ov e})^2\big(K(x,y)+K(y,x)\big)\Big]\cr
&\cdot \exp\big[({\pi g_2^2\ov 2\pi+g_2^2}+{g_1^2\ov 2})G_0(x,y)
-{g_1^2\ov 4}\big(G_0(x,x)+G_0(y,y)\big)\Big],\cr}\eqno(6.25)
$$
where $S_+(x)\equiv S_+(x,x)=\langle\pd (x)\psi(x)\rangle$ denotes the
chiral condensate.
Again the massless propagator must be regularized. We do this using the
prescription (6.10). Then
$$
S_+(x,y)=S_+(x)^{1\ov 4}S_+(y)^{1\ov 4}\;{\exp\Big[\ha({\pi m_\gam
\ov e})^2\big(G_{m_\gam}(x,y)+G_{m_\gam}(y,x)\big)\Big]\ov
\sqrt{2\pi}m_\gam^{g_1^2\ov 4\pi}\big(g(x)^{1\ov 8}g(y)^{1\ov 8}\hat s
\big)^{\ha(1+{g_1^2\ov 2\pi})}}.\eqno(6.26)
$$
Note that the coupling strength $g_1$ to the longitudinal current
enters the scaling exponent. On {\it flat space} $G_m$ reduces
to ${1\ov 2\pi}K_0(m\hat s)$ wich decays exponentially for large
separations. Hence we find
$$
\hat S_+(x,y)\sim {\hat S_+(x)^\ha \ov
\sqrt{2\pi\hat s}\big(m_\gam \hat s\big)^{g_1^2\ov 4\pi}}
\eqno(6.27)
$$
for large separations of $x$ and $y$. We have used that the chiral
condensate $\hat S_+(x)$ in (6.15) is constant, due to translational
invariance. For $g_1\es 0$ this simplifies to the Schwinger model result [3]
$$
\hat S_+(x,y)\sim \sqrt{m_\gam e^\gam\ov 2}{1\ov 2\pi \sqrt{\vert
x-y\vert}}.\eqno(6.28)
$$
Unlike the correlators of fields which in the bosonized
version are local in the massive boson field, this two-point
function does not decay exponentially. However the long
range correlations are suppressed by the coupling to the
longitudinal current.

\titlea{7.}{Thermodynamics of the ungauged Model}
In this chapter we derive the grand canonical potential, equation
of state and ground state energy for $A_\mu\es 0$. For the ungauged
model there is no Gauss constraint and the charge of the vacuum need
not vanish. Indeed, for $A_\mu\es 0$ the partition function
depends on the chemical potential and on the fermionic
boundary conditions. Technically this is due to the absence
of the toron integration which for the gauged model wiped
out any dependence on $\mu,\al$ and $\beta$.\par
The partition function of the ungauged model is given by
$$
Z=\int d^2h\cd \phi\cd\lam \;Z_F[\eta\es\be\es A\es 0]\;
e^{-S_B[A=0]},\eqno(7.1)
$$
where $Z_F$ is the fermionic generating functional (3.33)
and $S_B$ the bosonic action (3.35). The integration over the
harmonic fields is Gaussian and yields
$$
\int\limits_{-\infty}^\infty d^2h \Theta\Big[{-c_1\atop c_0}\Big]
\bar\Theta\Big[{-\bar c_1\atop \bar c_0}\Big]
e^{-(2\pi)^2\sqrt{\hat g}\hat g^{\mu\nu}h_\mu h_\nu}
={\Theta\Big[{u\atop w}\Big](\Lambda)\ov 4\pi \sqrt{1+g_0^2/2\pi}}
\eqno(7.2)
$$
where
$$
\Theta\Big[{u\atop w}\Big](\Lambda)=
\sum\limits_{n\in Z^2}\,e^{i\pi(n+u)\Lambda (n+u)+2\pi i (n+u)w}\eqno(7.3a)
$$
is the thetafunction with characteristics
$$
u=-\pmatrix{1\cr 1\cr}(\al_1+i\eta_1^{\;\nu}\beta_\nu)\mtx{and}
w=\pmatrix{1\cr -1\cr}(\al_0+i\eta_0^{\;\nu}\beta_\nu-\mu_0)\eqno(7.3b)
$$
and covariance
$$
\Lambda=\pmatrix{\tau&0\cr 0&-\bar \tau\cr}
+i{\pi g_0^2\tau_0\ov 2\pi+g_0^2}\pmatrix{g_0^2&-4\pi-g_0^2\cr
-4\pi-g_0^2&g_0^2\cr}.\eqno(7.3c)
$$
The remaining functional integrals in (7.1) are performed as in chapter
4. To obtain the partition function of the
Thirring model in the limit $g_i\es g$ we divide $Z$ by the
corresponding partition function ${\cal N}_0$ of the free bosons
(see 5.3b)). Using (4.4b) and (3.33) we obtain
$$
{Z\ov {\cal N}_0}={1\ov \vert\eta(\tau)\vert^2}
\sqrt{2\pi+g_2^2\ov 2\pi+g_0^2}\;
\Theta\Big[{u\atop w}\Big](\Lambda)\,e^{(1/24\pi+g_3^2)S_L}.\eqno(7.4)
$$
In the Thirring model limit $g_2\es g_0$ and the
square-root in this formula disappears.
\titlec{7.1} {Zero-temperature limit}
To investigate the thermodynamics of the model we assume spacetime to be flat
and
that $\tau\es i\beta/L$. Then
$$
\Omega=-{1\ov \beta}\log {Z\ov{\cal N}_0}\eqno(7.5)
$$
is the {\it grand canonical potential}. Let us now investigate the low
temperature limit of $\Omega$. For $\mu\es  0$ this yields the ground state
energy.\parn
To study this limit we observe that for $\tau\es i\beta/L$  the
covariance matrix $\Lambda$ in (7.3c) simplifies to
$$
i\pi\Lambda=-{\pi\beta\ov L}\Big[{\rm Id}+{g_0^2\ov 4\pi}{1\ov 2\pi+g_0^2}
\pmatrix{g_0^2&-4\pi-g_0^2\cr-4\pi-g_0^2&g_0^2\cr}\Big]\eqno(7.6a)
$$
and has eigenvalues
$$
\lam_1=-{\pi\beta\ov L}{2\pi+g_0^2\ov 2\pi}\mtx{and}
\lam_2=-{\pi\beta\ov L}{2\pi\ov 2\pi+g_0^2}\eqno(7.6b)
$$
with corresponding eigenvectors
$$
v_1=(-1,1)\mtx{and} v_2=(1,1).\eqno(7.6c)
$$
Also the $\eta$ tensor
(see 3.17b) and $\mu_0$ (see 3.11) in (7.3b) simplify to
$$
\eta_\mu^{\;\;\nu}=\pmatrix{0&\beta/L\cr-L/\beta&0\cr}\mtx{and}
\mu_0=-i{\beta\ov 2\pi}\mu.
$$
Now we can determine the low temperature limit of the grand potential
from (7.4) (with $S_L\es 0$) and (7.6). For that we note
that the saddle point approximation to the Gaussian sum
(7.3a) defining the theta-function becomes exact when
$\beta\to\infty$. Also, using that
$$
\log\vert\eta(\tau)\vert^2\longrightarrow -{\pi\beta\ov 6 L}
\quad\mtx{for}\beta\to\infty\eqno(7.7)
$$
we end up with
$$\eqalign{
\Omega(\beta\to\infty)=-{\pi\ov 6L}&-{4\pi\ov 2\pi+g_0^2}{\pi\ov L}
\big(\beta_1+{\mu L\ov 2\pi}\big)^2\cr
+{\pi\ov 2L}\min_{n\in Z^2}\Big[&
{2\pi+g_0^2\ov 2\pi}\big\{n_2-n_1-{4\pi\ov 2\pi+g_0^2}(\beta_1
+{\mu L\ov 2\pi})\big\}^2\cr &+{2\pi\ov 2\pi+g_0^2}\big\{
n_1+n_2-2\al_1\big\}^2\Big]\cr}\eqno(7.8)
$$
for the zero-temperature grand potential of the ungauged
model. The chemical potential and chiral twist enter
only through the combination $\beta_1\!+\!\mu L/2\pi$.
Up to the second term the potential is invariant under
$$
\al_1\longrightarrow \al_1+1\mtx{and}
\beta_1+{\mu L\ov 2\pi}\longrightarrow \beta_1+{\mu L\ov 2\pi}+
1+g_0^2/2\pi.\eqno(7.9)
$$
Let us now discuss the potential in the various {\it limiting cases}.\par
\noindent
First assume that there is {\it no chiral twist}, $\beta_1\es 0$, and
that the chemical potential vanishes. Then $\Omega(\beta\to\infty)$
coincides with the {\it groundstate energy}. The minimum in (7.8) is
attained for $n_1\es n_2\es [\ha\!+\!\al_1]$ and we find
$$
E_0(L,\al_1,\beta_1\es 0)=-{\pi\ov 6L}+{2\pi\ov L}{2\pi\ov 2\pi+g_0^2}
\big(\al_1-[\ha+\al_1]\big)^2.\eqno(7.10)
$$
Only for anti-periodic boundary conditions, that is for $\al_1\es 0$,
does this Casimir energy coincide with the corresponding result for
free fermions. For $g_0^2\!\geq\!4\pi$ the Casimir force is always
attractive whereas for $g_0^2\!<\! 4\pi$ it can be attractive or
repulsive, depending on the value of $\al_1$.
The result (7.10) is in agreement with the literature
[11]. For example, it coincides with De Vegas and Destri's result if
we make the identification $\omega_{\ss DD}
\es 2 \pi \al_{\ss 1}$ and $1/\beta_{\ss DD} \es 1+ g_0^2/2\pi$
in formula (42) of that paper.
\par\noindent
For {\it small twists and chemical potential} the minimum is assumed
for $n_i\es 0$ and the potential simplifies to
$$
\Omega(\beta\to\infty)=-{\pi\ov 6L}+{2\pi\ov L}{2\pi\ov 2\pi+g_0^2}
\al_1^2\eqno(7.11)
$$
and does not dependent on the chemical potential.\par\noindent
For {\it vanishing} $g_0$, that is for {\it free fermions}, the minimum
of (7.8) is attained for
$$
n_1=[\ha+\al_1-\beta_1-{\mu L\ov 2\pi}]\mtx{and}
n_2=[\ha+\al_1+\beta_1+{\mu L\ov 2\pi}],
$$
where $[x]$ denotes the biggest integer which is smaller or
equal to $x$. This then leads to the following zero temperature potential
$$\eqalign{
\Omega=&-{\pi\ov 6L} -{2\pi\ov L}(\beta_1+{\mu L\ov 2\pi})^2\cr
&+ {\pi\ov L}\Big\{\al_1-\beta_1-{\mu L\ov 2\pi}
-[\ha+\al_1-\beta_1-{\mu L\ov 2\pi}]\big\}^2\cr
&+{\pi\ov L}\Big\{\al_1+\beta_1+{\mu L\ov 2\pi}
-[\ha+\al_1+\beta_1+{\mu L\ov 2\pi}] \big\}^2 .\cr}\eqno(7.12)
$$
For $\mu\es \beta_1\es 0$ this reduces to the Casimir energy
for free fermions with left-right symmetric twists and agrees
with the results in [46].\par\noindent
Note, however, that for $\beta_{\ss 1}\! \neq \! 0$
we disagree with [11]. The difference is due to the second
term on the right in (7.8).
Let us give two arguments in favour of our result:\parn
The discrepancy arises from the prefactor appearing in the fermionic
determinant (3.19). As discussed earlier this prefactor
implies the breakdown of holomorphic factorization, a property which
has been presupposed in [11]. In appendix C we show that our
results can be reproduced by
starting with massive fermions and taking the limit $m\to 0$.
\par \noindent
The second argument goes as follows: Suppose that $\beta_{\ss 1}\es
\al_1\es 0$. Then (7.12) simplifies to
$$
\Omega (\beta\to\infty)=-{\pi\ov 6L} -{2\pi\ov L}\big({\mu L\ov 2\pi}\big)^2
+ {2\pi\ov L}\big({\mu L\ov 2\pi}-[\ha+{\mu L\ov 2\pi}] \big)^2
.\eqno(7.13)
$$
For massless fermions the Fermi energy is just $\mu$ and
at $T\es 0$ all electron states with energies less then $\mu$
and all positron states with energies less then $-\mu$ are filled.
The other states are empty. Since $d\Omega/d\mu$
is the expectation value of the electric charge in the
presence of $\mu$ we see that it must jump if $\mu$ crosses
an eigenvalue of the first quantized Dirac Hamiltonian $h$.
For vanishing twists the eigenvalues of $h$ are just $E_n\es(n-\ha)\pi/L $.
Indeed, from (7.13) one finds that the electric charge
$$
\<Q\>={d\Omega \ov d\mu}=2
\big[\ha+{\mu L\ov 2\pi}\big]=2n\quad\mtx{for} E_n\leq\mu<E_{n+1}\eqno(7.14)
$$
jumps at these values for $\mu$. Further observe,
that in the {\it thermodynamic limit} $L\to\infty$ the density
$$
{\Omega\ov L}\rightarrow - {2\pi\ov 2\pi+g_0^2}{\mu^2\ov 2\pi}, \eqno(7.15)
$$
reduces for $g_0\es 0$ to the standard result for free electrons.
\titlec{7.2}{Equation of state}
We wish to derive the equation of state for finite $T$
in the infinite volume limit $L\to\infty$.
This may be achieved by interchanging the roles played by $L$ and
$\beta$. More precisely, using that
$$
\Theta\Big[{u\atop w}\Big](\Lambda)=\sqrt{\det(i\Lambda^{-1})}\;
e^{2\pi i w\cdot u}\;\Theta\Big[{-w\atop u}\Big] (i\Lambda^{-1})\eqno(7.16)
$$
we find in analogy with the low temperature limit that for $L\to\infty$
the pressure is given by
$$\eqalign{
\beta p=\lim_{L\to\infty}{1\ov L}&\log {Z\ov {\cal N}_0}
={\pi\ov 6\beta}+{2\pi\ov \beta}{2\pi+g_0^2\ov 2\pi}\beta_0^2\cr
-{\pi\ov 2\beta}\min_{n\in Z^2}\Big[&
{2\pi+g_0^2\ov 2\pi}\big\{n_1+n_2+2\beta_0\big\}^2\cr &
+{2\pi\ov 2\pi+g_0^2}\big\{n_2-n_1+2\al_0+2i{\beta\mu\ov 2\pi}\big\}^2\Big]
.\cr}\eqno(7.17)
$$
Here the minimum of the real part has to be taken.
Again the minimization arises from the saddle point approximation
to the theta function which becomes exact when $L\to \infty$.
For {\it small twists} the miminum is assumed for $n_i\es 0$ and
then
$$
\beta p={\pi\ov 6\beta}-{2\pi\ov\beta}{2\pi\ov 2\pi+g_0^2}
\big(\al_0+i{\beta\mu\ov 2\pi}\big)^2\eqno(7.18)
$$
becomes independent on the chiral twist $\beta_0$.
As we have interchanged the roles of the temporal and spatial
twists this is consistent with
the earlier result that for small twists $\Omega$ is independent of
$\beta _{\ss 1}$. In particular, for $\al_{\ss 0}\es 0$, we are lead
to the following equation of state
$$
p(\beta,\mu,\al_0\es 0) ={\pi\ov 6\beta^2}+
{\mu^2\ov 2\pi}{2\pi\ov 2\pi+g_0^2},\eqno(7.19)
$$
which for small $\beta_0$ relates the pressure to the chemical
potential and temperature.
This result is consistent with the renormalization of the
electric charge which is conjugate to the chemical potential.
It shows in particular that the thermodynamic
behaviour of the Thirring model is
not just the one of free fermions as has been claimed in
[12]. Indeed, the zero point pressure is multiplied by a
factor $2\pi/(2\pi+g_0^2)$.
This modification arises from the coupling of the current to the
harmonic fields. It can not be seen if only the local part of the
auxillary field is considered, which is the
case if one quantizes the model on the infinite Euclidean space.
Furthermore, we see that the 'pressure' $p$ is real only
for $\al_{\ss 0}\es 0$. This phenomenon occurs also in the
Hamiltonian formalism [47]. However, finite temperature
physics dictates anti-periodic boundary conditions,
i.e $\al_{\ss 0} \es 0$, and then $p$ becomes real.

\titlea {8.}{Conformal structure of the ungauged model}
When we discussed the properties of the classical model (1.1)
in chapter 2 we have noticed that for $A_\mu\es 0$
it reduces to a conformal field theory on flat Minkowski spacetime. We
have found the results listed at the end of section 2.\pan
We determine the quantum corrections to these
classical results. As in the previous chapters we
do that within the Euclidean functional approach. Thus
we start from first principles and need not postulate the emerging
{\it Kac-Moody} and {\it Virasoro algebras} in advance [8,16].
When comparing the classical with the quantum results one should
keep in mind that  roles of $\psi_{\ss 0} ^{\ss \dagger }$
and $\psi_{\ss 1} ^{\ss \dagger }$ are interchanged
when one switches from Minkowski to Euclidean spacetime.
For further changes the reader is referred to appendix A.
\par
In what follows it is convenient to exploit the holomorphic
structure of the model. On the torus with flat metric $\hat g_{\mu\nu}$
the Cauchy-Riemann equations read
$$
(\eta_\mu^{\;\,\nu}\pa_\nu-i\pa_\mu)f= 0.\eqno(8.1)
$$
Then one chooses coordinates $x^{\pr a}\es e^a_{\;\,\mu}x^\mu$
and the corresponding complex coordinates $x\es x^{\pr 0}\!+\!ix^{\pr 1}$
such that (8.1) takes the standard form. More explicitly we chose
$$
x=i\bar\tau x^{\ss 0}+i x^{\ss 1} \mtx{so that}
\pa_x = {1\ov  2\tau_0}(\pa_{x^{\ss 0}} -\tau\pa_{x^{\ss 1}}).
\eqno(8.2)
$$
In this section $x$ and $\bar x$ always denote the complex coordinates
belonging to $x^\mu$. In these coordinates the free
Dirac operator and the corresponding Greensfunction are simple
$$
i\slash \partial  =2i\pmatrix{0&\pa_x\cr \pa_{\bar x}&0\cr}\mtx{and}
S(x^\al,y^\beta)={1\ov 2\pi i}\pmatrix{0&1/\xi\cr
1/\bar \xi &0\cr}+O(1),\eqno(8.3)
$$
where $\xi\es x\!-\!y$. The chiral components of the energy momentum
tensor and current are then given by
$$
T_{xx}={\tau_0\ov  2i}(\tau T^{\ss 00}+T^{\ss 01})=
{\tau_0\ov 2i}{d\hat g_{\mu\nu}\ov d\bar \tau}T^{\mu\nu}\mtx{and}
j_x={1\ov  2i} (\tau j^{\ss 0}-j^{\ss 1}).\eqno(8.4)
$$
Using that the energy momentum tensor is conserved and traceless
and that the vector and axial-vector currents are conserved it is easy to
check that these chiral components only depend on $x$ and not on
$\bar x$.
\titlec {8.1}{Virasoro and Kac-Moody algebras}
First we determine the
central charge from the short distance expansion of the $T_{xx}$
correlators. As in the classical theory (see (2.13)) the symmetric
energy momentum tensor measures the change of the effective action
$\Gamma=-\log Z$ under arbitrary variations of the metric. For the
torus there are two independent contributions.
One being due to variations of the modular parameter $\tau$
and its conjugate $\bar\tau$ which
depend implicitly on the metric. The other is due to the variations
of terms which depend explicitly on the metric. Since the chiral component
$T_{xx}$ is gotten by contracting $T^{\mu \nu}$ with
$ d\hat g_{\mu\nu}/d\bar\tau$ it follows that
$$
\<T_{xx}\>={i\tau_0\ov  \sqrt {g(x^\al)}}
\Big ({1\ov  L^2}{\pa\ov  \pa\bar\tau }  + {d\hat g_{\mu\nu}\ov
d\bar\tau} {\delta\ov  \delta g_{\mu \nu}(x^\al)} \Big )\;\Gamma
\lbrack g,\tau,\bar\tau \rbrack\equiv \delta_x
\Gamma\lbrack g,\tau,\bar\tau \rbrack.\eqno(8.5)
$$
It is always understood when doing metric variations,
that we take the flat spacetime limit afterwards.
The $\bar\tau$ variation is constant and may be
skipped in the short distance expansion. \pan
Taking several metric variations of the curvature dependent part
of $\log Z$ with $Z$ from (7.4), (5.3b) and (4.9) we find the following
short distance expansions for the three point correlation function
$$
\<T_{uu}\; T_{vv}\; T_{zz}\>\sim\;-
{3+ 24\pi g_3^2\ov (2\pi)^3}{1\ov (u-v)^2 (u-z)^2 (v-z)^2}
.\eqno(8.6)
$$
Comparing with the general expression [16] we read off the
{\it central charge} and the conformal weight of the energy momentum
tensor
$$
c=3+24 g_3^2 \pi \mtx{and} h_{T_{xx}}=\;2\;.\eqno(8.7)
$$
The first contribution is that of three free fields.
The $\gc-$dependent term already appeared
in the classical analysis and is related to the
coupling to the background curvature. It is well known from the
minimal conformal series. Note that the couplings $g_1$ and $g_2$
do not affect the central charge. In particular, if we subtract the
central charge of the auxiliary fields and set $g_3\es 0$
then the value of $c$ is the same as the one for the Thirring model, namely
$c\es 1$ [16].
\par
Next we determine the {\it Kac-Moody algebra} of the
$U(1)$ currents. To derive the correlation functions with current
insertions we couple the fermions to a gauge field, that is
consider the 'gauged' model without Maxwell term. For example,
$$
<j^{\mu}(x^\al)\;j^\nu(y^\beta)>\;=
{1\ov e^2\sqrt{g(x^\al)g(y^\beta )}}\;{\delta^2 \Gamma \lbrack g,A\rbrack
\ov \delta A_\nu(x^\al)\delta A_\mu(y^\beta)}\vert_{A=0}. \eqno(8.8)
$$
Using (4.6) on flat spacetime and without Maxwell term, together with
$$
\pa_\mu \phi=\eta_\mu^{\;\,\nu}A_\nu^{\ss T},\mtx{where}
A_\mu^{\ss T}=A_\mu-{2\pi\ov L}t_\mu-\nabla_\mu{1\ov\lap}\nabla^\nu A_\nu
\eqno(8.9)
$$
is the transversal part of $A_\mu$, one obtains the following short distance
expansion
$$
\<j_x\;j_y \>\; \sim -{1\ov 2\pi}{1\ov 2\pi +g_2^2}{1\ov (x-y)^2}\;.
\eqno(8.10)
$$
We read off the value $k$ of the {\it central extension} in the
$U(1)$-Kac-Moody algebra to be
$$
k ={2\pi \ov 2\pi + \gb ^2 }.\eqno(8.11)
$$
Finally we need to determine the conformal weight of the current. From
$$
\<j_x \; j_y \; T_{zz}\>\; \sim -{1\ov 4\pi^2}{1\ov 2\pi +g_2^2}{1\ov
(x-z)^2(y-z)^2}\eqno(8.12)
$$
we obtain $h_j=1$. To summarize, the symmetry algebra is
the semidirect product of a Virasoro algebra with central charge
(8.7) and a $U(1)$ Kac-Moody current algebra with central extension
(8.11).
\titlec{8.2}{Conformal weights}
To unravel the possible representations of the Virasoro algebra
realized in the model we must determine
the conformal weights of the fundamental fields.
The short distance expansions of the {\it fermionic two-point function}
with $T_{zz}$ follows from the metric variation of the Greensfunction
$$
\<\psi_0(x)\;\psi_1^{\ss\dagger}(y)\>=
S_{ij}(x,y)\cdot e^{[i g_1g_3\sigma(x)+\al G_R(x,x)]-[x\to y]
-2\al G(x,y)}\eqno(8.13a)
$$
where
$$
\al={1\ov 4}\Big(g_1^2-{2\pi g_2^2 \ov 2\pi+g^2_2}\Big).\eqno(8.13b)
$$
and $S_{ij}$ is the fermionic Green function in the external
gravitational field and harmonic gauge field but with $\phi$
and $\lambda$ set to zero. More precisely,
$$
\<\psi_{\ss 0}(x)\;\psi_{\ss 1}^{\ss\dagger}(y)\;T_{zz}\>=
{1\ov Z}\delta_z\Big(Z\<\psi_{\ss 0}(x)\;\psi_{\ss 1}^{\ss\dagger}(y)\>\Big).
$$
However, since $Z\sim\exp[F(\R^2)]$, its metric variation vanishes
after the flat spacetime limit has been taken.
We refer to appendix B for the variation of $S_{ij}$ and
$G(x,y)$. Collecting the most singular terms, we arrive at
$$\eqalign{
\<\psi_{\ss 0}&(x)\;\psi_{\ss 1}^{\ss\dagger}(y)\;T_{zz}\>\,\sim\,
{1\ov 2\pi i}{1\ov 4\pi}\Big[{1\ov (z\!-\!x)(z\!-\!y)}
\big({1\ov z\!-\!x}-{1\ov z\!-\!y}\big)\cr
&-{i g_1g_3\ov x\!-\!y}\big({1\ov (z\!-\!x)^2}-{1\ov(z\!-\!y)^2}\big)
+{\al\ov 2\pi}\big({1\ov z\!-\!x}-{1\ov z\!-\!y}\big)^2\Big]\;e^{2\al
G(x,y)}.\cr}
\eqno(8.14)
$$
Using that
$$
\pa_x e^{2\al G(x,y)}= -\pa_y e^{2\al G(x,y)}=-{\al\ov 2\pi}\al{1\ov
x-y}\;e^{2\al G(x,y)},\eqno(8.15)
$$
we find that the $2$-point function varies under a infinitesimal
 conformal transformation, paramatrized by $f(z)$, as
$$\eqalign{
{1\ov i}&\oint dz f(z)
\<\psi_{\ss 0}(x)\;\psi_{\ss 1}^{\ss\dagger}(y) \; T_{zz}\> =
\Big\{f(x)\pa_x+f(y)\pa_y\cr
& +\ha\big(1+{\al\ov 2\pi}\big)
\big[f^\pr (x) +f^\pr (y)\big]-{i\gaa \gc \ov 2}
\big[f^\pr (x) -f^\pr (y)\big]\Big\}
\;\<\psi_{\ss 0} (x)\psi_{\ss 1}^{\ss\dagger }(y) \>.\cr}\eqno(8.16)
$$
Note that the exponential factor has been absorbed to recover
the correlation function $\<\psi_ {\ss 0} (x)\psi_{\ss 1}
^{\ss\dagger }(y) \>$.
The short distance expansion with $T_{\bar z \bar z}$
is calculated similarly. Then one reads off the conformal weights
$$ \eqalign{
h_{ {\psi_0}}=&\;\ha+{1\ov  16 \pi} \gaa ^2
-{1\ov  16\pi}{2\pi\gb^2\ov  2\pi + \gb^2 }
-{i\gaa \gc \ov  2} \cr
h_{ {\psi_1^\dagger }}=&\; (h_{ {\psi_0}})^{\ss \dagger} \cr
\bar h_{ {\psi_0}}=&\;{1\ov  16 \pi} \gaa ^2
-{1\ov  16 \pi} {2 \pi \gb ^2 \ov  2 \pi + \gb^2 }
-{i\gaa \gc \ov  2}\;.\cr} \eqno(8.17)
$$
Thus we have reproduced the classical results supplemented
by additional $\gaa $ and $\gb $ dependent quantum corrections.
In the Thirring model limit $\gc \es 0$ and $\gaa \es \gb \es g$,
these terms add up to give the known anomalous dimension
appearing in the Thirring model [16].
The last classical term is a peculiar feature of the solution.
For the conformal weight to be real we are obliged to choose $\gc $
imaginary.\par
Let us now turn to the {\it auxiliary fields}. It is straightforward
to compute the correlators
$$\eqalign{
\<\lam_x\;T_{zz} \> \;\sim \;&{1\ov  4\pi}\gc {1 \ov  (x-z)^2} \cr
\<\lam _x \;\lam _y \; T_{zz} \> \;\sim\;&-{1\ov  32\pi ^2}
{1 \ov  (x-z) (y-z) } \cr
\<\phi _x \;\phi _y \; T_{zz} \> \;\sim \;&-{1\ov  16\pi }
{1 \ov  (x-z) (y-z) }.\cr}
\eqno(8.18)
$$
We see that the classical results are unchanged, that is
for $\gc \! \neq \! 0$ the scalar field $\lam $ is
not primary and for $\gc \es 0 $ we find the conformal weights
$h_{\lam } \es  h_{\phi} \es \; 0$.\pan
Finally we turn to {\it vertex operators} or
exponentials of the auxiliary fields. In contrast to $\lam$ and $\phi$
those are well defined even on the extended plane. Recalling
the regularization prescription (6.10) we find
$$\eqalign{
\<:e^{\al_1\phi(x)}:\,:e^{\al_2\phi(y)}:T_{zz}\>\; \sim &
- {1\ov 16\pi}{1\ov 2\pi\!+\!g_2^2}\Big[{\al_1\ov z\!-\!x}+
{\al_2\ov z\!-\!y}\Big]^2\cr &
\<:e^{\al_1\phi(x)}:\,:e^{\al_2\phi(y)}:\> \cr}\eqno(8.19)
$$
and hence
$$\eqalign{
{1\ov i}\int_C&\;f(z)\<:e^{\al_1\phi(x)}:\,:e^{\al_2\phi(y)}:T_{zz}\>\; \sim
\Big[f(x)\pa_x +f(y)\pa_y \cr &
-{1\ov 8(2\pi\!+\!g_2^2)}\big(\al_1^2 f^\pr (x)
+\al_2^2 f^\pr (y)\big)\Big]\<:e^{\al_1\phi(x)}:\,:e^{\al_2\phi(y)}:\>\; .
\cr}\eqno(8.20)
$$
{}From this we read off the conformal weights of the vertex operators
$$
h_i= \bar h_i=-{\al_i^2\ov 8 (2 \pi +\gb ^2 )}\; . \eqno(8.21)
$$
Note that $\al_i$ must be imaginary to get a positive weight.
A similar analysis for the $\lam$-field yields
$$\eqalign{
{1\ov i}&\int_C\;f(z)\<:e^{\al_1\lam(x)}:\,:e^{\al_2\lam(y)}:T_{zz}\>\;
\sim \Big[f(x)\pa_x+f(y)\pa_y \cr&
-{\al_1\ov 2}({\al_1\ov 8\pi}\!+\!g_3)f^\pr (x)
-{\al_2\ov 2}({\al_2\ov 8\pi}\!+\!g_3)f^\pr (y) \Big]
\<:e^{\al_1\lam(x)}:\,:e^{\al_2\lam(y)}:\>\cr}\eqno(8.22)
$$
and hence
$$
h_i= -\ha\al_i\,({\al_i\ov 8\pi}+g_3)\; .\eqno(8.23)
$$
Here both $\al _i $ and $g_3$ must be imaginary
for the weights to be positive. Note that contrary to $\lam$
the fields $:e^{\al\lam(x)}:$ remain primary when the $\lam{\cal
R }$
coupling is switched on. This coupling results only
in a shift of the conformal weights.
\titlec{8.3}{$U(1)$-charges}
To see how the left and right Kac Moody currents act on the fermionic fields
we notice that after the integration over the auxiliary fields
the $A$-dependence of the fermionic Greenfunction factorizes as
$$
\<\psi_0(x)\psi_1^{\ss \dagger}(y)\>_{\ss A}=e^{\ha m_\gam
\int\varphi\lap\varphi}
\cdot e^{-eg(x)}\;\<\psi_0(x)\psi_1^{\ss \dagger}(y)\>_{\ss A=0}\;e^{-e\bar
g(y)},
$$
where
$$
g(x)=-i\al(x)+\gam_5\beta \varphi(x),\qquad
\beta={2\pi\ov 2\pi+g_2^2}.
$$
Also, using that on flat spacetime
$$\eqalign{
\phi(x)&=-i\int\pa_z G(x,z)A^z +i\int \pa_{\bar z}G(x,z)A^{\bar z}\cr
\al(x)&=\int\pa_z G(x,z)A^z+\int\pa_{\bar z}G(x,z)A^{\bar z},\cr}\eqno(8.24)
$$
one ends up with
$$\eqalign{
\<\psi_0(x)\psi_1(y)^\dagger j_z\>\;
&={1\ov 4\pi i}\Big[{4\pi+g_2^2\ov 2\pi+g_2^2}\,{1\ov z\!-\!x}
+{g_2^2\ov 2\pi+g_2^2}\,{1\ov z\!-\!y}\Big]
\<\psi_0(x)\psi_1(y)^\dagger\> \cr
\<\psi_0(x)\psi_1(y)^\dagger j_{\bar z}\>\;
&={1\ov 4\pi i}\Big[{g_2^2\ov 2\pi+g_2^2}\,{1\ov \bar z\!-\!\bar x}
+{4\pi+g_2^2\ov 2\pi+g_2^2}\,{1\ov \bar z\!-\!\bar y}\Big]
\<\psi_0(x)\psi_1(y)^\dagger\> \cr}\eqno(8.25)
$$
and thus obtains the following the $U(1)$ charges
$$
q_{\psi_0}=\ha\big(1+{2\pi\ov 2 \pi+\gb^2}\big)\mtx ,
\bar q_{\psi_0}=\ha\big(1-{2\pi\ov 2\pi+\gb^2}\big).
\eqno(8.26)
$$
We have used the convention where the electric
charge $q\!+\!\bar q$ is unity. In the Thirring model limit we
can compare (8.26) with the results obtained in [16].
For that we need to rescale the currents such that the central
extension (8.11) of the Kac-Moody algebra becomes unity
$$
j_z \to \sqrt {1+\gb^2/2\pi} \; j_z \; . \eqno(8.27)
$$
Now it is easy to see that we agree with [16]
if we make the identification
$$
\bar g_{\ss {Fu}}={\gb^2\ov 4\pi}{1\ov \sqrt{1+\gb^2/2\pi}}\; . \eqno(8.28)
$$
Let us summarize our results. The classical
conformal and axial transformations of all fields besides
$\phi$ and $\lambda$ are deformed. The longitudinal part
of the current-current interaction in (1.1) changes
the conformal weights of the fermion field only. The transversal
part affects all weigths and $U(1)$-charges. The background
charge changes the conformal weight of the vertex operators
belonging to the scalar field.\pan
Of course, the same structure is found in the other chiral sector.

\titlec {8.4}{Finite size effects}
When quantizing a conformal field theory on a spacetime
with finite volume one introduces a length scale.
The presence of this length scale in turn breaks the conformal
invariance and gives rise to finite size effects. It has
been conjectured [13] that the finite size effects
are proportional to the central charge. For example
when one stretches space time, $x^\al\to a x^\al$, then
the change of the effective action is proportional to $c$:
$$
\Gamma_{ax}-\Gamma_x=-{c\ov 6}\log a\cdot\chi,\eqno(8.29)
$$
where $\chi$ is the Euler number of the euclidean space time.
In [48] this conjecture has been proven for
a class of conformal field theories on spaces with boundaries.
The only important assumption has been that the regurlarization
respects general covariance. In this subsection we shall
show that the conjecture does not hold for the model (1.1)
on Riemanniann surfaces.\par
Unfortunately, the only global conformal transformations on the
torus are translations which do not give rise to finite size
effects. Also, the Euler number vanishes and according to
(8.29) the finite size effects are insensitve to the value of $c$.
For that reason we quantize the ungauged model (1.1) on the
sphere where the global conformal group is the Moebius group.\par
An effective method to compute finite size effects has been
developped in [48]. It is based on the following
observation: Any conformal transformation $z\to w(z)$ is a composition
of a diffeomorphism (defined by the same $w$) and a compensating
Weyl transformation $g_{\mu\nu}\to e^{2\sigma}g_{\mu\nu}$
with
$$
e^{2\sigma}={dw(z)\ov dz}{d\bar w(\bar z)\ov d\bar z},\qquad
z=x^0+ix^1.\eqno(8.30)
$$
Therefore, chosing a diffeomorphism invariant regularization
one has
$$
0=\delta \Gamma_{Diff}=\delta\Gamma_{Conf}-\delta\Gamma_{Weyl}.\eqno(8.31)
$$
Now we apply the techniques of the previous sections to derive
the change $\delta \Gamma_{Weyl}$ of the effective action
on the sphere under Weyl transformations. This change is given
by the trace anomaly.\pan
The change of the effective action under Weyl rescalings is
$$
\delta\Gamma_{Weyl}=-\log {\int{\cal D}(\lam\phi)\det (i\slash D )
\exp (-S_B\lbrack A=0,g\rbrack )\ov
\int{\cal D}(\lam\phi)\det (i\hat{\slash D})\exp (-S_B\lbrack A=0,\hat g
\rbrack )}\;, \eqno(8.32)
$$
where $S_B$ is the bosonic action (3.35) with vanishing
gauge field. Also, since on the sphere there are no harmonic
vector fields the term $\sim h^2$ in $S_B$ is not present.
Thus the calculation on the sphere is actually simpler
as on the torus (see 7.1) since there is no integration
over the harmonic fields. As on the torus we must impose
the conditions (3.38) in order to eliminate the additional
degrees of freedom we introduced in the derivative coupling
representation. Thus we obtain
$$
\delta\Gamma_{Weyl}=\log {\hat V\ov V}-{S_L\ov 24\pi}+
+{g_3^2\ov 4}\int \R{1\ov \lap}\R+
\log{\det^\pr \triangle\ov \det^\pr \hat\triangle}.\eqno(8.33)
$$
Here we used that (3.30a) in the trivial sector still holds
on the sphere. Also we used the scaling law (4.4b). $S_L$
is the Liouville action (3.30b) in which we can not put
$\hat\R$ to zero, since
$$
\int\sqrt{g}\R=8\pi=4\pi\chi\eqno(8.34)
$$
for any curvature and thus in particular for $\hat\R$.
As for the fermions (see 3.26) one introduces the $1$-parametric
family of Laplacians
$$
\lap_\tau=e^{-2\tau\sigma}\hat\lap\eqno(8.35)
$$
interpolating between $\hat\lap$ and $\lap$. The $\tau$ derivative
of the corresponding determinant is given by the trace anomaly [32,48].
The explicit calculation yields
$$
\log{\det}^\pr{\lap\over\hat\lap}=2\int_0^1 d\tau
\int\sqrt{g^\tau}\Big(-{1\ov 4\pi} a_1^\tau-P^\tau\Big)\sigma,\eqno(8.36)
$$
Again $g^\tau$ is the determinant and $a_1^\tau={1\ov 6}\R^\tau$ the
relevant Seeley-deWitt coefficient of the deformed metric
$g^\tau_{\mu\nu}\es e^{2\tau\sigma}\hat g_{\mu\nu}$.
$P^\tau$ is the projection onto the zero-mode of $\lap^\tau$.
Using that the normalized zeromode is constant and
$\sim 1/\sqrt{V^\tau}$, one finds
$$
\log{\det}^\pr{\lap\over\hat\lap}=\log{V\ov\hat V}+{1\ov 12\pi}
S_L.\eqno(8.37)
$$
The $\sim \log V$ term cancels against the same term in (8.33) and
we end up with
$$
\delta\Gamma={\gc^2\ov 4}\int\sqrt{g}\R\ilap (\R-{8\pi\ov V})
-{3\ov 24\pi }\int\sqrt{\hat g}\hat\R\si+{3\ov 24\pi}\int\sqrt{\hat g}
\si\hat\lap\si.\eqno(8.38)
$$
Now we can see why
the finite size conjecture generally fails to be true, although
it holds for theories without background charge on domains with boundaries
[48]. Take the simple case of a dilatation $w(z)\es a z$. Then, the
conformal angle is a constant $\si\es\log a$ and $(\R-8\pi/V)\es 0$. Thus
the first term in (8.38) vanishes and the finite
size effect does not depend on $\gc ^2$. It is
given by
$$
\delta\Gamma=-{3\ov 24\pi}\log a\int\sqrt{\hat g}\hat\R=-\log a\eqno(8.39)
$$
and does not agrees with (8.29) since $c$ in (8.7) depends on
$g_3$. Thus we have disproved the conjecture. On other
Riemannian surfaces one would find the same result:
the effective action scales as in (8.7) where $c$ is the
central charge of the model without background charge. It
is evident that the finite size scaling comes from the middle term
$\sim \log a\int \sqrt{\hat g}\hat R$ in (8.38).
\par
It is interesting to compare the finite size scaling on
Riemannian surfaces with the one on domains with boundaries.
In the presence of boundaries (8.36) is modified to
$$
\log{\det}^\pr{\lap\over\hat\lap}=-{1\ov 2\pi}\int_0^1 d\tau\Big(
\int\sqrt{g^\tau}a_1^\tau\;\sigma
+\oint\sqrt{\tilde g^\tau}b_1^\tau\;\sigma\Big),\eqno(8.40)
$$
where the second integral is over the boundary of spacetime
and $\tilde g_{\mu\nu}$ the induced metric on this boundary.
On a domain we can always put $\hat R$ to zero and the
middle term in (8.40) does not contribute to the scaling.
The scaling comes from the surface term in (8.40). Diffeomorphism
invariance implies that the bulk term determines the surface
term (up to diffeomorphism invariant surface terms). This
is how the central charge, defined by the short distance expansion
of the $T_{zz}$-correlators and thus by the bulk term, reemerges
in the scaling law (8.7), which is determined by the
surface term.
\acknow{This work has been supported by the Swiss National Science
Foundation. We would like to thank K. Gawedzki, C. Kiefer and E. Seiler
for discussions.}

\titlea {}{Appendix A: Conventions}

In this appendix we set up our notation and give a list of
useful formulae. Let $g_{\mu \nu} $ be the metric of
spacetime. The sign convention for the curvature tensors
is such that
$$
{\cal R}^{\al}_{\beta\gamma\delta}=
\Gamma^{\al}_{\delta\beta,\gamma}-\Gamma^{\al}_{\gamma\beta,\delta}
+\Gamma^{\si}_{\delta\beta}\Gamma^{\al}_{\gamma\si}
-\Gamma^{\si}_{\gamma\beta}\Gamma^{\al}_{\delta\si}\mtx{and}
{\cal R}_{\beta\delta}={\cal R}^{\al}_{\beta\al\delta}.\eqno(A.1)
$$
In $2$ dimensions the only independent component is $\R _{{\ss 0101}}$.
In order to couple fermions to gravity we must introduce a local
Lorentz frame (or tetrad), $e_{\mu a} $, relating the
Lorentz and spacetime indices:
$$\eqalign{
e_{\al a} e_\beta^{\;\,a} = g_{\al \beta} \mtx ,
e_{\al a} e^\al_{\;\,b}=\eta_{ab} \mtx ,
\eta_{ab}=\pmatrix{1&0\cr 0&-1\cr}  \cr}.
\eqno(A.2)
$$
The latin and greek indices are Lorentz and spacetime indices,
respectively. All physical laws should be general- and Lorentz covariant.
If $g_{\al \beta}$ has Euclidean signature then $\eta_{ab}$
in (A.2) is changed to $\delta_{ab}$.\par\noindent
The 'curved' gamma matrices are related to the flat ones as
$$
\gammu=\beinmui\gamtia. \eqno(A.3)
$$
We us the following chiral representation for the flat $\gamma$'s:
$$
\hat\gamma^{{\ss 0}}_{{\ss M}}=\pmatrix{0&1\cr 1&0\cr}
\qquad,\qquad \hat\gamma^{{\ss 1}}_{{\ss M}}=\pmatrix{0&-1\cr 1&0\cr}
\eqno(A.4)
$$
and in Euclidean spacetime we may choose
$$
\hat\gamma^{{\ss 0}}_{{\ss E}}=\hat\gamma^{{\ss 0}}_{{\ss M}}
\qquad,\qquad \hat\gamma^{{\ss 1}}_{{\ss E}}=i\hat\gamma^{{\ss 1}}_{{\ss M}}.
\eqno(A.5)
$$
We may also define
$$
\fiveti=\;\hat\gamma^{{\ss 0}}_{{\ss M}}\hat\gamma^{{\ss 1}}_{{\ss M}}=
-i\hat\gamma^{{\ss 0}}_{{\ss E}}\hat\gamma^{{\ss 1}}_{{\ss E}}
=\pmatrix{1&0\cr 0&-1\cr}\eqno(A.6a)
$$
The relations
$$
\gamtia_{{\ss M}}\fiveti=\epsilon^a_{\;\,b}\gamtib \quad,\quad
\gamtia_{{\ss E}}\fiveti=-i\epsilon^a_{\;\,b}\gamtib ,\mtx{where}
\epsilon_{ab}=\pmatrix{0&1\cr -1&0\cr}\eqno(A.6b)
$$
are particular to $2$ dimensions and play an important
role in the body of the paper. Note that depending whether
one is in Minkowskian or Euclidian spacetime the Lorentzindex
$a$ is raised with $\eta^{ab}$ or $\delta^{ab}$.
The curved space analogue of (A.6a) reads
$$
\gamfive={1\over 2}\eta_{\mu\nu}\gammu_{{\ss M}}\gamnu_{{\ss M }}
={1\over 2i}\eta_{\mu\nu}\gammu_{{\ss E }}\gamnu_{{\ss E}}=\fiveti,
\eqno(A.7)
$$
where $\eta_{\mu\nu}\es \sqrt{\vert g\vert}\epsilon_{\mu\nu}$ is
the antisymmetric tensor (whereas the flat metric has Lorentz-
indices, the antisymmetric tensor has space-time indices).
To implement local Lorentz invariance one needs to introduce
a connection $\omega_{\mu ab}$.
For example, in the Lagrangean the Lorentz-covariant
derivative acting on the spinors read
$$
D_\mu=\dmu + i \spinmu, \eqno(A.8)
$$
where the spin connection $\spinmu$ is defined by
$$
D_\mu\equiv \partial_{\mu }e_\nu^{\;\,a}\!-\!\Gamma^\lam _{\mu\nu}
e_\lam^{\;\,a}\!+\!\omega_{\mu ab}e_\nu^{\;\, b} =0,\quad
\spinmu=\ha\omega_{\mu ab} \Sigma^{ab},\quad
\Sigma^{ab}={1\over 4i}[\gamtia,\gamtib]. \eqno(A.9)
$$
In $2$ dimensions this reduces to
$$
\spinmu^{{\ss M}}={1\ov 2i}\omega_{\mu{\ss 01}}\gamfive\quad\mtx{or}\quad
\spinmu^{{\ss E}}=\ha\omega_{\mu{\ss 01}}\gamfive.
\eqno(A.10)
$$
Finally we list some useful scaling relations. If the $2$-bein
scales as $e_\mu^{\;\,a}\es e^\sigma\hat e_\mu^{\;\,a}$ then the above
introduced quantities scale as
$$\eqalign{
\g=e^{2\si}\gind\quad,\quad
\sqrt{g}=& e^{2\si}\sqrt{\hat g}\quad,\quad
\R=e^{-2\si} (\Rind-2\lapind\si)\cr
\omega_{\mu a b}=\hat \omega_{\mu ab}&-\pa_a\si
\hat e_{\mu b}+\pa_b \si\hat e_{\mu a},\cr
\Gamma^\al_{\mu\nu}=\hat\Gamma^\al_{\mu\nu}+\Big(\pa_\mu\si\delta^\al_\nu
&+\pa_\nu\si\delta^\al_\mu-\pa_\beta\si\hat g^{\beta\al}
\hat g_{\mu\nu}\Big),\cr
\lap= e^{-2\si} \lapind \quad,\quad \slash\partial &+i\slash \omega=
e^{-{3\over 2} \si} (\hat {\slash \partial } + i \hat{ \slash \omega })
e^{{1\over 2}\si }.\cr} \eqno(A.11)
$$

\titlea {}{Appendix B: Variational formulae}
\vskip 0.5truecm

\noindent
In the following $D_\mu $ denotes the spacetime and Lorentz covariant
derivative. How it acts on spacetime and Lorentz tensors follows
from the first formula in (A.9).
\par \noindent
Using the definition of the Christoffel symbol and
(A.2) it is straight forward to show that
$$\eqalign{
\delta g_{\mu\nu}=\delta e_\mu^{\;\,a}e _{\nu a}+e_\mu^{\;\,a}\delta
e_{\nu a}\quad &,\quad
\delta\sqrt{g}=\ha \sqrt{g} g^{\mu \nu} \delta g_{\mu \nu}\cr
\delta\gamma^\mu=-\gamma^\nu e^\mu_{\;\,a}\delta e_\nu^{\;\,a}\quad,\quad
\delta\eta_\mu^{\;\;\nu}&=\ha (\eta^{\al\nu}\delta g_{\mu\al}
-\eta_\mu^{\;\;\si}g^{\nu\rho}\delta g_{\si\rho})\cr
\delta\Gamma^{\al}_{\mu\nu}=\ha g^{\al \beta }(D_\nu \delta g_{\beta\mu}
+&D_\mu\delta g_{\beta \nu} -D_\beta \delta g_{\mu \nu}).\cr}
\eqno(B.1)
$$
For some formulae related to the variation of the tetrad let us refer to
[49]
$$\eqalign{
&\delta e^\mu_{\;\;a}=\ha e_{\nu a}\delta g^{\mu\nu}-t_a^{\;\;b}
e^\mu_{\;\,b}\mtx ,
\delta e_\mu^{\;\,a}=\ha e^{\nu a}\delta g_{\mu \nu}-t^a_{\;\;b}
e_\mu^{\;\, b}\cr&\qquad\mtx{where}
t^a_{\;\;b}=\ha (e^{\nu a}\delta e_{\nu b}-e^\nu_{\;\, b}\delta
e_\nu^{\;\, a}). \cr} \eqno(B.2)
$$
Then using (A.9) it is easy to see that
$$
\delta\omega_{\mu ab}=D_\mu t_{ab} - \al _{\mu ab}\;\quad
\al_{\mu ab}=\ha e^\al_{\;\,a}e^\beta_{\;\, b}(D_\al\delta g_{\beta \mu}-
D_\beta\delta g_{\al\mu}).\eqno(B.3)
$$
When performing the variation of curvature dependent expressions we have
used the identities
$$ \eqalign{
& g^{\mu\nu}\delta \R_{\mu\nu}=\omega^{\al}_{\;;\al}\;,\mtx{where}
\omega^{\al}=g^{\mu \nu} \delta\Gamma^{\al}_{\mu\nu}-g^{\al\nu}
\delta\Gamma^\mu_{\mu\nu}\cr &\mtx{and}
\int\sqrt{g}\,\omega^{\al}A_{\al}=\int\sqrt {g}
\{g^{\al\beta}\nabla_{\mu} A^{\mu}-\nabla^{\al}A^{\beta}\}
\delta g_{\al\beta}\;.\cr}\eqno(B.4)
$$
Depending on the topology of spacetime,
the induced curvature $\hat \R $ appearing in (A.11)
may be different from zero. In this case it is not possible to express
the conformal angle $\si $ in terms of the curvature scalar.
Nevertheless, to perform variations of $\si$-dependent expressions, the
identity
$$
\delta (\sqrt{g}\R )=-2\delta(\sqrt{g}\lap\si)\eqno(B.5)
$$
proves to be useful.\par\noindent
Taking the variations of the equations
$$
\sqrt{g}\Box G(x,y) =-\de(x-y) \mtx{and}\sqrt{g}\,i\slash D S(x,y)=\de^2(x-y)
\eqno(B.6)
$$
for the scalar and fermionic Greensfunctions we may derive
(up to contact terms) the following variational formulae
$$\eqalign{
\de G=&\int\;\big(-\ha g^{\mu\nu}g^{\al\beta}
+g^{\al\mu}g^{\beta\nu}\big)\pa_{\al} G(x,u)\,\pa_{\beta}
G(u,y)\sqrt{g}\delta g_{\mu\nu}\cr
\de S=&{i\ov 4}\int\;\Big(2S(x,u)\gamma^\mu D^\nu
S(u,y)\!-\!D_\al [S(x,u)\gamma^\delta\eta_\delta^{\;\,\mu}
\eta^{\nu\al} S(u,y)]\Big)\sqrt{g}\delta g_{\mu\nu},\cr}
\eqno(B.9)
$$
where all arguments and derivatives which are not made explicit
in the integral refer to the coordinate $u$ over which is integrated.
Finally, we need the following formula for the variation
of the inverse Laplacian
$$
\delta\left(\ilap f\right) =\ilap\left(\delta f-\delta (\lap )
\ilap f\right)-{1\over 2V}\int\sqrt{g}g^{\mu\nu}\delta g_{\mu\nu}\ilap f
, \eqno(B.10)
$$
where $V$ is the volume of spacetime and $f$ an arbitrary function. To
prove this identity we note that for $f\in (\hbox{Kern}\triangle)^{\bot}$ we
have
$$
\lap\ilap f=f.
$$
Varying this equation yields
$$
\lap (\delta \ilap f ) =\delta f-(\delta \lap ) \ilap f
$$
which may be inverted to give
$$
\delta \left(\ilap f \right )  = \ilap \left (\delta f
- \delta (\lap )
\ilap f \right) + {1\over V} \int \sqrt{g}
\delta \left(\ilap f \right ) \; .\eqno(B.11)
$$
Varying the identity
$$
{1\over V} \int \sqrt{g}\ilap f =\; 0
$$
allows to replace the last term of (B.11) to obtain the required
result (B.10).

\titlea{Appendix C:}{Canonical approach to the partition function}
In this appendix we compute the partition function for massive
Dirac fermions in the canonical formalism. In the limit $m\to 0$
we confirm some of the results in sections $3$ and $7$.
For massive fermions one cannot consistently impose chirally twisted
boundary conditions. However, from the explicit eigenvalues (3.17)
one sees at once that the chiral twist $\beta_1$ and the
chemical potential are equivalent. One can easily verify that
this equivalence holds also for massless fermions in the canonical approach
and that $\beta_1\sim \mu L/2\pi$. Let us therefore compute the partition
function
$$
Z(\beta)=Tr\big[e^{-\beta:(H-\mu Q):}\big]\eqno(C.1)
$$
for massive Dirac fermions with chemical potential $\mu$ on a
cylinder with (nonchiral) twisted boundary conditions
$$
\psi(x+L,t)=-e^{-2i\pi\al_1}\psi(x,t).\eqno(C.2)
$$
For massive particles it is more convenient to use the Dirac representation
$$
\gam^0=\sigma_3\quad\gam^1=-i\sigma_2,\quad\gam^5=\gam^0\gam^1 =-\sigma_1.
\eqno(C.3)
$$
The Dirac field is expanded in terms of the eigenmodes of the
first quantized Hamiltonian
$$
h=\pmatrix{m&i\pa_x\cr i\pa_x&-m\cr}\eqno(C.4)
$$
as
$$
\Psi(x,t)=\sum_n \psi_{n,+}b_n+\sum_n \psi_{n,-}d^{\ss\dagger}_n,\eqno(C.5a)
$$
where the $\psi_{n,+}$ and $\psi_{n,-}$ are the positive and
negative energy modes,
$$\eqalign{
&\psi_{n,+}=e^{-i\om_nt-i\lam_n x}c_n,\quad
\psi_{n,-}=e^{i\om_nt-i\lam_n x}\gam_1 c_n,\cr
&\qquad c_n={1\ov\sqrt{2\om_n (\om_n+m)L}}\pmatrix{\om_n+m\cr \lam_n\cr}.\cr}
\eqno(C.5b)$$
The momenta $\lam_n$ and energies $\om_n$ are determined by the
boundary condition (C.2) to be
$$
\lam_n={2\pi\ov L}\big(n-\ha-\al_1\big)\mtx{and} \om_n=\sqrt{m^2+\lam_n^2}.
\eqno(C.5c)
$$
After normal ordering the 'positron' operators with respect to
the Fock vacuum defined by $H$ we find
$$
(H-\mu Q)=
\sum_n(\om_n-\mu)b^{\ss\dagger}_nb_n+\sum_n (\om_n+\mu)d^{\ss\dagger}_n d_n
-\sum_n(\om_n+\mu),\eqno(C.6)
$$
where the last $c$-number term represents the infinite vacuum
contribution which must be regularized.
To do that we employ the zeta function regularization.
That is we define the zeta-function for $s\!>\!1$ by the sum
$$
\zeta(s)=\sum_n(\om_n+\mu)^{-s}
$$
which in turn defines an analytic function on the whole complex $s$-plane
up to a simple pole at $s\es 1$. The analytic continuation is constructed
by a Poisson resummation
$$
\sum_{n}(\om_n+\mu)^{-s}={L^s\ov 2\pi}\sum_{n}F(n)\eqno(C.7a)
$$
where
$$
F(\xi)= e^{2\pi i\xi(\ha-\al_1)}\int
dy\; e^{i\xi y}\big[\sqrt{\tilde m^2+y^2}\ +\tilde\mu\big]^{-s},\eqno(C.7b)
$$
and $\tilde m \es Lm$, $\tilde\mu\es L\mu$. Taking the Mellin transform
of (C.7b) we find
$$
\eqalign{
F(\xi)=& e^{2\pi i\xi(\ha-\al_1)}{1\ov\Gamma(s)}\int dy\ e^{i\xi y}\int
dt\ t^{s-1}e^{-t\sqrt{\tilde m^2+y^2}-t\tilde\mu}\cr
=&{-2\ov\Gamma(s)}e^{2\pi i\xi(\ha-\al_1)}\int dt\ t^{s-1}e^{-t\tilde\mu}\
{d\ov dt}K_0(\tilde\mu\sqrt{\xi^2+t^2})\cr
=&{2\tilde m\ov\Gamma(s)}e^{2\pi i\xi(\ha-\al_1)}\int dt\ t^s e^{-t\tilde\mu}
{K_1(\tilde\mu\sqrt{\xi^2+t^2})\ov \sqrt{\xi^2+t^2}}. \cr}\eqno(C.7c)
$$
$F$ diverges at $0$ since the Kelvin function
$K_1(z)\sim 1/z$ for small $z$. It follows that the $n\es0$
term in (C.7a) diverges. This divergence is regularized by substracting
the ground state energy of the infinite volume system. Indeed,
because of the exponential decay of the Bessel function for large
arguments, only the $n\es 0$ term contributes for infinite volume. So we
find for the regularized sum
$$
\sum_n(\om_n+\mu)^{-s}={\tilde m L^s\ov\Gamma(s)\pi}\sum_{n\neq 0}\!\int
dt\ e^{2\pi in(\ha-\al_1)} t^s e^{-t\tilde\mu}{K_1(\tilde m\sqrt{n^2+t^2})
\ov \sqrt{n^2+t^2}}.\eqno(C.8)
$$
Noe we perform the limit $m\rightarrow 0$. Only the most
singular term in the expansion of the Bessel function contributes,
hence
$$\eqalign{
\sum_n(\om_n+\mu)^{-s}=&{L^s\ov\Gamma(s)\pi}\sum_{n\neq 0} \int
dt\ e^{2\pi in(\ha-\al_1)} t^s e^{-t\tilde\mu}{1\ov (n^2+t^2)}\cr
=&{sL^s\ov \pi}\sum_{n\neq 0} e^{2\pi in(\ha-\al_1)}\sqrt{\tilde\mu}
n^{s-\ha}S_{-s-\ha;\ha}(\tilde\mu n),\cr}\eqno(C.9)
$$
where $S_{a;b}(z)$ is the Lommel function [50]. In particular for
$s\es-1$ this function is  $S=1/z$ so that finally
$$
\sum_n(\om_n+\mu)^{reg}=-{1\ov \pi L}\sum_{n\neq 0}{(-)^n\ov n^2} e^{-2\pi
in\al_1} ={\pi\ov 6L}-{2\pi\ov L}(\al_1-[\al_1\!+\!\ha])^2.
\eqno(C.10)
$$
Inserting this into (C.6) then yields the regularized expression
$$
:H-\mu Q:=\sum_n(\om_n-\mu)b^{\ss\dagger}_nb_n +\sum_n (\om_n+\mu)
b^{\ss\dagger}_nd_n-{\pi\ov 6L}+{2\pi\ov L}(\al_1-[\al_1\!+\!\ha])^2.
\eqno(C.11)
$$
For small $\mu$ the normal ordering is $\mu$-independent so that
$$
\langle 0\vert :H-\mu Q:\vert 0\rangle=
-{\pi\ov 6L}+{2\pi\ov L}(\al_1-[\al_1+\ha])^2=
\langle 0\vert :H:\vert 0\rangle
\eqno(C.12)
$$
is independent of $\mu$ and coincides with the Casimir energy
[46]. \par
Finally we compute the partition function. Using (C.12) we
easily find
$$\eqalign{
Z(\beta)&=\tr\big[e^{-\beta:(H-\mu Q):}\big]=q^{[\al_1^2-{1\ov 12}]}\cr
&=\prod_{n>[\ha+\al_1]}^\infty
(1+q^{(n-\ha-\al_1)}e^{\beta\mu})\prod_{n>-[\ha+\al_1]}^\infty
(1+q^{(n-\ha+\al_1)}e^{\beta\mu})\cdot\cr
&\;\prod_{n>[\ha-\al_1]}^\infty
(1+q^{(n-\ha+\al_1)}e^{-\beta\mu})\prod_{n>-[\ha-\al_1]}^\infty
(1+q^{(n-\ha-\al_1)}e^{-\beta\mu})\cr
&\;= {1\over |\eta(\tau)|^2}
\Theta\Big[{-\al_1\atop
i\mu {\beta\ov 2\pi}}\Big](0,\tau)\
\bar\Theta\Big[{ -\al_1   \atop
-i\mu {\beta\ov 2\pi} }\Big](0,\tau),\cr}\eqno(C.13)
$$
where we have used the product representation of the theta functions
in the last idendity and that $q=e^{2\pi i\tau}=e^{-2\pi\beta/L}$.
A non-vanishing chiral twist $\beta_1$ can now we included
by shifting the chemical potential. Thus we have confirmed
the formula (3.19) in our paper.\par
Note that for $\mu\neq 0$ the zero-temperature limit of the
grand potential is not equal to the vacuum expectation
value of $:H-\mu Q:$. For $\mu\neq 0$ all states up to
the $\mu$-dependent fermi energy are filled. For example,
for $\om_1<\mu<\om_2$ the limit $\lim_{\beta\to\infty} \Omega$
reduces to the expectation value of $:H-\mu Q:$ in the one-electron
state.

\begrefchapter{References}
\refno{1.} W. Dittrich and M. Reuter, Effective Lagrangians in QED,
Lecture notes in Physics, Springer, Heidelberg, 1984.
\refno{2.} E.V. Shuryak, The QCD vacuum, hadrons and superdense
matter, World Scientific, Singapore, 1988.
\refno{3.} I. Sachs and A. Wipf, Helv. Phys. Acta {\bf 65} (1992) 653.
\refno{4.} see e.g. H. Leutwyler and A. Smilga, Spectrum of Dirac operator
and role of winding number in QCD, preprint BUTP-92/10.
\refno{5.} N.D. Birrell and P.C.W. Davies, Phys. Rev. {\bf D18} (1978) 4408.
\refno{6.} G.W. Gibbons and M.J. Perry, Proc. R. Soc. London {\bf A 358}
(1978) 467.
\refno{7.} C.G. Callen, R.F. Dashen and D.J. Gross, Phys. Lett. {\bf 63B}
(1976) 334.
\refno{8.} K. Johnson, Nuovo Cim. {\bf 20} (1964) 773.
\refno{9.} C. Jayewardena, Helv. Phys. Acta {\bf 61} (1988) 636.
\refno{10.} D.Z. Freedman and K. Pilch, Phys. Lett. {\bf 213B} (1988) 331;
D.Z. Freedman and K. Pilch, Ann. Phys. {\bf 192} (1989) 331;
S. Wu, Comm. Math. Phys. {\bf 124} (1989) 133.
\refno{11.} C. Destri and J.J. deVega, Phys. Lett. {\bf 223B} (1989) 365.
\refno{12.} H. Yokota, Prog. Theor. Phys. {\bf 77} (1987) 1450.
\refno{13.} J.L. Cardy, Fields, Strings and Statistical Mechanics, Les
Houches, 1988; I. Affleck, Phys. Rev. Lett. {\bf 56} (1986) 746;
H. Bl{\"o}te, J. Cardy and M. Nightingale, Phys. Rev. Lett. {56} (1986) 742.
\refno{14.} B. Klaiber, Lecture notes in Phys. XA, Gordon and
Breach, New York, 1968.
\refno{15.} C.G. Callan, S. Coleman and R. Jackiw, Ann. Phys. {\bf 59}
(1970) 42.
\refno{16.} P. Furlan, G.M. Sotkov and I.T. Todorov, Riv. Nuovo Cim.
{\bf 12} (1989) 1.
\refno{17.} J. Bagger, M. Goulian, in Proc. 18th Int. Conf. Diff. Geom.
Meth. in Physics, Plenum Press, 1990.
\refno{18.} W.E. Thirring, Ann. Phys. {\bf 3} (1958) 91; Nuovo Cim. {\bf 9}
(1958) 1007; V. Glaser, Nuovo Cim. {\bf 9} (1958) 990.
\refno{19.} R. Casalbuoni, Il Nuovo Cim. {\bf A33} (1976) 115.
\refno{20.} L. O'Raifeartaigh and A. Wipf, Phys. Lett. {251B} (1990) 361;
L. Feher, L. O'Raifeartaigh, P. Ruelle, I. Tsutsui and A. Wipf,
Phys. Rep. {\bf 222} (1992) 1.
\refno{21.} L. Dolan and R. Jackiw, Phys. Rev. {\bf D9} (1974) 3320.
\refno{22.} L. Alvarez-Gaume and E. Witten, Nucl. Phys. {\bf B234} (1983)
269; H. Leutwyler, Phys. Lett. {\bf 153B} (1985) 65.
\refno{23.} F. Lenz, H.W.L. Naus, K. Ohta and M. Thies, Quantum Mechanics
of Gauge Fixing, Univ. Erlangen preprint, 1993.
\refno{24.} A.V. Smilga, Are $Z_N$ bubbles really there ?, Bern
preprint BUTP-93/3.
\refno{25.} L. Susskind, Phys. Rev. {\bf D20} (1979) 2610; N. Weiss,
Phys. Rev. {\bf D24} (1981) 475; {\bf D25} (1982) 2667.
\refno{26.} C. Itzykson and J.M. Drouffe, Statistical field theory,
Cambridge Univ. Press, 1989.
\refno{27.} N.D. Birrell and P.C.W. Davies, Quantum fields in curved
space, Cambridge Univ. Press, 1982.
\refno{28.} H. Joos, Nucl. Phys. {\bf B17} (Proc. Suppl.) (1990) 704;
Helv. Phys. Acta {\bf 63} (1990) 670.
\refno{29.} A. Actor, Fortschritte der Phys. {\bf 35} (1987) 793;
K. Kirsten, J. Phys. A {\bf 24} (1991) 3281.
\refno{30.}S. Blau, M. Visser and A. Wipf, Int. J. Mod. Phys. {\bf A4}
(1989) 1467.
\refno{31.} A. Weil, Elliptic functions according to Eisenstein
and Kronecker, Springer, Berlin, 1976.
\refno{32.} see e.g. A.M. Polyakov, Gauge fields and Strings, Harwood
Academic Publishers, 1987.
\refno{33.} A.J. da Silva, M. Gomes and R. K{\"o}berle, Phys. Rev.
{\bf D34} (1986) 504; M. Gomes and A.J. da Silva, Phys. Rev. {\bf D34}
(1986) 3916.
\refno{34.} K. Gawedzky, Conformal field theory, to appear in
Birkh{\"a}user.
\refno{35.} C. Wiesendanger and A. Wipf, Running coupling constants
from finite size effects, preprint ETH-TH/93-10 and ZU-TH 6/93.
\refno{36.} S. Christensen and S. Fulling, Phys. Rev. {\bf D15} (1977) 2088.
\refno{37.} J. Schwinger, Phys. Rev. {\bf 128} (1962) 2425.
\refno{38.} S. Blau, M. Visser and A.Wipf, Int. J. Mod. Phys. {\bf A4}
(1992) 5406.
\refno{39.} J. Ambjorn and S. Wolfram, Ann. Phys. {\bf 147} (1983) 1.
\refno{40.} R.L. Stratonovich, Sov. Phys. Dokl. {\bf 2} (1958) 416;
J. Hubbard, Phys. Rev. Lett. {\bf 3} (1958) 77.
\refno{41.} S. Coleman, Phys. Rev. {\bf 11} (1975) 2088;
J. Froehlich and E. Seiler, Helv. Phys. Acta {\bf 49} (1976) 889;.
Ch.-H. Tze, in Chiral Solitons, ed. by. K.F. Liu, World Scientific,
Singapore, 1987.
\refno{42.} M. L{\"u}scher, Ann. Phys. {\bf 142} (1982) 359.
\refno{43.} S.M. Christensen, Phys. Rev. {\bf D14} (1976) 2490.
\refno{44.} P. Gilkey, Invariance theory, the heat equation and Athiyah
Singer Index theorem, Publish or Perish, 1984; A.E.M. van deVen,
Nucl. Phys. {\bf B250} (1985) 593.
\refno{45.} K. Fredenhagen and M. Marcu, Commun. Math. Phys. {\bf 92}
(1983) 81.
\refno{46.} C. Kiefer and A. Wipf, Functional Schr{\"o}dinger equation
for fermions in external gauge fields, preprint ETH-TH/93-17 and
ZU-TH 8/93; S. Iso and H. Murayama, Progr. Theor. Phys. {\bf 84} (1990) 142.
\refno{47.} D. L{\"u}st and S. Theisen, Lectures on string theory,
Lecture notes in physics, Springer, 1989
\refno{48.} A. Dettki and A. Wipf, Nucl. Phys. {\bf B377} (1992) 252.
\refno{49.} H. Leutwyler and S. Mallik, Z. Phys. {\bf C33} (1986) 205.
\refno{50.} I.S. Gradshteyn and I.M. Ryzhik, Table of Integrals, Series
and Products, Academic, London, 1980.

\vfill\eject

\end


\magnification=\magstep1
\font \authfont               = cmr10 scaled\magstep4
\font \fivesans               = cmss10 at 5pt
\font \headfont               = cmbx12 scaled\magstep4
\font \markfont               = cmr10 scaled\magstep1
\font \ninebf                 = cmbx9
\font \ninei                  = cmmi9
\font \nineit                 = cmti9
\font \ninerm                 = cmr9
\font \ninesans               = cmss10 at 9pt
\font \ninesl                 = cmsl9
\font \ninesy                 = cmsy9
\font \ninett                 = cmtt9
\font \sevensans              = cmss10 at 7pt
\font \sixbf                  = cmbx6
\font \sixi                   = cmmi6
\font \sixrm                  = cmr6
\font \sixsans                = cmss10 at 6pt
\font \sixsy                  = cmsy6
\font \smallescriptfont       = cmr5 at 7pt
\font \smallescriptscriptfont = cmr5
\font \smalletextfont         = cmr5 at 10pt
\font \subhfont               = cmr10 scaled\magstep4
\font \tafonts                = cmbx7  scaled\magstep2
\font \tafontss               = cmbx5  scaled\magstep2
\font \tafontt                = cmbx10 scaled\magstep2
\font \tams                   = cmmib10
\font \tamss                  = cmmib10
\font \tamt                   = cmmib10 scaled\magstep2
\font \tass                   = cmsy7  scaled\magstep2
\font \tasss                  = cmsy5  scaled\magstep2
\font \tast                   = cmsy10 scaled\magstep2
\font \tasys                  = cmex10 scaled\magstep1
\font \tasyt                  = cmex10 scaled\magstep2
\font \tbfonts                = cmbx7  scaled\magstep1
\font \tbfontss               = cmbx5  scaled\magstep1
\font \tbfontt                = cmbx10 scaled\magstep1
\font \tbms                   = cmmib10 scaled 833
\font \tbmss                  = cmmib10 scaled 600
\font \tbmt                   = cmmib10 scaled\magstep1
\font \tbss                   = cmsy7  scaled\magstep1
\font \tbsss                  = cmsy5  scaled\magstep1
\font \tbst                   = cmsy10 scaled\magstep1
\font \tenbfne                = cmb10
\font \tensans                = cmss10
\font \tpfonts                = cmbx7  scaled\magstep3
\font \tpfontss               = cmbx5  scaled\magstep3
\font \tpfontt                = cmbx10 scaled\magstep3
\font \tpmt                   = cmmib10 scaled\magstep3
\font \tpss                   = cmsy7  scaled\magstep3
\font \tpsss                  = cmsy5  scaled\magstep3
\font \tpst                   = cmsy10 scaled\magstep3
\font \tpsyt                  = cmex10 scaled\magstep3
\vsize=22.5true cm
\hsize=13.8true cm
\hfuzz=2pt
\tolerance=500
\abovedisplayskip=3 mm plus6pt minus 4pt
\belowdisplayskip=3 mm plus6pt minus 4pt
\abovedisplayshortskip=0mm plus6pt minus 2pt
\belowdisplayshortskip=2 mm plus4pt minus 4pt
\predisplaypenalty=0
\clubpenalty=10000
\widowpenalty=10000
\frenchspacing
\newdimen\oldparindent\oldparindent=1.5em
\parindent=1.5em
\skewchar\ninei='177 \skewchar\sixi='177
\skewchar\ninesy='60 \skewchar\sixsy='60
\hyphenchar\ninett=-1
\def\newline{\hfil\break}%
\catcode`@=11
\def\folio{\ifnum\pageno<\z@
\uppercase\expandafter{\romannumeral-\pageno}%
\else\number\pageno \fi}
\catcode`@=12 
  \mathchardef\Gamma="0100
  \mathchardef\Delta="0101
  \mathchardef\Theta="0102
  \mathchardef\Lambda="0103
  \mathchardef\Xi="0104
  \mathchardef\Pi="0105
  \mathchardef\Sigma="0106
  \mathchardef\Upsilon="0107
  \mathchardef\Phi="0108
  \mathchardef\Psi="0109
  \mathchardef\Omega="010A
  \mathchardef\bfGamma="0\the\bffam 00
  \mathchardef\bfDelta="0\the\bffam 01
  \mathchardef\bfTheta="0\the\bffam 02
  \mathchardef\bfLambda="0\the\bffam 03
  \mathchardef\bfXi="0\the\bffam 04
  \mathchardef\bfPi="0\the\bffam 05
  \mathchardef\bfSigma="0\the\bffam 06
  \mathchardef\bfUpsilon="0\the\bffam 07
  \mathchardef\bfPhi="0\the\bffam 08
  \mathchardef\bfPsi="0\the\bffam 09
  \mathchardef\bfOmega="0\the\bffam 0A

\def\sq{\hbox{\rlap{$\sqcap$}$\sqcup$}}

\def\utw{\smash{\rlap{\lower5pt\hbox{$\sim$}}}}
\def\udtw{\smash{\rlap{\lower6pt\hbox{$\approx$}}}}

\def\diameter{{\ifmmode\mathchoice
{\ooalign{\hfil\hbox{$\displaystyle/$}\hfil\crcr
{\hbox{$\displaystyle\mathchar"20D$}}}}
{\ooalign{\hfil\hbox{$\textstyle/$}\hfil\crcr
{\hbox{$\textstyle\mathchar"20D$}}}}
{\ooalign{\hfil\hbox{$\scriptstyle/$}\hfil\crcr
{\hbox{$\scriptstyle\mathchar"20D$}}}}
{\ooalign{\hfil\hbox{$\scriptscriptstyle/$}\hfil\crcr
{\hbox{$\scriptscriptstyle\mathchar"20D$}}}}
\else{\ooalign{\hfil/\hfil\crcr\mathhexbox20D}}%
\fi}}


\def\bbbc{{\mathchoice {\setbox0=\hbox{$\displaystyle\rm C$}\hbox{\hbox
to0pt{\kern0.4\wd0\vrule height0.9\ht0\hss}\box0}}
{\setbox0=\hbox{$\textstyle\rm C$}\hbox{\hbox
to0pt{\kern0.4\wd0\vrule height0.9\ht0\hss}\box0}}
{\setbox0=\hbox{$\scriptstyle\rm C$}\hbox{\hbox
to0pt{\kern0.4\wd0\vrule height0.9\ht0\hss}\box0}}
{\setbox0=\hbox{$\scriptscriptstyle\rm C$}\hbox{\hbox
to0pt{\kern0.4\wd0\vrule height0.9\ht0\hss}\box0}}}}
\def\bbbe{{\mathchoice {\setbox0=\hbox{\smalletextfont e}\hbox{\raise
0.1\ht0\hbox to0pt{\kern0.4\wd0\vrule width0.3pt height0.7\ht0\hss}\box0}}
{\setbox0=\hbox{\smalletextfont e}\hbox{\raise
0.1\ht0\hbox to0pt{\kern0.4\wd0\vrule width0.3pt height0.7\ht0\hss}\box0}}
{\setbox0=\hbox{\smallescriptfont e}\hbox{\raise
0.1\ht0\hbox to0pt{\kern0.5\wd0\vrule width0.2pt height0.7\ht0\hss}\box0}}
{\setbox0=\hbox{\smallescriptscriptfont e}\hbox{\raise
0.1\ht0\hbox to0pt{\kern0.4\wd0\vrule width0.2pt height0.7\ht0\hss}\box0}}}}
\def\bbbq{{\mathchoice {\setbox0=\hbox{$\displaystyle\rm Q$}\hbox{\raise
0.15\ht0\hbox to0pt{\kern0.4\wd0\vrule height0.8\ht0\hss}\box0}}
{\setbox0=\hbox{$\textstyle\rm Q$}\hbox{\raise
0.15\ht0\hbox to0pt{\kern0.4\wd0\vrule height0.8\ht0\hss}\box0}}
{\setbox0=\hbox{$\scriptstyle\rm Q$}\hbox{\raise
0.15\ht0\hbox to0pt{\kern0.4\wd0\vrule height0.7\ht0\hss}\box0}}
{\setbox0=\hbox{$\scriptscriptstyle\rm Q$}\hbox{\raise
0.15\ht0\hbox to0pt{\kern0.4\wd0\vrule height0.7\ht0\hss}\box0}}}}
\def\bbbt{{\mathchoice {\setbox0=\hbox{$\displaystyle\rm
T$}\hbox{\hbox to0pt{\kern0.3\wd0\vrule height0.9\ht0\hss}\box0}}
{\setbox0=\hbox{$\textstyle\rm T$}\hbox{\hbox
to0pt{\kern0.3\wd0\vrule height0.9\ht0\hss}\box0}}
{\setbox0=\hbox{$\scriptstyle\rm T$}\hbox{\hbox
to0pt{\kern0.3\wd0\vrule height0.9\ht0\hss}\box0}}
{\setbox0=\hbox{$\scriptscriptstyle\rm T$}\hbox{\hbox
to0pt{\kern0.3\wd0\vrule height0.9\ht0\hss}\box0}}}}
\def\bbbs{{\mathchoice
{\setbox0=\hbox{$\displaystyle     \rm S$}\hbox{\raise0.5\ht0\hbox
to0pt{\kern0.35\wd0\vrule height0.45\ht0\hss}\hbox
to0pt{\kern0.55\wd0\vrule height0.5\ht0\hss}\box0}}
{\setbox0=\hbox{$\textstyle        \rm S$}\hbox{\raise0.5\ht0\hbox
to0pt{\kern0.35\wd0\vrule height0.45\ht0\hss}\hbox
to0pt{\kern0.55\wd0\vrule height0.5\ht0\hss}\box0}}
{\setbox0=\hbox{$\scriptstyle      \rm S$}\hbox{\raise0.5\ht0\hbox
to0pt{\kern0.35\wd0\vrule height0.45\ht0\hss}\raise0.05\ht0\hbox
to0pt{\kern0.5\wd0\vrule height0.45\ht0\hss}\box0}}
{\setbox0=\hbox{$\scriptscriptstyle\rm S$}\hbox{\raise0.5\ht0\hbox
to0pt{\kern0.4\wd0\vrule height0.45\ht0\hss}\raise0.05\ht0\hbox
to0pt{\kern0.55\wd0\vrule height0.45\ht0\hss}\box0}}}}
\def\bbbz{{\mathchoice {\hbox{$\sans\textstyle Z\kern-0.4em Z$}}
{\hbox{$\sans\textstyle Z\kern-0.4em Z$}}
{\hbox{$\sans\scriptstyle Z\kern-0.3em Z$}}
{\hbox{$\sans\scriptscriptstyle Z\kern-0.2em Z$}}}}
\def\qed{\ifmmode\sq\else{\unskip\nobreak\hfil
\penalty50\hskip1em\null\nobreak\hfil\sq
\parfillskip=0pt\finalhyphendemerits=0\endgraf}\fi}
\newfam\sansfam
\textfont\sansfam=\tensans\scriptfont\sansfam=\sevensans
\scriptscriptfont\sansfam=\fivesans
\def\sans{\fam\sansfam\tensans}
\def\stackfigbox{\if
Y\FIG\global\setbox\figbox=\vbox{\unvbox\figbox\box1}%
\else\global\setbox\figbox=\vbox{\box1}\global\let\FIG=Y\fi}
\def\placefigure{\dimen0=\ht1\advance\dimen0by\dp1
\advance\dimen0by5\baselineskip
\advance\dimen0by0.4true cm
\ifdim\dimen0>\vsize\pageinsert\box1\vfill\endinsert
\else
\if Y\FIG\stackfigbox\else
\dimen0=\pagetotal\ifdim\dimen0<\pagegoal
\advance\dimen0by\ht1\advance\dimen0by\dp1\advance\dimen0by1.4true cm
\ifdim\dimen0>\pagegoal\stackfigbox
\else\box1\vskip4true mm\fi
\else\box1\vskip4true mm\fi\fi\fi}
%
\def\begfig#1cm#2\endfig{\par
\setbox1=\vbox{\dimen0=#1true cm\advance\dimen0
by1true cm\kern\dimen0#2}\placefigure}
\def\begdoublefig#1cm #2 #3 \enddoublefig{\begfig#1cm%
\vskip-.8333\baselineskip\line{\vtop{\hsize=0.46\hsize#2}\hfill
\vtop{\hsize=0.46\hsize#3}}\endfig}
\def\begfigsidebottom#1cm#2cm#3\endfigsidebottom{\dimen0=#2true cm
\ifdim\dimen0<0.4\hsize\message{Room for legend to narrow;
begfigsidebottom changed to begfig}\begfig#1cm#3\endfig\else
\par\def\figure##1##2{\vbox{\noindent\petit{\bf
Fig.\ts##1\unskip.\ }\ignorespaces ##2\par}}%
\dimen0=\hsize\advance\dimen0 by-.8true cm\advance\dimen0 by-#2true cm
\setbox1=\vbox{\hbox{\hbox to\dimen0{\vrule height#1true cm\hrulefill}%
\kern.8true cm\vbox{\hsize=#2true cm#3}}}\placefigure\fi}
\def\begfigsidetop#1cm#2cm#3\endfigsidetop{\dimen0=#2true cm
\ifdim\dimen0<0.4\hsize\message{Room for legend to narrow; begfigsidetop
changed to begfig}\begfig#1cm#3\endfig\else
\par\def\figure##1##2{\vbox{\noindent\petit{\bf
Fig.\ts##1\unskip.\ }\ignorespaces ##2\par}}%
\dimen0=\hsize\advance\dimen0 by-.8true cm\advance\dimen0 by-#2true cm
\setbox1=\vbox{\hbox{\hbox to\dimen0{\vrule height#1true cm\hrulefill}%
\kern.8true cm\vbox to#1true cm{\hsize=#2 true cm#3\vfill
}}}\placefigure\fi}
\def\figure#1#2{\vskip1true cm\setbox0=\vbox{\noindent\petit{\bf
Fig.\ts#1\unskip.\ }\ignorespaces #2\smallskip
\count255=0\global\advance\count255by\prevgraf}%
\ifnum\count255>1\box0\else
\centerline{\petit{\bf Fig.\ts#1\unskip.\
}\ignorespaces#2}\smallskip\fi}

\def\begtab#1cm#2\endtab{\par
   \ifvoid\topins\midinsert\medskip\vbox{#2\kern#1true cm}\endinsert
   \else\topinsert\vbox{#2\kern#1true cm}\endinsert\fi}
\def\begpet{\vskip6pt\bgroup\petit}
\def\endpet{\vskip6pt\egroup}
\newcount\frpages
\newcount\frpagegoal
\def\freepage#1{\global\frpagegoal=#1\relax\global\frpages=0\relax
\loop\global\advance\frpages by 1\relax
\message{Doing freepage \the\frpages\space of
\the\frpagegoal}\null\vfill\eject
\ifnum\frpagegoal>\frpages\repeat}
\newdimen\refindent
\def\begrefchapter#1{\titlea{}{\ignorespaces#1}%
\bgroup\petit
\setbox0=\hbox{1000.\enspace}\refindent=\wd0}
\def\ref{\goodbreak
\hangindent\oldparindent\hangafter=1
\noindent\ignorespaces}
\def\refno#1{\goodbreak
\hangindent\refindent\hangafter=1
\noindent\hbox to\refindent{#1\hss}\ignorespaces}
\def\vec#1{{\textfont1=\tams\scriptfont1=\tamss
\textfont0=\tenbf\scriptfont0=\sevenbf
\mathchoice{\hbox{$\displaystyle#1$}}{\hbox{$\textstyle#1$}}
{\hbox{$\scriptstyle#1$}}{\hbox{$\scriptscriptstyle#1$}}}}
\def\petit{\def\rm{\fam0\ninerm}%
\textfont0=\ninerm \scriptfont0=\sixrm \scriptscriptfont0=\fiverm
 \textfont1=\ninei \scriptfont1=\sixi \scriptscriptfont1=\fivei
 \textfont2=\ninesy \scriptfont2=\sixsy \scriptscriptfont2=\fivesy
 \def\it{\fam\itfam\nineit}%
 \textfont\itfam=\nineit
 \def\sl{\fam\slfam\ninesl}%
 \textfont\slfam=\ninesl
 \def\bf{\fam\bffam\ninebf}%
 \textfont\bffam=\ninebf \scriptfont\bffam=\sixbf
 \scriptscriptfont\bffam=\fivebf
 \def\sans{\fam\sansfam\ninesans}%
 \textfont\sansfam=\ninesans \scriptfont\sansfam=\sixsans
 \scriptscriptfont\sansfam=\fivesans
 \def\tt{\fam\ttfam\ninett}%
 \textfont\ttfam=\ninett
 \normalbaselineskip=11pt
 \setbox\strutbox=\hbox{\vrule height7pt depth2pt width0pt}%
 \normalbaselines\rm
\def\vec##1{{\textfont1=\tbms\scriptfont1=\tbmss
\textfont0=\ninebf\scriptfont0=\sixbf
\mathchoice{\hbox{$\displaystyle##1$}}{\hbox{$\textstyle##1$}}
{\hbox{$\scriptstyle##1$}}{\hbox{$\scriptscriptstyle##1$}}}}}
\nopagenumbers
%
\let\header=Y
\let\FIG=N
\newbox\figbox
\output={\if N\header\headline={\hfil}\fi\plainoutput\global\let\header=Y
\if Y\FIG\topinsert\unvbox\figbox\endinsert\global\let\FIG=N\fi}
\let\lasttitle=N
\def\bookauthor#1{\vfill\eject
     \bgroup
     \baselineskip=22pt
     \lineskip=0pt
     \pretolerance=10000
     \authfont
     \rightskip 0pt plus 6em
     \centerpar{#1}\vskip2true cm\egroup}
\def\bookhead#1#2{\bgroup
     \baselineskip=36pt
     \lineskip=0pt
     \pretolerance=10000
     \headfont
     \rightskip 0pt plus 6em
     \centerpar{#1}\vskip1true cm
     \baselineskip=22pt
     \subhfont\centerpar{#2}\vfill
     \parindent=0pt
     \baselineskip=16pt
     \leftskip=2.2true cm
     \markfont Springer-Verlag\newline
     Berlin Heidelberg New York\newline
     London Paris Tokyo Singapore\bigskip\bigskip
     [{\it This is page III of your manuscript and will be reset by
     Springer.}]
     \egroup\let\header=N\eject}
\def\centerpar#1{{\parfillskip=0pt
\rightskip=0pt plus 1fil
\leftskip=0pt plus 1fil
\advance\leftskip by\oldparindent
\advance\rightskip by\oldparindent
\def\newline{\break}%
\noindent\ignorespaces#1\par}}
\def\part#1#2{\vfill\supereject\let\header=N
\centerline{\subhfont#1}%
\vskip75pt
     \bgroup
\textfont0=\tpfontt \scriptfont0=\tpfonts \scriptscriptfont0=\tpfontss
\textfont1=\tpmt \scriptfont1=\tbmt \scriptscriptfont1=\tams
\textfont2=\tpst \scriptfont2=\tpss \scriptscriptfont2=\tpsss
\textfont3=\tpsyt \scriptfont3=\tasys \scriptscriptfont3=\tenex
     \baselineskip=20pt
     \lineskip=0pt
     \pretolerance=10000
     \tpfontt
     \centerpar{#2}
     \vfill\eject\egroup\ignorespaces}
\newtoks\AUTHOR
\newtoks\HEAD
\catcode`\@=\active
\def\author#1{\bgroup
\baselineskip=22pt
\lineskip=0pt
\pretolerance=10000
\markfont
\centerpar{#1}\bigskip\egroup
{\def@##1{}%
\setbox0=\hbox{\petit\kern2.5true cc\ignorespaces#1\unskip}%
\ifdim\wd0>\hsize
\message{The names of the authors exceed the headline, please use a }%
\message{short form with AUTHORRUNNING}\gdef\leftheadline{%
\hbox to2.5true cc{\folio\hfil}AUTHORS suppressed due to excessive
length\hfil}%
\global\AUTHOR={AUTHORS were to long}\else
\xdef\leftheadline{\hbox to2.5true
cc{\noexpand\folio\hfil}\ignorespaces#1\hfill}%
\global\AUTHOR={\def@##1{}\ignorespaces#1\unskip}\fi
}\let\INS=E}
\def\address#1{\bgroup
\centerpar{#1}\bigskip\egroup
\catcode`\@=12
\vskip2cm\noindent\ignorespaces}
\let\INS=N%
\def@#1{\if N\INS\unskip\ $^{#1}$\else\if
E\INS\noindent$^{#1}$\let\INS=Y\ignorespaces
\else\par
\noindent$^{#1}$\ignorespaces\fi\fi}%
\catcode`\@=12
\headline={\petit\def\newline{ }\def\fonote#1{}\ifodd\pageno
\rightheadline\else\leftheadline\fi}
\def\rightheadline{\hfil Missing CONTRIBUTION
title\hbox to2.5true cc{\hfil\folio}}
\def\leftheadline{\hbox to2.5true cc{\folio\hfil}Missing name(s) of the
author(s)\hfil}
\nopagenumbers
\let\header=Y

\let\lasttitle=N
 \def\contribution#1{\vfill\supereject
 \ifodd\pageno\else\null\vfill\supereject\fi
 \let\header=N\bgroup
 \textfont0=\tafontt \scriptfont0=\tafonts \scriptscriptfont0=\tafontss
 \textfont1=\tamt \scriptfont1=\tams \scriptscriptfont1=\tams
 \textfont2=\tast \scriptfont2=\tass \scriptscriptfont2=\tasss
 \par\baselineskip=16pt
     \lineskip=16pt
     \tafontt
     \raggedright
     \pretolerance=10000
     \noindent
     \centerpar{\ignorespaces#1}%
     \vskip12pt\egroup
     \nobreak
     \parindent=0pt
     \everypar={\global\parindent=1.5em
     \global\let\lasttitle=N\global\everypar={}}%
     \global\let\lasttitle=A%
     \setbox0=\hbox{\petit\def\newline{ }\def\fonote##1{}\kern2.5true
     cc\ignorespaces#1}\ifdim\wd0>\hsize
     \message{Your CONTRIBUTIONtitle exceeds the headline,
please use a short form
with CONTRIBUTIONRUNNING}\gdef\rightheadline{\hfil CONTRIBUTION title
suppressed due to excessive length\hbox to2.5true cc{\hfil\folio}}%
\global\HEAD={HEAD was to long}\else
\gdef\rightheadline{\hfill\ignorespaces#1\unskip\hbox to2.5true
cc{\hfil\folio}}\global\HEAD={\ignorespaces#1\unskip}\fi
\catcode`\@=\active
     \ignorespaces}
 \def\contributionnext#1{\vfill\supereject
 \let\header=N\bgroup
 \textfont0=\tafontt \scriptfont0=\tafonts \scriptscriptfont0=\tafontss
 \textfont1=\tamt \scriptfont1=\tams \scriptscriptfont1=\tams
 \textfont2=\tast \scriptfont2=\tass \scriptscriptfont2=\tasss
 \par\baselineskip=16pt
     \lineskip=16pt
     \tafontt
     \raggedright
     \pretolerance=10000
     \noindent
     \centerpar{\ignorespaces#1}%
     \vskip12pt\egroup
     \nobreak
     \parindent=0pt
     \everypar={\global\parindent=1.5em
     \global\let\lasttitle=N\global\everypar={}}%
     \global\let\lasttitle=A%
     \setbox0=\hbox{\petit\def\newline{ }\def\fonote##1{}\kern2.5true
     cc\ignorespaces#1}\ifdim\wd0>\hsize
     \message{Your CONTRIBUTIONtitle exceeds the headline,
please use a short form
with CONTRIBUTIONRUNNING}\gdef\rightheadline{\hfil CONTRIBUTION title
suppressed due to excessive length\hbox to2.5true cc{\hfil\folio}}%
\global\HEAD={HEAD was to long}\else
\gdef\rightheadline{\hfill\ignorespaces#1\unskip\hbox to2.5true
cc{\hfil\folio}}\global\HEAD={\ignorespaces#1\unskip}\fi
\catcode`\@=\active
     \ignorespaces}
\def\motto#1#2{\bgroup\petit\leftskip=6.5true cm\noindent\ignorespaces#1
\if!#2!\else\medskip\noindent\it\ignorespaces#2\fi\bigskip\egroup
\let\lasttitle=M
\parindent=0pt
\everypar={\global\parindent=\oldparindent
\global\let\lasttitle=N\global\everypar={}}%
\global\let\lasttitle=M%
\ignorespaces}
\def\abstract#1{\bgroup\petit\noindent
{\bf Abstract: }\ignorespaces#1\vskip28pt\egroup
\let\lasttitle=N
\parindent=0pt
\everypar={\global\parindent=\oldparindent
\global\let\lasttitle=N\global\everypar={}}%
\ignorespaces}
\def\titlea#1#2{\if N\lasttitle\else\vskip-28pt
     \fi
     \vskip18pt plus 4pt minus4pt
     \bgroup
\textfont0=\tbfontt \scriptfont0=\tbfonts \scriptscriptfont0=\tbfontss
\textfont1=\tbmt \scriptfont1=\tbms \scriptscriptfont1=\tbmss
\textfont2=\tbst \scriptfont2=\tbss \scriptscriptfont2=\tbsss
\textfont3=\tasys \scriptfont3=\tenex \scriptscriptfont3=\tenex
     \baselineskip=16pt
     \lineskip=0pt
     \pretolerance=10000
     \noindent
     \tbfontt
     \rightskip 0pt plus 6em
     \setbox0=\vbox{\vskip23pt\def\fonote##1{}%
     \noindent
     \if!#1!\ignorespaces#2
     \else\setbox0=\hbox{\ignorespaces#1\unskip\ }\hangindent=\wd0
     \hangafter=1\box0\ignorespaces#2\fi
     \vskip18pt}%
     \dimen0=\pagetotal\advance\dimen0 by-\pageshrink
     \ifdim\dimen0<\pagegoal
     \dimen0=\ht0\advance\dimen0 by\dp0\advance\dimen0 by
     3\normalbaselineskip
     \advance\dimen0 by\pagetotal
     \ifdim\dimen0>\pagegoal\eject\fi\fi
     \noindent
     \if!#1!\ignorespaces#2
     \else\setbox0=\hbox{\ignorespaces#1\unskip\ }\hangindent=\wd0
     \hangafter=1\box0\ignorespaces#2\fi
     \vskip18pt plus4pt minus4pt\egroup
     \nobreak
     \parindent=0pt
     \everypar={\global\parindent=\oldparindent
     \global\let\lasttitle=N\global\everypar={}}%
     \global\let\lasttitle=A%
     \ignorespaces}
 \def\titleb#1#2{\if N\lasttitle\else\vskip-28pt
     \fi
     \vskip18pt plus 4pt minus4pt
     \bgroup
\textfont0=\tenbf \scriptfont0=\sevenbf \scriptscriptfont0=\fivebf
\textfont1=\tams \scriptfont1=\tamss \scriptscriptfont1=\tbmss
     \lineskip=0pt
     \pretolerance=10000
     \noindent
     \bf
     \rightskip 0pt plus 6em
     \setbox0=\vbox{\vskip23pt\def\fonote##1{}%
     \noindent
     \if!#1!\ignorespaces#2
     \else\setbox0=\hbox{\ignorespaces#1\unskip\enspace}\hangindent=\wd0
     \hangafter=1\box0\ignorespaces#2\fi
     \vskip10pt}%
     \dimen0=\pagetotal\advance\dimen0 by-\pageshrink
     \ifdim\dimen0<\pagegoal
     \dimen0=\ht0\advance\dimen0 by\dp0\advance\dimen0 by
     3\normalbaselineskip
     \advance\dimen0 by\pagetotal
     \ifdim\dimen0>\pagegoal\eject\fi\fi
     \noindent
     \if!#1!\ignorespaces#2
     \else\setbox0=\hbox{\ignorespaces#1\unskip\enspace}\hangindent=\wd0
     \hangafter=1\box0\ignorespaces#2\fi
     \vskip8pt plus4pt minus4pt\egroup
     \nobreak
     \parindent=0pt
     \everypar={\global\parindent=\oldparindent
     \global\let\lasttitle=N\global\everypar={}}%
     \global\let\lasttitle=B%
     \ignorespaces}
 \def\titlec#1#2{\if N\lasttitle\else\vskip-23pt
     \fi
     \vskip18pt plus 4pt minus4pt
     \bgroup
\textfont0=\tenbfne \scriptfont0=\sevenbf \scriptscriptfont0=\fivebf
\textfont1=\tams \scriptfont1=\tamss \scriptscriptfont1=\tbmss
     \tenbfne
     \lineskip=0pt
     \pretolerance=10000
     \noindent
     \rightskip 0pt plus 6em
     \setbox0=\vbox{\vskip23pt\def\fonote##1{}%
     \noindent
     \if!#1!\ignorespaces#2
     \else\setbox0=\hbox{\ignorespaces#1\unskip\enspace}\hangindent=\wd0
     \hangafter=1\box0\ignorespaces#2\fi
     \vskip6pt}%
     \dimen0=\pagetotal\advance\dimen0 by-\pageshrink
     \ifdim\dimen0<\pagegoal
     \dimen0=\ht0\advance\dimen0 by\dp0\advance\dimen0 by
     2\normalbaselineskip
     \advance\dimen0 by\pagetotal
     \ifdim\dimen0>\pagegoal\eject\fi\fi
     \noindent
     \if!#1!\ignorespaces#2
     \else\setbox0=\hbox{\ignorespaces#1\unskip\enspace}\hangindent=\wd0
     \hangafter=1\box0\ignorespaces#2\fi
     \vskip6pt plus4pt minus4pt\egroup
     \nobreak
     \parindent=0pt
     \everypar={\global\parindent=\oldparindent
     \global\let\lasttitle=N\global\everypar={}}%
     \global\let\lasttitle=C%
     \ignorespaces}
 \def\titled#1{\if N\lasttitle\else\vskip-\baselineskip
     \fi
     \vskip12pt plus 4pt minus 4pt
     \bgroup
\textfont1=\tams \scriptfont1=\tamss \scriptscriptfont1=\tbmss
     \bf
     \noindent
     \ignorespaces#1\ \ignorespaces\egroup
     \ignorespaces}
\let\ts=\thinspace
\def\footnoterule{\kern-3pt\hrule width 2true cm\kern2.6pt}
\newcount\footcount \footcount=0
\def\advftncnt{\advance\footcount by1\global\footcount=\footcount}
\def\fonote#1{\advftncnt$^{\the\footcount}$\begingroup\petit
\parfillskip=0pt plus 1fil
\def\textindent##1{\hangindent0.5\oldparindent\noindent\hbox
to0.5\oldparindent{##1\hss}\ignorespaces}%
\vfootnote{$^{\the\footcount}$}{#1\vskip-9.69pt}\endgroup}
\def\item#1{\par\noindent
\hangindent6.5 mm\hangafter=0
\llap{#1\enspace}\ignorespaces}

\def\titleao#1{\vfill\supereject
\ifodd\pageno\else\null\vfill\supereject\fi
\let\header=N
     \bgroup
\textfont0=\tafontt \scriptfont0=\tafonts \scriptscriptfont0=\tafontss
\textfont1=\tamt \scriptfont1=\tams \scriptscriptfont1=\tamss
\textfont2=\tast \scriptfont2=\tass \scriptscriptfont2=\tasss
\textfont3=\tasyt \scriptfont3=\tasys \scriptscriptfont3=\tenex
     \baselineskip=18pt
     \lineskip=0pt
     \pretolerance=10000
     \tafontt
     \centerpar{#1}%
     \vskip75pt\egroup
     \nobreak
     \parindent=0pt
     \everypar={\global\parindent=\oldparindent
     \global\let\lasttitle=N\global\everypar={}}%
     \global\let\lasttitle=A%
     \ignorespaces}
\def\acknow{\titleao{Acknowledgements}}





\def\leaderfill{\kern0.5em\leaders\hbox to 0.5em{\hss.\hss}\hfill\kern
0.5em}
\newdimen\chapindent
\newdimen\sectindent
\newdimen\subsecindent
\newdimen\thousand
\setbox0=\hbox{\bf 10. }\chapindent=\wd0
\setbox0=\hbox{10.10 }\sectindent=\wd0
\setbox0=\hbox{10.10.1 }\subsecindent=\wd0
\setbox0=\hbox{\enspace 100}\thousand=\wd0
\def\contpart#1#2{\medskip\noindent
\vbox{\kern10pt\leftline{\textfont1=\tams
\scriptfont1=\tamss\scriptscriptfont1=\tbmss\bf
\advance\chapindent by\sectindent
\hbox to\chapindent{\ignorespaces#1\hss}\ignorespaces#2}\kern8pt}%
\let\lasttitle=Y\par}
\def\contcontribution#1#2{\if N\lasttitle\bigskip\fi
\let\lasttitle=N\line{{\textfont1=\tams
\scriptfont1=\tamss\scriptscriptfont1=\tbmss\bf#1}%
\if!#2!\hfill\else\leaderfill\hbox to\thousand{\hss#2}\fi}\par}
\def\conttitlea#1#2#3{\line{\hbox to
\chapindent{\strut\bf#1\hss}{\textfont1=\tams
\scriptfont1=\tamss\scriptscriptfont1=\tbmss\bf#2}%
\if!#3!\hfill\else\leaderfill\hbox to\thousand{\hss#3}\fi}\par}
\def\conttitleb#1#2#3{\line{\kern\chapindent\hbox
to\sectindent{\strut#1\hss}{#2}%
\if!#3!\hfill\else\leaderfill\hbox to\thousand{\hss#3}\fi}\par}
\def\conttitlec#1#2#3{\line{\kern\chapindent\kern\sectindent
\hbox to\subsecindent{\strut#1\hss}{#2}%
\if!#3!\hfill\else\leaderfill\hbox to\thousand{\hss#3}\fi}\par}
\long\def\lemma#1#2{\removelastskip\vskip\baselineskip\noindent{\tenbfne
Lemma\if!#1!\else\ #1\fi\ \ }{\it\ignorespaces#2}\vskip\baselineskip}
\long\def\proposition#1#2{\removelastskip\vskip\baselineskip\noindent{\tenbfne
Proposition\if!#1!\else\ #1\fi\ \ }{\it\ignorespaces#2}\vskip\baselineskip}
\long\def\theorem#1#2{\removelastskip\vskip\baselineskip\noindent{\tenbfne
Theorem\if!#1!\else\ #1\fi\ \ }{\it\ignorespaces#2}\vskip\baselineskip}
\long\def\corollary#1#2{\removelastskip\vskip\baselineskip\noindent{\tenbfne
Corollary\if!#1!\else\ #1\fi\ \ }{\it\ignorespaces#2}\vskip\baselineskip}
\long\def\example#1#2{\removelastskip\vskip\baselineskip\noindent{\tenbfne
Example\if!#1!\else\ #1\fi\ \ }\ignorespaces#2\vskip\baselineskip}
\long\def\exercise#1#2{\removelastskip\vskip\baselineskip\noindent{\tenbfne
Exercise\if!#1!\else\ #1\fi\ \ }\ignorespaces#2\vskip\baselineskip}
\long\def\problem#1#2{\removelastskip\vskip\baselineskip\noindent{\tenbfne
Problem\if!#1!\else\ #1\fi\ \ }\ignorespaces#2\vskip\baselineskip}
\long\def\solution#1#2{\removelastskip\vskip\baselineskip\noindent{\tenbfne
Solution\if!#1!\else\ #1\fi\ \ }\ignorespaces#2\vskip\baselineskip}


\long\def\definition#1#2{\removelastskip\vskip\baselineskip\noindent{\tenbfne
Definition\if!#1!\else\
#1\fi\ \ }\ignorespaces#2\vskip\baselineskip}
\def\frame#1{\bigskip\vbox{\hrule\hbox{\vrule\kern5pt
\vbox{\kern5pt\advance\hsize by-10.8pt
\centerline{\vbox{#1}}\kern5pt}\kern5pt\vrule}\hrule}\bigskip}
\def\frameddisplay#1#2{$$\vcenter{\hrule\hbox{\vrule\kern5pt
\vbox{\kern5pt\hbox{$\displaystyle#1$}%
\kern5pt}\kern5pt\vrule}\hrule}\eqno#2$$}
\def\typeset{\petit\noindent This book was processed by the author using
the \TeX\ macro package from Springer-Verlag.\par}
\outer\def\byebye{\bigskip\bigskip\typeset
\footcount=1\ifx\speciali\undefined\else
\loop\smallskip\noindent special character No\number\footcount:
\csname special\romannumeral\footcount\endcsname
\advance\footcount by 1\global\footcount=\footcount
\ifnum\footcount<11\repeat\fi
\gdef\leftheadline{\hbox to2.5true cc{\folio\hfil}\ignorespaces
\the\AUTHOR\unskip: \the\HEAD\hfill}\vfill\supereject\end}




/Mathdict 150 dict def
Mathdict begin
/Mlmarg		1.0 72 mul def
/Mrmarg		1.0 72 mul def
/Mbmarg		1.0 72 mul def
/Mtmarg		1.0 72 mul def
/Mwidth		8.5 72 mul def
/Mheight	11 72 mul def
/Mtransform	{  } bind def
/Mnodistort	true def
/Mfixwid	true	def
/Mfixdash	false def
/Mrot		0	def
/Mpstart {
MathPictureStart
} bind def
/Mpend {
MathPictureEnd
} bind def
/Mscale {
0 1 0 1
5 -1 roll
MathScale
} bind def
/ISOLatin1Encoding dup where
{ pop pop }
{
[
/.notdef /.notdef /.notdef /.notdef /.notdef /.notdef /.notdef /.notdef
/.notdef /.notdef /.notdef /.notdef /.notdef /.notdef /.notdef /.notdef
/.notdef /.notdef /.notdef /.notdef /.notdef /.notdef /.notdef /.notdef
/.notdef /.notdef /.notdef /.notdef /.notdef /.notdef /.notdef /.notdef
/space /exclam /quotedbl /numbersign /dollar /percent /ampersand /quoteright
/parenleft /parenright /asterisk /plus /comma /minus /period /slash
/zero /one /two /three /four /five /six /seven
/eight /nine /colon /semicolon /less /equal /greater /question
/at /A /B /C /D /E /F /G
/H /I /J /K /L /M /N /O
/P /Q /R /S /T /U /V /W
/X /Y /Z /bracketleft /backslash /bracketright /asciicircum /underscore
/quoteleft /a /b /c /d /e /f /g
/h /i /j /k /l /m /n /o
/p /q /r /s /t /u /v /w
/x /y /z /braceleft /bar /braceright /asciitilde /.notdef
/.notdef /.notdef /.notdef /.notdef /.notdef /.notdef /.notdef /.notdef
/.notdef /.notdef /.notdef /.notdef /.notdef /.notdef /.notdef /.notdef
/dotlessi /grave /acute /circumflex /tilde /macron /breve /dotaccent
/dieresis /.notdef /ring /cedilla /.notdef /hungarumlaut /ogonek /caron
/space /exclamdown /cent /sterling /currency /yen /brokenbar /section
/dieresis /copyright /ordfeminine /guillemotleft
/logicalnot /hyphen /registered /macron
/degree /plusminus /twosuperior /threesuperior
/acute /mu /paragraph /periodcentered
/cedilla /onesuperior /ordmasculine /guillemotright
/onequarter /onehalf /threequarters /questiondown
/Agrave /Aacute /Acircumflex /Atilde /Adieresis /Aring /AE /Ccedilla
/Egrave /Eacute /Ecircumflex /Edieresis /Igrave /Iacute /Icircumflex /Idieresis
/Eth /Ntilde /Ograve /Oacute /Ocircumflex /Otilde /Odieresis /multiply
/Oslash /Ugrave /Uacute /Ucircumflex /Udieresis /Yacute /Thorn /germandbls
/agrave /aacute /acircumflex /atilde /adieresis /aring /ae /ccedilla
/egrave /eacute /ecircumflex /edieresis /igrave /iacute /icircumflex /idieresis
/eth /ntilde /ograve /oacute /ocircumflex /otilde /odieresis /divide
/oslash /ugrave /uacute /ucircumflex /udieresis /yacute /thorn /ydieresis
] def
} ifelse
/MFontDict 50 dict def
/MStrCat
{
exch
dup length
2 index length add
string
dup 3 1 roll
copy
length
exch dup
4 2 roll exch
putinterval
} def
/MCreateEncoding
{
1 index
255 string cvs
(-) MStrCat
1 index MStrCat
cvn exch
(Encoding) MStrCat
cvn dup where
{
exch get
}
{
pop
StandardEncoding
} ifelse
3 1 roll
dup MFontDict exch known not
{
1 index findfont
dup length dict
begin
{1 index /FID ne
{def}
{pop pop}
ifelse} forall
/Encoding 3 index
def
currentdict
end
1 index exch definefont pop
MFontDict 1 index
null put
}
if
exch pop
exch pop
} def
/ISOLatin1 { (ISOLatin1) MCreateEncoding } def
/ISO8859 { (ISOLatin1) MCreateEncoding } def
/Mcopyfont {
dup
maxlength
dict
exch
{
1 index
/FID
eq
{
pop pop
}
{
2 index
3 1 roll
put
}
ifelse
}
forall
} def
/Plain	/Courier findfont Mcopyfont definefont pop
/Bold	/Courier-Bold findfont Mcopyfont definefont pop
/Italic /Courier-Oblique findfont Mcopyfont definefont pop
/MathPictureStart {
gsave
Mtransform
Mlmarg
Mbmarg
translate
Mwidth
Mlmarg Mrmarg add
sub
/Mwidth exch def
Mheight
Mbmarg Mtmarg add
sub
/Mheight exch def
/Mtmatrix
matrix currentmatrix
def
/Mgmatrix
matrix currentmatrix
def
} bind def
/MathPictureEnd {
grestore
} bind def
/MFill {
0 0 		moveto
Mwidth 0 	lineto
Mwidth Mheight 	lineto
0 Mheight 	lineto
fill
} bind def
/MPlotRegion {
3 index
Mwidth mul
2 index
Mheight mul
translate
exch sub
Mheight mul
/Mheight
exch def
exch sub
Mwidth mul
/Mwidth
exch def
} bind def
/MathSubStart {
Momatrix
Mgmatrix Mtmatrix
Mwidth Mheight
7 -2 roll
moveto
Mtmatrix setmatrix
currentpoint
Mgmatrix setmatrix
9 -2 roll
moveto
Mtmatrix setmatrix
currentpoint
2 copy translate
/Mtmatrix matrix currentmatrix def
3 -1 roll
exch sub
/Mheight exch def
sub
/Mwidth exch def
} bind def
/MathSubEnd {
/Mheight exch def
/Mwidth exch def
/Mtmatrix exch def
dup setmatrix
/Mgmatrix exch def
/Momatrix exch def
} bind def
/Mdot {
moveto
0 0 rlineto
stroke
} bind def
/Mtetra {
moveto
lineto
lineto
lineto
fill
} bind def
/Metetra {
moveto
lineto
lineto
lineto
closepath
gsave
fill
grestore
0 setgray
stroke
} bind def
/Mistroke {
flattenpath
0 0 0
{
4 2 roll
pop pop
}
{
4 -1 roll
2 index
sub dup mul
4 -1 roll
2 index
sub dup mul
add sqrt
4 -1 roll
add
3 1 roll
}
{
stop
}
{
stop
}
pathforall
pop pop
currentpoint
stroke
moveto
currentdash
3 -1 roll
add
setdash
} bind def
/Mfstroke {
stroke
currentdash
pop 0
setdash
} bind def
/Mrotsboxa {
gsave
dup
/Mrot
exch def
Mrotcheck
Mtmatrix
dup
setmatrix
7 1 roll
4 index
4 index
translate
rotate
3 index
-1 mul
3 index
-1 mul
translate
/Mtmatrix
matrix
currentmatrix
def
grestore
Msboxa
3  -1 roll
/Mtmatrix
exch def
/Mrot
0 def
} bind def
/Msboxa {
newpath
5 -1 roll
Mvboxa
pop
Mboxout
6 -1 roll
5 -1 roll
4 -1 roll
Msboxa1
5 -3 roll
Msboxa1
Mboxrot
[
7 -2 roll
2 copy
[
3 1 roll
10 -1 roll
9 -1 roll
]
6 1 roll
5 -2 roll
]
} bind def
/Msboxa1 {
sub
2 div
dup
2 index
1 add
mul
3 -1 roll
-1 add
3 -1 roll
mul
} bind def
/Mvboxa	{
Mfixwid
{
Mvboxa1
}
{
dup
Mwidthcal
0 exch
{
add
}
forall
exch
Mvboxa1
4 index
7 -1 roll
add
4 -1 roll
pop
3 1 roll
}
ifelse
} bind def
/Mvboxa1 {
gsave
newpath
[ true
3 -1 roll
{
Mbbox
5 -1 roll
{
0
5 1 roll
}
{
7 -1 roll
exch sub
(m) stringwidth pop
.3 mul
sub
7 1 roll
6 -1 roll
4 -1 roll
Mmin
3 -1 roll
5 index
add
5 -1 roll
4 -1 roll
Mmax
4 -1 roll
}
ifelse
false
}
forall
{ stop } if
counttomark
1 add
4 roll
]
grestore
} bind def
/Mbbox {
1 dict begin
0 0 moveto
/temp (T) def
{	gsave
currentpoint newpath moveto
temp 0 3 -1 roll put temp
false charpath flattenpath currentpoint
pathbbox
grestore moveto lineto moveto} forall
pathbbox
newpath
end
} bind def
/Mmin {
2 copy
gt
{ exch } if
pop
} bind def
/Mmax {
2 copy
lt
{ exch } if
pop
} bind def
/Mrotshowa {
dup
/Mrot
exch def
Mrotcheck
Mtmatrix
dup
setmatrix
7 1 roll
4 index
4 index
translate
rotate
3 index
-1 mul
3 index
-1 mul
translate
/Mtmatrix
matrix
currentmatrix
def
Mgmatrix setmatrix
Mshowa
/Mtmatrix
exch def
/Mrot 0 def
} bind def
/Mshowa {
4 -2 roll
moveto
2 index
Mtmatrix setmatrix
Mvboxa
7 1 roll
Mboxout
6 -1 roll
5 -1 roll
4 -1 roll
Mshowa1
4 1 roll
Mshowa1
rmoveto
currentpoint
Mfixwid
{
Mshowax
}
{
Mshoway
}
ifelse
pop pop pop pop
Mgmatrix setmatrix
} bind def
/Mshowax {
0 1
4 index length
-1 add
{
2 index
4 index
2 index
get
3 index
add
moveto
4 index
exch get
Mfixdash
{
Mfixdashp
}
if
show
} for
} bind def
/Mfixdashp {
dup
length
1
gt
1 index
true exch
{
45
eq
and
} forall
and
{
gsave
(--)
stringwidth pop
(-)
stringwidth pop
sub
2 div
0 rmoveto
dup
length
1 sub
{
(-)
show
}
repeat
grestore
}
if
} bind def
/Mshoway {
3 index
Mwidthcal
5 1 roll
0 1
4 index length
-1 add
{
2 index
4 index
2 index
get
3 index
add
moveto
4 index
exch get
[
6 index
aload
length
2 add
-1 roll
{
pop
Strform
stringwidth
pop
neg
exch
add
0 rmoveto
}
exch
kshow
cleartomark
} for
pop
} bind def
/Mwidthcal {
[
exch
{
Mwidthcal1
}
forall
]
[
exch
dup
Maxlen
-1 add
0 1
3 -1 roll
{
[
exch
2 index
{
1 index
Mget
exch
}
forall
pop
Maxget
exch
}
for
pop
]
Mreva
} bind def
/Mreva	{
[
exch
aload
length
-1 1
{1 roll}
for
]
} bind def
/Mget	{
1 index
length
-1 add
1 index
ge
{
get
}
{
pop pop
0
}
ifelse
} bind def
/Maxlen	{
[
exch
{
length
}
forall
Maxget
} bind def
/Maxget	{
counttomark
-1 add
1 1
3 -1 roll
{
pop
Mmax
}
for
exch
pop
} bind def
/Mwidthcal1 {
[
exch
{
Strform
stringwidth
pop
}
forall
]
} bind def
/Strform {
/tem (x) def
tem 0
3 -1 roll
put
tem
} bind def
/Mshowa1 {
2 copy
add
4 1 roll
sub
mul
sub
-2 div
} bind def
/MathScale {
Mwidth
Mheight
Mlp
translate
scale
/yscale exch def
/ybias exch def
/xscale exch def
/xbias exch def
/Momatrix
xscale yscale matrix scale
xbias ybias matrix translate
matrix concatmatrix def
/Mgmatrix
matrix currentmatrix
def
} bind def
/Mlp {
3 copy
Mlpfirst
{
Mnodistort
{
Mmin
dup
} if
4 index
2 index
2 index
Mlprun
11 index
11 -1 roll
10 -4 roll
Mlp1
8 index
9 -5 roll
Mlp1
4 -1 roll
and
{ exit } if
3 -1 roll
pop pop
} loop
exch
3 1 roll
7 -3 roll
pop pop pop
} bind def
/Mlpfirst {
3 -1 roll
dup length
2 copy
-2 add
get
aload
pop pop pop
4 -2 roll
-1 add
get
aload
pop pop pop
6 -1 roll
3 -1 roll
5 -1 roll
sub
div
4 1 roll
exch sub
div
} bind def
/Mlprun {
2 copy
4 index
0 get
dup
4 1 roll
Mlprun1
3 copy
8 -2 roll
9 -1 roll
{
3 copy
Mlprun1
3 copy
11 -3 roll
/gt Mlpminmax
8 3 roll
11 -3 roll
/lt Mlpminmax
8 3 roll
} forall
pop pop pop pop
3 1 roll
pop pop
aload pop
5 -1 roll
aload pop
exch
6 -1 roll
Mlprun2
8 2 roll
4 -1 roll
Mlprun2
6 2 roll
3 -1 roll
Mlprun2
4 2 roll
exch
Mlprun2
6 2 roll
} bind def
/Mlprun1 {
aload pop
exch
6 -1 roll
5 -1 roll
mul add
4 -2 roll
mul
3 -1 roll
add
} bind def
/Mlprun2 {
2 copy
add 2 div
3 1 roll
exch sub
} bind def
/Mlpminmax {
cvx
2 index
6 index
2 index
exec
{
7 -3 roll
4 -1 roll
} if
1 index
5 index
3 -1 roll
exec
{
4 1 roll
pop
5 -1 roll
aload
pop pop
4 -1 roll
aload pop
[
8 -2 roll
pop
5 -2 roll
pop
6 -2 roll
pop
5 -1 roll
]
4 1 roll
pop
}
{
pop pop pop
} ifelse
} bind def
/Mlp1 {
5 index
3 index	sub
5 index
2 index mul
1 index
le
1 index
0 le
or
dup
not
{
1 index
3 index	div
.99999 mul
8 -1 roll
pop
7 1 roll
}
if
8 -1 roll
2 div
7 -2 roll
pop sub
5 index
6 -3 roll
pop pop
mul sub
exch
} bind def
/intop 0 def
/inrht 0 def
/inflag 0 def
/outflag 0 def
/xadrht 0 def
/xadlft 0 def
/yadtop 0 def
/yadbot 0 def
/Minner {
outflag
1
eq
{
/outflag 0 def
/intop 0 def
/inrht 0 def
} if
5 index
gsave
Mtmatrix setmatrix
Mvboxa pop
grestore
3 -1 roll
pop
dup
intop
gt
{
/intop
exch def
}
{ pop }
ifelse
dup
inrht
gt
{
/inrht
exch def
}
{ pop }
ifelse
pop
/inflag
1 def
} bind def
/Mouter {
/xadrht 0 def
/xadlft 0 def
/yadtop 0 def
/yadbot 0 def
inflag
1 eq
{
dup
0 lt
{
dup
intop
mul
neg
/yadtop
exch def
} if
dup
0 gt
{
dup
intop
mul
/yadbot
exch def
}
if
pop
dup
0 lt
{
dup
inrht
mul
neg
/xadrht
exch def
} if
dup
0 gt
{
dup
inrht
mul
/xadlft
exch def
} if
pop
/outflag 1 def
}
{ pop pop}
ifelse
/inflag 0 def
/inrht 0 def
/intop 0 def
} bind def
/Mboxout {
outflag
1
eq
{
4 -1
roll
xadlft
leadjust
add
sub
4 1 roll
3 -1
roll
yadbot
leadjust
add
sub
3 1
roll
exch
xadrht
leadjust
add
add
exch
yadtop
leadjust
add
add
/outflag 0 def
/xadlft 0 def
/yadbot 0 def
/xadrht 0 def
/yadtop 0 def
} if
} bind def
/leadjust {
(m) stringwidth pop
.5 mul
} bind def
/Mrotcheck {
dup
90
eq
{
yadbot
/yadbot
xadrht
def
/xadrht
yadtop
def
/yadtop
xadlft
def
/xadlft
exch
def
}
if
dup
cos
1 index
sin
Checkaux
dup
cos
1 index
sin neg
exch
Checkaux
3 1 roll
pop pop
} bind def
/Checkaux {
4 index
exch
4 index
mul
3 1 roll
mul add
4 1 roll
} bind def
/Mboxrot {
Mrot
90 eq
{
brotaux
4 2
roll
}
if
Mrot
180 eq
{
4 2
roll
brotaux
4 2
roll
brotaux
}
if
Mrot
270 eq
{
4 2
roll
brotaux
}
if
} bind def
/brotaux {
neg
exch
neg
} bind def
/Mabsproc {
0
matrix defaultmatrix
dtransform idtransform
dup mul exch
dup mul
add sqrt
} bind def
/Mabswid {
Mabsproc
setlinewidth
} bind def
/Mabsdash {
exch
[
exch
{
Mabsproc
}
forall
]
exch
setdash
} bind def
/MBeginOrig { Momatrix concat} bind def
/MEndOrig { Mgmatrix setmatrix} bind def
/sampledsound where
{ pop}
{ /sampledsound {
exch
pop
exch
5 1 roll
mul
4 idiv
mul
2 idiv
exch pop
exch
/Mtempproc exch def
{ Mtempproc pop}
repeat
} bind def
} ifelse
/g { setgray} bind def
/k { setcmykcolor} bind def
/m { moveto} bind def
/p { gsave} bind def
/r { setrgbcolor} bind def
/w { setlinewidth} bind def
/C { curveto} bind def
/F { fill} bind def
/L { lineto} bind def
/P { grestore} bind def
/s { stroke} bind def
/setcmykcolor where
{ pop}
{ /setcmykcolor {
4 1
roll
[
4 1
roll
]
{
1 index
sub
1
sub neg
dup
0
lt
{
pop
0
}
if
dup
1
gt
{
pop
1
}
if
exch
} forall
pop
setrgbcolor
} bind def
} ifelse
/Mcharproc
{
currentfile
(x)
readhexstring
pop
0 get
exch
div
} bind def
/Mshadeproc
{
dup
3 1
roll
{
dup
Mcharproc
3 1
roll
} repeat
1 eq
{
setgray
}
{
3 eq
{
setrgbcolor
}
{
setcmykcolor
} ifelse
} ifelse
} bind def
/Mrectproc
{
3 index
2 index
moveto
2 index
3 -1
roll
lineto
dup
3 1
roll
lineto
lineto
fill
} bind def
/Mcolorimage
{
7 1
roll
pop
pop
matrix
invertmatrix
concat
2 exch exp
1 sub
3 1 roll
1 1
2 index
{
1 1
4 index
{
dup
1 sub
exch
2 index
dup
1 sub
exch
7 index
9 index
Mshadeproc
Mrectproc
} for
pop
} for
pop pop pop pop
} bind def
/Mimage
{
pop
matrix
invertmatrix
concat
2 exch exp
1 sub
3 1 roll
1 1
2 index
{
1 1
4 index
{
dup
1 sub
exch
2 index
dup
1 sub
exch
7 index
Mcharproc
setgray
Mrectproc
} for
pop
} for
pop pop pop
} bind def

MathPictureStart
/Courier findfont 10  scalefont  setfont
0.0241558 0.966234 -0.15684 0.966234 [
[(-1)] .68841  .01208  .0059  1 Msboxa
[(-0.5)] .52012  .03954  .00042  1 Msboxa
[(0)] .35409  .06664  -0.00498 1 Msboxa
[(0.5)] .19026  .09338  -0.0103 1 Msboxa
[(1)] .02859  .11976  -0.01554 1 Msboxa
[(    M)(Log[-])(    T)] .35439  .00625  -0.00498 1 Msboxa
[(0)] .33892  .61894  .00538  -1 Msboxa
[(2)] .26344  .57058  .0078  -1 Msboxa
[(4)] .18586  .52087  .0103  -1 Msboxa
[(6)] .10606  .46975  .01288  -1 Msboxa
[(8)] .02397  .41715  .01554  -1 Msboxa
[(g2)] .18523  .58126  .0103  -1 Msboxa
[(0)] .01638  .14023  1 -0.16287 Msboxa
[(0.1)] .01496  .23149  1 -0.15913 Msboxa
[(0.2)] .01352  .3233  1 -0.15537 Msboxa
[(O)(-)(M)] -0.0453 .27913  1 -0.15757 Msboxa
[ 0 0 0 0 ]
[ 1 .63102  0 0 ]
] MathScale
1 setlinecap
1 setlinejoin
newpath
[ ] 0 setdash
0 g
p
p
.002  w
.68848  .02416  m
.0284  .13184  L
s
P
p
.002  w
.68848  .02416  m
.68851  .03019  L
s
P
[(-1)] .68841  .01208  .0059  1 Mshowa
p
.002  w
.52013  .05162  m
.52013  .05766  L
s
P
[(-0.5)] .52012  .03954  .00042  1 Mshowa
p
.002  w
.35403  .07872  m
.354  .08475  L
s
P
[(0)] .35409  .06664  -0.00498 1 Mshowa
p
.002  w
.19014  .10545  m
.19007  .11149  L
s
P
[(0.5)] .19026  .09338  -0.0103 1 Mshowa
p
.002  w
.0284  .13184  m
.02831  .13787  L
s
P
[(1)] .02859  .11976  -0.01554 1 Mshowa
p
.001  w
.65463  .02968  m
.65464  .0333  L
s
P
p
.001  w
.62087  .03519  m
.62088  .03881  L
s
P
p
.001  w
.5872  .04068  m
.58721  .0443  L
s
P
p
.001  w
.55362  .04616  m
.55362  .04978  L
s
P
p
.001  w
.48673  .05707  m
.48673  .06069  L
s
P
p
.001  w
.45342  .0625  m
.45342  .06612  L
s
P
p
.001  w
.4202  .06792  m
.42019  .07154  L
s
P
p
.001  w
.38707  .07333  m
.38706  .07695  L
s
P
p
.001  w
.32108  .08409  m
.32105  .08771  L
s
P
p
.001  w
.28821  .08945  m
.28818  .09308  L
s
P
p
.001  w
.25543  .0948  m
.2554  .09842  L
s
P
p
.001  w
.22274  .10013  m
.22271  .10376  L
s
P
p
.001  w
.15762  .11076  m
.15758  .11438  L
s
P
p
.001  w
.12519  .11605  m
.12514  .11967  L
s
P
p
.001  w
.09284  .12132  m
.09279  .12495  L
s
P
p
.001  w
.06058  .12659  m
.06053  .13021  L
s
P
[(    M)(Log[-])(    T)] .35439  .00625  -0.00498 1 Mshowa
P
p
p
.002  w
.33898  .60686  m
.02416  .40508  L
s
P
p
.002  w
.33898  .60686  m
.33902  .60082  L
s
P
[(0)] .33892  .61894  .00538  -1 Mshowa
p
.002  w
.26354  .55851  m
.26359  .55247  L
s
P
[(2)] .26344  .57058  .0078  -1 Mshowa
p
.002  w
.18598  .5088  m
.18604  .50276  L
s
P
[(4)] .18586  .52087  .0103  -1 Mshowa
p
.002  w
.10622  .45767  m
.1063  .45164  L
s
P
[(6)] .10606  .46975  .01288  -1 Mshowa
p
.002  w
.02416  .40508  m
.02425  .39904  L
s
P
[(8)] .02397  .41715  .01554  -1 Mshowa
p
.001  w
.32406  .5973  m
.32408  .59367  L
s
P
p
.001  w
.30905  .58768  m
.30908  .58406  L
s
P
p
.001  w
.29397  .57801  m
.29399  .57439  L
s
P
p
.001  w
.27879  .56829  m
.27882  .56466  L
s
P
p
.001  w
.2482  .54868  m
.24823  .54505  L
s
P
p
.001  w
.23277  .53879  m
.23281  .53517  L
s
P
p
.001  w
.21726  .52885  m
.2173  .52522  L
s
P
p
.001  w
.20167  .51885  m
.2017  .51523  L
s
P
p
.001  w
.17021  .49869  m
.17025  .49506  L
s
P
p
.001  w
.15434  .48852  m
.15439  .4849  L
s
P
p
.001  w
.13839  .4783  m
.13844  .47467  L
s
P
p
.001  w
.12235  .46801  m
.1224  .46439  L
s
P
p
.001  w
.08999  .44727  m
.09004  .44365  L
s
P
p
.001  w
.07367  .43682  m
.07372  .43319  L
s
P
p
.001  w
.05726  .4263  m
.05732  .42267  L
s
P
p
.001  w
.04076  .41572  m
.04081  .4121  L
s
P
[(g2)] .18523  .58126  .0103  -1 Mshowa
P
p
p
.002  w
.0284  .13184  m
.02416  .40508  L
s
P
p
.002  w
.0283  .13829  m
.03426  .13732  L
s
P
[(0)] .01638  .14023  1 -0.16287 Mshowa
p
.002  w
.02688  .22959  m
.03285  .22864  L
s
P
[(0.1)] .01496  .23149  1 -0.15913 Mshowa
p
.002  w
.02546  .32144  m
.03142  .32052  L
s
P
[(0.2)] .01352  .3233  1 -0.15537 Mshowa
p
.001  w
.02802  .1565  m
.0316  .15592  L
s
P
p
.001  w
.02774  .17474  m
.03131  .17417  L
s
P
p
.001  w
.02745  .19301  m
.03103  .19243  L
s
P
p
.001  w
.02717  .21129  m
.03075  .21072  L
s
P
p
.001  w
.0266  .24792  m
.03018  .24735  L
s
P
p
.001  w
.02631  .26627  m
.02989  .2657  L
s
P
p
.001  w
.02603  .28464  m
.02961  .28408  L
s
P
p
.001  w
.02574  .30303  m
.02932  .30247  L
s
P
p
.001  w
.02517  .33988  m
.02875  .33932  L
s
P
p
.001  w
.02488  .35834  m
.02846  .35778  L
s
P
p
.001  w
.0246  .37681  m
.02818  .37627  L
s
P
p
.001  w
.02431  .39532  m
.02789  .39477  L
s
P
[(O)(-)(M)] -0.0453 .27913  1 -0.15757 Mshowa
P
0 0 m
1 0 L
1 .63102  L
0 .63102  L
closepath
clip
newpath
p
.002  w
.97185  .2505  m
.97584  .51654  L
s
.97584  .51654  m
.69013  .30388  L
s
.69013  .30388  m
.68848  .02416  L
s
.68848  .02416  m
.97185  .2505  L
s
.34038  .34671  m
.0284  .13184  L
s
.0284  .13184  m
.02416  .40508  L
s
.02416  .40508  m
.33898  .60686  L
s
.33898  .60686  m
.34038  .34671  L
s
.97185  .2505  m
.97584  .51654  L
s
.97584  .51654  m
.33898  .60686  L
s
.33898  .60686  m
.34038  .34671  L
s
.34038  .34671  m
.97185  .2505  L
s
.68848  .02416  m
.0284  .13184  L
s
.0284  .13184  m
.02416  .40508  L
s
.02416  .40508  m
.69013  .30388  L
s
.69013  .30388  m
.68848  .02416  L
s
P
p
.695  .632  .782  r
.0015  w
.38343  .59414  .36228  .57006  .31774  .57591  .33902  .60062  Metetra
.697  .635  .783  r
.42801  .58646  .40698  .56325  .36228  .57006  .38343  .59414  Metetra
.707  .64  .78  r
.47276  .57401  .45185  .55201  .40698  .56325  .42801  .58646  Metetra
.726  .648  .771  r
.51764  .55004  .49687  .52991  .45185  .55201  .47276  .57401  Metetra
.748  .655  .759  r
.5626  .50791  .54199  .4905  .49687  .52991  .51764  .55004  Metetra
.764  .662  .75  r
.60757  .44802  .58713  .43364  .54199  .4905  .5626  .50791  Metetra
.77  .667  .748  r
.65252  .38254  .63227  .37014  .58713  .43364  .60757  .44802  Metetra
.764  .67  .758  r
.69754  .32985  .67746  .31738  .63227  .37014  .65252  .38254  Metetra
.743  .672  .781  r
.74276  .29957  .72283  .28578  .67746  .31738  .69754  .32985  Metetra
.715  .672  .807  r
.78822  .28577  .76844  .27097  .72283  .28578  .74276  .29957  Metetra
.699  .671  .82  r
.8339  .27783  .81426  .26272  .76844  .27097  .78822  .28577  Metetra
.696  .671  .822  r
.87974  .27081  .86025  .25561  .81426  .26272  .8339  .27783  Metetra
.696  .671  .822  r
.92576  .26381  .90641  .24855  .86025  .25561  .87974  .27081  Metetra
.696  .671  .822  r
.97194  .25678  .95274  .24147  .90641  .24855  .92576  .26381  Metetra
.692  .606  .755  r
.36228  .57006  .34104  .5332  .29641  .53773  .31774  .57591  Metetra
.693  .609  .757  r
.40698  .56325  .38583  .52817  .34104  .5332  .36228  .57006  Metetra
.7  .616  .759  r
.45185  .55201  .43081  .51952  .38583  .52817  .40698  .56325  Metetra
.716  .627  .757  r
.49687  .52991  .47594  .50157  .43081  .51952  .45185  .55201  Metetra
.74  .644  .754  r
.54199  .4905  .52121  .46855  .47594  .50157  .49687  .52991  Metetra
.763  .663  .752  r
.58713  .43364  .56653  .41928  .52121  .46855  .54199  .4905  Metetra
.777  .678  .755  r
.63227  .37014  .61187  .36133  .56653  .41928  .58713  .43364  Metetra
.774  .686  .766  r
.67746  .31738  .65725  .30922  .61187  .36133  .63227  .37014  Metetra
.75  .684  .788  r
.72283  .28578  .70277  .27454  .65725  .30922  .67746  .31738  Metetra
.717  .676  .809  r
.76844  .27097  .74851  .25686  .70277  .27454  .72283  .28578  Metetra
.7  .672  .82  r
.81426  .26272  .79446  .24758  .74851  .25686  .76844  .27097  Metetra
.696  .671  .822  r
.86025  .25561  .8406  .24029  .79446  .24758  .81426  .26272  Metetra
.696  .671  .822  r
.90641  .24855  .88691  .23318  .8406  .24029  .86025  .25561  Metetra
.696  .671  .822  r
.95274  .24147  .93339  .22604  .88691  .23318  .90641  .24855  Metetra
.69  .605  .756  r
.34104  .5332  .31966  .49622  .27493  .49968  .29641  .53773  Metetra
.691  .608  .758  r
.38583  .52817  .36455  .49261  .31966  .49622  .34104  .5332  Metetra
.696  .613  .759  r
.43081  .51952  .40963  .48612  .36455  .49261  .38583  .52817  Metetra
.711  .624  .759  r
.47594  .50157  .45488  .4721  .40963  .48612  .43081  .51952  Metetra
.733  .642  .758  r
.52121  .46855  .50028  .44568  .45488  .4721  .47594  .50157  Metetra
.758  .665  .758  r
.56653  .41928  .54577  .40497  .50028  .44568  .52121  .46855  Metetra
.779  .687  .763  r
.61187  .36133  .5913  .35424  .54577  .40497  .56653  .41928  Metetra
.782  .701  .774  r
.65725  .30922  .63687  .30417  .5913  .35424  .61187  .36133  Metetra
.76  .698  .793  r
.70277  .27454  .68255  .26617  .63687  .30417  .65725  .30922  Metetra
.725  .684  .81  r
.74851  .25686  .72843  .2441  .68255  .26617  .70277  .27454  Metetra
.702  .674  .819  r
.79446  .24758  .77451  .23266  .72843  .2441  .74851  .25686  Metetra
.696  .672  .822  r
.8406  .24029  .8208  .22488  .77451  .23266  .79446  .24758  Metetra
.696  .671  .822  r
.88691  .23318  .86725  .21768  .8208  .22488  .8406  .24029  Metetra
.696  .671  .822  r
.93339  .22604  .91389  .21048  .86725  .21768  .88691  .23318  Metetra
.688  .614  .768  r
.31966  .49622  .2981  .46466  .25328  .46757  .27493  .49968  Metetra
.688  .615  .769  r
.36455  .49261  .3431  .46173  .2981  .46466  .31966  .49622  Metetra
.693  .62  .769  r
.40963  .48612  .38829  .45647  .3431  .46173  .36455  .49261  Metetra
.707  .629  .768  r
.45488  .4721  .43366  .44518  .38829  .45647  .40963  .48612  Metetra
.727  .645  .766  r
.50028  .44568  .47919  .42396  .43366  .44518  .45488  .4721  Metetra
.752  .667  .767  r
.54577  .40497  .52484  .39067  .47919  .42396  .50028  .44568  Metetra
.775  .692  .771  r
.5913  .35424  .57056  .34718  .52484  .39067  .54577  .40497  Metetra
.785  .71  .781  r
.63687  .30417  .61633  .30049  .57056  .34718  .5913  .35424  Metetra
.771  .711  .796  r
.68255  .26617  .66218  .26038  .61633  .30049  .63687  .30417  Metetra
.737  .695  .811  r
.72843  .2441  .70819  .2334  .66218  .26038  .68255  .26617  Metetra
.709  .678  .819  r
.77451  .23266  .75441  .21843  .70819  .2334  .72843  .2441  Metetra
.698  .672  .822  r
.8208  .22488  .80083  .20947  .75441  .21843  .77451  .23266  Metetra
.696  .671  .822  r
.86725  .21768  .84744  .20206  .80083  .20947  .8208  .22488  Metetra
.696  .671  .822  r
.91389  .21048  .89422  .1948  .84744  .20206  .86725  .21768  Metetra
.686  .626  .783  r
.2981  .46466  .27636  .4385  .23142  .44127  .25328  .46757  Metetra
.686  .627  .784  r
.3431  .46173  .32148  .43571  .27636  .4385  .2981  .46466  Metetra
.691  .629  .782  r
.38829  .45647  .36678  .43095  .32148  .43571  .3431  .46173  Metetra
.704  .637  .779  r
.43366  .44518  .41228  .42134  .36678  .43095  .38829  .45647  Metetra
.722  .65  .776  r
.47919  .42396  .45794  .40381  .41228  .42134  .43366  .44518  Metetra
.745  .669  .776  r
.52484  .39067  .50374  .37629  .45794  .40381  .47919  .42396  Metetra
.768  .693  .779  r
.57056  .34718  .54964  .33917  .50374  .37629  .52484  .39067  Metetra
.783  .713  .787  r
.61633  .30049  .59561  .29664  .54964  .33917  .57056  .34718  Metetra
.777  .72  .799  r
.66218  .26038  .64164  .25613  .59561  .29664  .61633  .30049  Metetra
.75  .706  .811  r
.70819  .2334  .68781  .22495  .64164  .25613  .66218  .26038  Metetra
.719  .686  .818  r
.75441  .21843  .73416  .20552  .68781  .22495  .70819  .2334  Metetra
.701  .674  .821  r
.80083  .20947  .78071  .19433  .73416  .20552  .75441  .21843  Metetra
.696  .672  .822  r
.84744  .20206  .82747  .18636  .78071  .19433  .80083  .20947  Metetra
.696  .671  .822  r
.89422  .1948  .8744  .179  .82747  .18636  .84744  .20206  Metetra
.685  .639  .798  r
.27636  .4385  .25443  .41622  .20936  .4191  .23142  .44127  Metetra
.685  .638  .797  r
.32148  .43571  .29967  .41326  .25443  .41622  .27636  .4385  Metetra
.69  .64  .794  r
.36678  .43095  .3451  .40859  .29967  .41326  .32148  .43571  Metetra
.701  .644  .79  r
.41228  .42134  .39073  .39996  .3451  .40859  .36678  .43095  Metetra
.718  .655  .786  r
.45794  .40381  .43653  .38497  .39073  .39996  .41228  .42134  Metetra
.738  .671  .784  r
.50374  .37629  .48248  .36175  .43653  .38497  .45794  .40381  Metetra
.76  .691  .786  r
.54964  .33917  .52855  .32991  .48248  .36175  .50374  .37629  Metetra
.776  .711  .791  r
.59561  .29664  .5747  .29166  .52855  .32991  .54964  .33917  Metetra
.779  .722  .8  r
.64164  .25613  .62093  .25223  .5747  .29166  .59561  .29664  Metetra
.761  .715  .81  r
.68781  .22495  .66726  .21829  .62093  .25223  .64164  .25613  Metetra
.731  .696  .817  r
.73416  .20552  .71375  .19436  .66726  .21829  .68781  .22495  Metetra
.707  .679  .82  r
.78071  .19433  .76044  .1799  .71375  .19436  .73416  .20552  Metetra
.698  .673  .822  r
.82747  .18636  .80733  .17069  .76044  .1799  .78071  .19433  Metetra
.696  .671  .822  r
.8744  .179  .85442  .16307  .80733  .17069  .82747  .18636  Metetra
.684  .649  .809  r
.25443  .41622  .2323  .39643  .18711  .39955  .20936  .4191  Metetra
.684  .648  .808  r
.29967  .41326  .27768  .39315  .2323  .39643  .25443  .41622  Metetra
.689  .648  .804  r
.3451  .40859  .32325  .38836  .27768  .39315  .29967  .41326  Metetra
.699  .651  .799  r
.39073  .39996  .369  .3803  .32325  .38836  .3451  .40859  Metetra
.714  .659  .794  r
.43653  .38497  .41494  .36707  .369  .3803  .39073  .39996  Metetra
.732  .672  .791  r
.48248  .36175  .46104  .34703  .41494  .36707  .43653  .38497  Metetra
.751  .689  .791  r
.52855  .32991  .50727  .31941  .46104  .34703  .48248  .36175  Metetra
.768  .707  .794  r
.5747  .29166  .55361  .28518  .50727  .31941  .52855  .32991  Metetra
.776  .72  .8  r
.62093  .25223  .60002  .24776  .55361  .28518  .5747  .29166  Metetra
.767  .719  .808  r
.66726  .21829  .64653  .21258  .60002  .24776  .62093  .25223  Metetra
.743  .704  .815  r
.71375  .19436  .69317  .18488  .64653  .21258  .66726  .21829  Metetra
.717  .686  .819  r
.76044  .1799  .74  .16659  .69317  .18488  .71375  .19436  Metetra
.701  .675  .821  r
.80733  .17069  .78704  .1553  .74  .16659  .76044  .1799  Metetra
.696  .672  .822  r
.85442  .16307  .83427  .14707  .78704  .1553  .80733  .17069  Metetra
.684  .657  .818  r
.2323  .39643  .21  .37818  .16466  .3816  .18711  .39955  Metetra
.684  .655  .816  r
.27768  .39315  .25551  .37453  .21  .37818  .2323  .39643  Metetra
.689  .655  .811  r
.32325  .38836  .30121  .36953  .25551  .37453  .27768  .39315  Metetra
.698  .657  .805  r
.369  .3803  .34711  .3618  .30121  .36953  .32325  .38836  Metetra
.711  .663  .8  r
.41494  .36707  .39318  .3498  .34711  .3618  .369  .3803  Metetra
.727  .673  .797  r
.46104  .34703  .43943  .33213  .39318  .3498  .41494  .36707  Metetra
.744  .686  .796  r
.50727  .31941  .48582  .30785  .43943  .33213  .46104  .34703  Metetra
.76  .702  .797  r
.55361  .28518  .53233  .27718  .48582  .30785  .50727  .31941  Metetra
.771  .715  .801  r
.60002  .24776  .57893  .2422  .53233  .27718  .55361  .28518  Metetra
.769  .719  .806  r
.64653  .21258  .62562  .20701  .57893  .2422  .60002  .24776  Metetra
.752  .71  .813  r
.69317  .18488  .67244  .17666  .62562  .20701  .64653  .21258  Metetra
.727  .693  .817  r
.74  .16659  .71941  .15458  .67244  .17666  .69317  .18488  Metetra
.707  .679  .82  r
.78704  .1553  .76658  .14049  .71941  .15458  .74  .16659  Metetra
.698  .673  .822  r
.83427  .14707  .81396  .13111  .76658  .14049  .78704  .1553  Metetra
.684  .663  .823  r
.21  .37818  .1875  .36086  .14203  .36461  .16466  .3816  Metetra
.685  .661  .821  r
.25551  .37453  .23316  .35683  .1875  .36086  .21  .37818  Metetra
.689  .66  .816  r
.30121  .36953  .279  .3516  .23316  .35683  .25551  .37453  Metetra
.697  .661  .811  r
.34711  .3618  .32503  .34405  .279  .3516  .30121  .36953  Metetra
.708  .665  .806  r
.39318  .3498  .37125  .33294  .32503  .34405  .34711  .3618  Metetra
.722  .673  .802  r
.43943  .33213  .41764  .31705  .37125  .33294  .39318  .3498  Metetra
.737  .684  .8  r
.48582  .30785  .46419  .29541  .41764  .31705  .43943  .33213  Metetra
.753  .698  .799  r
.53233  .27718  .51087  .2678  .46419  .29541  .48582  .30785  Metetra
.764  .71  .801  r
.57893  .2422  .55765  .23535  .51087  .2678  .53233  .27718  Metetra
.767  .717  .805  r
.62562  .20701  .60453  .20099  .55765  .23535  .57893  .2422  Metetra
.758  .713  .81  r
.67244  .17666  .65152  .16911  .60453  .20099  .62562  .20701  Metetra
.737  .699  .815  r
.71941  .15458  .69865  .14379  .65152  .16911  .67244  .17666  Metetra
.715  .684  .819  r
.76658  .14049  .74597  .12652  .69865  .14379  .71941  .15458  Metetra
.701  .675  .821  r
.81396  .13111  .79349  .11535  .74597  .12652  .76658  .14049  Metetra
.684  .666  .827  r
.1875  .36086  .16483  .3441  .11921  .34818  .14203  .36461  Metetra
.685  .665  .824  r
.23316  .35683  .21062  .33971  .16483  .3441  .1875  .36086  Metetra
.689  .664  .819  r
.279  .3516  .2566  .33424  .21062  .33971  .23316  .35683  Metetra
.696  .664  .814  r
.32503  .34405  .30277  .32678  .2566  .33424  .279  .3516  Metetra
.706  .667  .809  r
.37125  .33294  .34913  .31633  .30277  .32678  .32503  .34405  Metetra
.718  .673  .805  r
.41764  .31705  .39567  .3018  .34913  .31633  .37125  .33294  Metetra
.732  .683  .803  r
.46419  .29541  .44238  .28226  .39567  .3018  .41764  .31705  Metetra
.746  .694  .802  r
.51087  .2678  .48922  .25723  .44238  .28226  .46419  .29541  Metetra
.758  .705  .802  r
.55765  .23535  .53619  .22721  .48922  .25723  .51087  .2678  Metetra
.764  .713  .804  r
.60453  .20099  .58325  .19418  .53619  .22721  .55765  .23535  Metetra
.76  .713  .808  r
.65152  .16911  .63042  .1617  .58325  .19418  .60453  .20099  Metetra
.744  .703  .812  r
.69865  .14379  .67772  .1339  .63042  .1617  .65152  .16911  Metetra
.723  .689  .817  r
.74597  .12652  .72518  .11347  .67772  .1339  .69865  .14379  Metetra
.706  .677  .82  r
.79349  .11535  .77285  .1  .72518  .11347  .74597  .12652  Metetra
.685  .669  .829  r
.16483  .3441  .14196  .32766  .0962  .33206  .11921  .34818  Metetra
.686  .667  .826  r
.21062  .33971  .1879  .32294  .14196  .32766  .16483  .3441  Metetra
.69  .666  .822  r
.2566  .33424  .23402  .31725  .1879  .32294  .21062  .33971  Metetra
.696  .667  .817  r
.30277  .32678  .28034  .30984  .23402  .31725  .2566  .33424  Metetra
.704  .669  .812  r
.34913  .31633  .32684  .29986  .28034  .30984  .30277  .32678  Metetra
.715  .674  .808  r
.39567  .3018  .37353  .28638  .32684  .29986  .34913  .31633  Metetra
.727  .681  .805  r
.44238  .28226  .42038  .26852  .37353  .28638  .39567  .3018  Metetra
.74  .691  .803  r
.48922  .25723  .46739  .24565  .42038  .26852  .44238  .28226  Metetra
.752  .701  .803  r
.53619  .22721  .51453  .21787  .46739  .24565  .48922  .25723  Metetra
.76  .709  .804  r
.58325  .19418  .56178  .18642  .51453  .21787  .53619  .22721  Metetra
.76  .711  .806  r
.63042  .1617  .60914  .15405  .56178  .18642  .58325  .19418  Metetra
.75  .705  .81  r
.67772  .1339  .65661  .12456  .60914  .15405  .63042  .1617  Metetra
.731  .693  .814  r
.72518  .11347  .70423  .1013  .65661  .12456  .67772  .1339  Metetra
.713  .681  .818  r
.77285  .1  .75205  .08522  .70423  .1013  .72518  .11347  Metetra
.686  .671  .83  r
.14196  .32766  .1189  .3114  .07301  .3161  .0962  .33206  Metetra
.687  .669  .827  r
.1879  .32294  .16498  .30638  .1189  .3114  .14196  .32766  Metetra
.69  .668  .823  r
.23402  .31725  .21125  .30048  .16498  .30638  .1879  .32294  Metetra
.696  .668  .819  r
.28034  .30984  .25771  .29308  .21125  .30048  .23402  .31725  Metetra
.703  .67  .815  r
.32684  .29986  .30436  .28346  .25771  .29308  .28034  .30984  Metetra
.712  .674  .811  r
.37353  .28638  .3512  .2708  .30436  .28346  .32684  .29986  Metetra
.723  .68  .808  r
.42038  .26852  .39821  .25429  .3512  .2708  .37353  .28638  Metetra
.735  .688  .805  r
.46739  .24565  .44538  .23324  .39821  .25429  .42038  .26852  Metetra
.747  .697  .804  r
.51453  .21787  .49269  .20746  .44538  .23324  .46739  .24565  Metetra
.756  .704  .804  r
.56178  .18642  .54012  .17766  .49269  .20746  .51453  .21787  Metetra
.759  .708  .805  r
.60914  .15405  .58766  .14588  .54012  .17766  .56178  .18642  Metetra
.753  .706  .808  r
.65661  .12456  .63531  .11542  .58766  .14588  .60914  .15405  Metetra
.738  .697  .812  r
.70423  .1013  .6831  .08981  .63531  .11542  .65661  .12456  Metetra
.719  .685  .816  r
.75205  .08522  .73107  .07107  .6831  .08981  .70423  .1013  Metetra
.686  .672  .83  r
.1189  .3114  .09566  .29523  .04962  .3002  .07301  .3161  Metetra
.688  .671  .828  r
.16498  .30638  .14188  .28993  .09566  .29523  .1189  .3114  Metetra
.691  .67  .825  r
.21125  .30048  .1883  .28383  .14188  .28993  .16498  .30638  Metetra
.695  .67  .821  r
.25771  .29308  .2349  .27642  .1883  .28383  .21125  .30048  Metetra
.702  .671  .817  r
.30436  .28346  .2817  .26707  .2349  .27642  .25771  .29308  Metetra
.71  .674  .813  r
.3512  .2708  .32869  .25507  .2817  .26707  .30436  .28346  Metetra
.72  .679  .81  r
.39821  .25429  .37585  .23966  .32869  .25507  .3512  .2708  Metetra
.731  .686  .807  r
.44538  .23324  .42318  .22013  .37585  .23966  .39821  .25429  Metetra
.742  .693  .805  r
.49269  .20746  .47066  .19612  .42318  .22013  .44538  .23324  Metetra
.751  .701  .804  r
.54012  .17766  .51827  .16794  .47066  .19612  .49269  .20746  Metetra
.756  .705  .804  r
.58766  .14588  .56599  .13706  .51827  .16794  .54012  .17766  Metetra
.754  .705  .806  r
.63531  .11542  .61383  .10619  .56599  .13706  .58766  .14588  Metetra
.743  .699  .809  r
.6831  .08981  .66179  .07878  .61383  .10619  .63531  .11542  Metetra
.726  .688  .814  r
.73107  .07107  .70992  .05754  .66179  .07878  .6831  .08981  Metetra
.687  .673  .83  r
.09566  .29523  .07222  .27907  .02603  .2843  .04962  .3002  Metetra
.688  .672  .828  r
.14188  .28993  .11859  .27352  .07222  .27907  .09566  .29523  Metetra
.691  .671  .825  r
.1883  .28383  .16515  .26725  .11859  .27352  .14188  .28993  Metetra
.695  .671  .822  r
.2349  .27642  .21191  .25982  .16515  .26725  .1883  .28383  Metetra
.701  .672  .818  r
.2817  .26707  .25885  .25067  .21191  .25982  .2349  .27642  Metetra
.708  .674  .815  r
.32869  .25507  .30599  .23919  .25885  .25067  .2817  .26707  Metetra
.717  .678  .811  r
.37585  .23966  .3533  .22467  .30599  .23919  .32869  .25507  Metetra
.727  .684  .808  r
.42318  .22013  .40079  .20642  .3533  .22467  .37585  .23966  Metetra
.738  .691  .806  r
.47066  .19612  .44844  .18396  .40079  .20642  .42318  .22013  Metetra
.747  .697  .804  r
.51827  .16794  .49623  .15734  .44844  .18396  .47066  .19612  Metetra
.753  .702  .804  r
.56599  .13706  .54414  .12751  .49623  .15734  .51827  .16794  Metetra
.754  .704  .804  r
.61383  .10619  .59216  .09668  .54414  .12751  .56599  .13706  Metetra
.746  .699  .807  r
.66179  .07878  .6403  .068  .59216  .09668  .61383  .10619  Metetra
.732  .691  .811  r
.70992  .05754  .6886  .04454  .6403  .068  .66179  .07878  Metetra
P
p
.002  w
.97185  .2505  m
.97584  .51654  L
s
.97584  .51654  m
.69013  .30388  L
s
.69013  .30388  m
.68848  .02416  L
s
.68848  .02416  m
.97185  .2505  L
s
.68848  .02416  m
.0284  .13184  L
s
.0284  .13184  m
.02416  .40508  L
s
.02416  .40508  m
.69013  .30388  L
s
.69013  .30388  m
.68848  .02416  L
s
P
p
p
.002  w
.68848  .02416  m
.0284  .13184  L
s
P
p
.002  w
.68848  .02416  m
.68851  .03019  L
s
P
[(-1)] .68841  .01208  .0059  1 Mshowa
p
.002  w
.52013  .05162  m
.52013  .05766  L
s
P
[(-0.5)] .52012  .03954  .00042  1 Mshowa
p
.002  w
.35403  .07872  m
.354  .08475  L
s
P
[(0)] .35409  .06664  -0.00498 1 Mshowa
p
.002  w
.19014  .10545  m
.19007  .11149  L
s
P
[(0.5)] .19026  .09338  -0.0103 1 Mshowa
p
.002  w
.0284  .13184  m
.02831  .13787  L
s
P
[(1)] .02859  .11976  -0.01554 1 Mshowa
p
.001  w
.65463  .02968  m
.65464  .0333  L
s
P
p
.001  w
.62087  .03519  m
.62088  .03881  L
s
P
p
.001  w
.5872  .04068  m
.58721  .0443  L
s
P
p
.001  w
.55362  .04616  m
.55362  .04978  L
s
P
p
.001  w
.48673  .05707  m
.48673  .06069  L
s
P
p
.001  w
.45342  .0625  m
.45342  .06612  L
s
P
p
.001  w
.4202  .06792  m
.42019  .07154  L
s
P
p
.001  w
.38707  .07333  m
.38706  .07695  L
s
P
p
.001  w
.32108  .08409  m
.32105  .08771  L
s
P
p
.001  w
.28821  .08945  m
.28818  .09308  L
s
P
p
.001  w
.25543  .0948  m
.2554  .09842  L
s
P
p
.001  w
.22274  .10013  m
.22271  .10376  L
s
P
p
.001  w
.15762  .11076  m
.15758  .11438  L
s
P
p
.001  w
.12519  .11605  m
.12514  .11967  L
s
P
p
.001  w
.09284  .12132  m
.09279  .12495  L
s
P
p
.001  w
.06058  .12659  m
.06053  .13021  L
s
P
[(    M)(Log[-])(    T)] .35439  .00625  -0.00498 1 Mshowa
P
p
p
.002  w
.33898  .60686  m
.02416  .40508  L
s
P
p
.002  w
.33898  .60686  m
.33902  .60082  L
s
P
[(0)] .33892  .61894  .00538  -1 Mshowa
p
.002  w
.26354  .55851  m
.26359  .55247  L
s
P
[(2)] .26344  .57058  .0078  -1 Mshowa
p
.002  w
.18598  .5088  m
.18604  .50276  L
s
P
[(4)] .18586  .52087  .0103  -1 Mshowa
p
.002  w
.10622  .45767  m
.1063  .45164  L
s
P
[(6)] .10606  .46975  .01288  -1 Mshowa
p
.002  w
.02416  .40508  m
.02425  .39904  L
s
P
[(8)] .02397  .41715  .01554  -1 Mshowa
p
.001  w
.32406  .5973  m
.32408  .59367  L
s
P
p
.001  w
.30905  .58768  m
.30908  .58406  L
s
P
p
.001  w
.29397  .57801  m
.29399  .57439  L
s
P
p
.001  w
.27879  .56829  m
.27882  .56466  L
s
P
p
.001  w
.2482  .54868  m
.24823  .54505  L
s
P
p
.001  w
.23277  .53879  m
.23281  .53517  L
s
P
p
.001  w
.21726  .52885  m
.2173  .52522  L
s
P
p
.001  w
.20167  .51885  m
.2017  .51523  L
s
P
p
.001  w
.17021  .49869  m
.17025  .49506  L
s
P
p
.001  w
.15434  .48852  m
.15439  .4849  L
s
P
p
.001  w
.13839  .4783  m
.13844  .47467  L
s
P
p
.001  w
.12235  .46801  m
.1224  .46439  L
s
P
p
.001  w
.08999  .44727  m
.09004  .44365  L
s
P
p
.001  w
.07367  .43682  m
.07372  .43319  L
s
P
p
.001  w
.05726  .4263  m
.05732  .42267  L
s
P
p
.001  w
.04076  .41572  m
.04081  .4121  L
s
P
[(g2)] .18523  .58126  .0103  -1 Mshowa
P
p
p
.002  w
.0284  .13184  m
.02416  .40508  L
s
P
p
.002  w
.0283  .13829  m
.03426  .13732  L
s
P
[(0)] .01638  .14023  1 -0.16287 Mshowa
p
.002  w
.02688  .22959  m
.03285  .22864  L
s
P
[(0.1)] .01496  .23149  1 -0.15913 Mshowa
p
.002  w
.02546  .32144  m
.03142  .32052  L
s
P
[(0.2)] .01352  .3233  1 -0.15537 Mshowa
p
.001  w
.02802  .1565  m
.0316  .15592  L
s
P
p
.001  w
.02774  .17474  m
.03131  .17417  L
s
P
p
.001  w
.02745  .19301  m
.03103  .19243  L
s
P
p
.001  w
.02717  .21129  m
.03075  .21072  L
s
P
p
.001  w
.0266  .24792  m
.03018  .24735  L
s
P
p
.001  w
.02631  .26627  m
.02989  .2657  L
s
P
p
.001  w
.02603  .28464  m
.02961  .28408  L
s
P
p
.001  w
.02574  .30303  m
.02932  .30247  L
s
P
p
.001  w
.02517  .33988  m
.02875  .33932  L
s
P
p
.001  w
.02488  .35834  m
.02846  .35778  L
s
P
p
.001  w
.0246  .37681  m
.02818  .37627  L
s
P
p
.001  w
.02431  .39532  m
.02789  .39477  L
s
P
[(O)(-)(M)] -0.0453 .27913  1 -0.15757 Mshowa
P
MathPictureEnd
end
showpage
